\documentclass[twocolumn]{aastex701}

\usepackage{longtable}
\usepackage{booktabs}
\usepackage{amsmath}

\begin{document}

\title{A new measurement of the FRB DM-galaxy cross correlation and a first joint analysis with the kinematic SZ effect} 

\author[orcid=0009-0008-5043-6220]{Samuel McCarty}
\affiliation{Center for Astrophysics $|$ Harvard \& Smithsonian, 60 Garden Street, Cambridge MA 02138, USA}
\email[show]{samuel.mccarty@cfa.harvard.edu}  

\author[orcid=0000-0002-7587-6352]{Liam Connor} 
\affiliation{Center for Astrophysics $|$ Harvard \& Smithsonian, 60 Garden Street, Cambridge MA 02138, USA}
\email{liam.connor@cfa.harvard.edu}

\author[orcid=0000-0002-4477-3625]{Kritti Sharma} 
\affiliation{Cahill Center for Astronomy and Astrophysics, MC 249-17 California Institute of Technology, Pasadena CA 91125, USA}
\email{}

\author[orcid=0000-0003-4992-7854]{Simone Ferraro}
\affiliation{Lawrence Berkeley National Laboratory, One Cyclotron Road, Berkeley, CA 94720, USA}
\affiliation{Berkeley Center for Cosmological Physics, Department of Physics,
University of California, Berkeley, CA 94720, USA}
\email{sferraro@lbl.gov}

\author[orcid=0000-0002-7252-5485]{Vikram Ravi} 
\affiliation{Cahill Center for Astronomy and Astrophysics, MC 249-17 California Institute of Technology, Pasadena CA 91125, USA}
\affiliation{Owens Valley Radio Observatory, California Institute of Technology, Big Pine CA 93513, USA}
\email{}

\author[0000-0001-8356-2014]{Elisabeth Krause} 
\affiliation{Department of Astronomy/Steward Observatory, University of Arizona, 933 North Cherry Avenue, Tucson, AZ 85721, USA}
\affiliation{Department of Physics, University of Arizona, 1118 E Fourth Street, Tucson, AZ 85721, USA}
\email{}

\author[orcid=0000-0002-2312-3121]{Boryana Hadzhiyska} 
\affiliation{Institute of Astronomy, Madingley Road, Cambridge, CB3 0HA, UK}
\affiliation{Kavli Institute for Cosmology Cambridge, Madingley Road, Cambridge, CB3 0HA, UK}
\email{}

\author[0000-0003-3244-2711]{Vishnu Balakrishnan} 
\affiliation{Center for Astrophysics $|$ Harvard \& Smithsonian, 60 Garden Street, Cambridge MA 02138, USA}
\email{}  

\author[0000-0002-4119-9963]{Casey Law} 
\affiliation{Cahill Center for Astronomy and Astrophysics, MC 249-17 California Institute of Technology, Pasadena CA 91125, USA}
\email{}  

\author[0000-0001-5504-229X]{Pranav Sanghavi} 
\affiliation{Center for Astrophysics $|$ Harvard \& Smithsonian, 60 Garden Street, Cambridge MA 02138, USA}
\email{}  

\author[0000-0002-6823-2073]{Kaitlyn Shin} 
\affiliation{Cahill Center for Astronomy and Astrophysics, MC 249-17 California Institute of Technology, Pasadena CA 91125, USA}
\email{}  

\begin{abstract}

Understanding the distribution of cosmic baryons is crucial for advancing cosmology and galaxy formation. Localized Fast Radio Bursts (FRBs) are a promising new probe of the diffuse gas, but measurements thus far have been limited by small sample size. In this work, we measure the FRB dispersion measure (DM)-galaxy cross correlation with 130 localized FRBs, including new spectroscopic host redshifts from the DSA-110, and the DESI Legacy Survey Bright Galaxy Sample. We detect this signal at the highest significance to date ($6.5\sigma$) in both configuration and harmonic space.  By comparing to simulations and a halo model, we demonstrate that this statistic can constrain the strength of baryonic feedback. Our measurements disfavor a no-feedback scenario in which gas traces the underlying dark matter at $\sim$\,$9\sigma$. In addition to small scale feedback, this represents one of the first direct measurements of the clustering of electrons on scales larger than a few Mpc. We compare our measurement to recent kinematic Sunyaev-Zeldovich effect (kSZ) studies, which show good agreement despite the differences in observational methods and sample selection. Finally, we make a first attempt at breaking the kSZ optical depth degeneracy with FRBs, jointly measuring the growth rate of large-scale structure $f\sigma_8$. With future samples, FRB two-point statistics will provide precision constraints on the distribution of cosmic baryons and complement other probes. 

\end{abstract}

\section{Introduction} 

The distribution of cosmic baryons outside of galaxies is a leading open question in astrophysics and cosmology. Greater than 90\% of the baryonic matter in the Universe exists as a hot, diffuse, and difficult to detect gas \citep{Cen_1999,Connor2025}. This gas is intimately tied to galaxy formation and evolution through accretion and feedback \citep{McCarthy_2010,Tumlinson_2017,peeples2019understandingcircumgalacticmediumcritical}. Thus, measurements of these baryons can serve as a benchmark for galaxy formation simulations and their subgrid models. Further, astrophysical processes such as Active Galactic Nuclei (AGN) feedback redistribute the baryons relative to the underlying dark matter. This manifests as a suppression of the matter power spectrum on the scales of dark matter halos, which hinders cosmological measurements of the small scale structure \citep{van_Daalen_2011,vanDaalen2020}. Pushing to smaller scales is essential for maximizing the constraining power of cosmological measurements and testing deviation from the prevailing cosmological model. At present, baryonic feedback is a key uncertainty in weak lensing analyses, and can bias inference of cosmological parameters \citep{Chisari_2019,Amon_2022}. Using independent probes of the baryons to calibrate feedback is a promising approach to maximize the utility of upcoming Stage IV weak lensing experiments by the \textit{Euclid} satellite \citep{EuclidI}, 
the Vera C. Rubin Observatory \citep{Ivezi_2019},
and the \textit{Nancy Roman Space Telescope} \citep{spergel2013widefieldinfraredsurveytelescopeastrophysics}. 

Over the past two decades, several observational techniques to study the cosmic gas around galaxies (the circumgalactic medium, CGM; intra-group medium, IGrM; and intra-cluster medium, ICM) have matured. The thermal and kinematic Sunyaev-Zeldovich effects (tSZ, kSZ; \cite{SZoriginal}), are sensitive to the free electron column, weighted by temperature and bulk velocity respectively. These tools have revealed halo gas in unprecedented detail \citep{Schaan_2021, Amodeo_2021, Das_2023, bigwood2024weaklensingcombinedkinetic, hadzhiyska2025evidencelargebaryonicfeedback,Guachalla_2025,qu2026precisionkinematicsunyaevzeldovichmeasurements,hadzhiyska2026precisionkinematicsunyaevzeldovichmeasurements}. X-ray observations are a gold standard for measuring the gas in the inner regions of massive systems, with recent eROSITA data providing tight constraints on halo gas fractions \citep{Zhang_2024,Popesso_2026}. Between kSZ and X-ray studies, a new picture has emerged of strong AGN feedback \citep{siegel2025jointxraykineticsunyaevzeldovich,siegel2025suppressionmatterpowerspectrum,hadzhiyska2025missingbaryonsrecoveredmeasurement}, although significant uncertainties remain. Each probe is subject to limitations: modeling the tSZ signal is challenging due to numerous systematics. The kSZ effect is limited by uncertainty in the halo velocity, and the signal is intrinsically faint compared to primary CMB fluctuations. X-ray studies are prone to selection effects and assumptions about gas metallicity and phase. Each is further restricted in the scales and halo masses they can measure.

A new probe, Fast Radio Bursts (FRBs), provide a relatively simple tool for measuring cosmic baryons. FRBs are luminous, millisecond-duration radio pulses, typically of extragalactic origin, whose signal is dispersed as it travels through ionized cosmic gas \citep{Lorimer_2007, Petroff_2019, Cordes_2019}. The dispersion measure (DM) of an FRB encodes the line-of-sight electron column and can thus can be used to study intervening gas. One-point statistics of the DM have successfully constrained the baryon content of the IGM and CGM as well as baryonic feedback \citep{Macquart_2020,Connor2025,reischke2025measurementbaryonicfeedbackfast,sharma2025hydrodynamicalsimulationsbasedmodelconnects,sharma2026signaturessuppressedmatterclustering}, and targeted studies provide unique measurements of individual systems in contrast to the stacking that is typically required for other probes \citep{Connor_2023,Anna_Thomas_2025,lanman2026constraininggasmassfractions,mccarty2026cgmlocaluniversefrbs}.

\begin{figure*}
    \centering
    \includegraphics[width=0.85\textwidth]{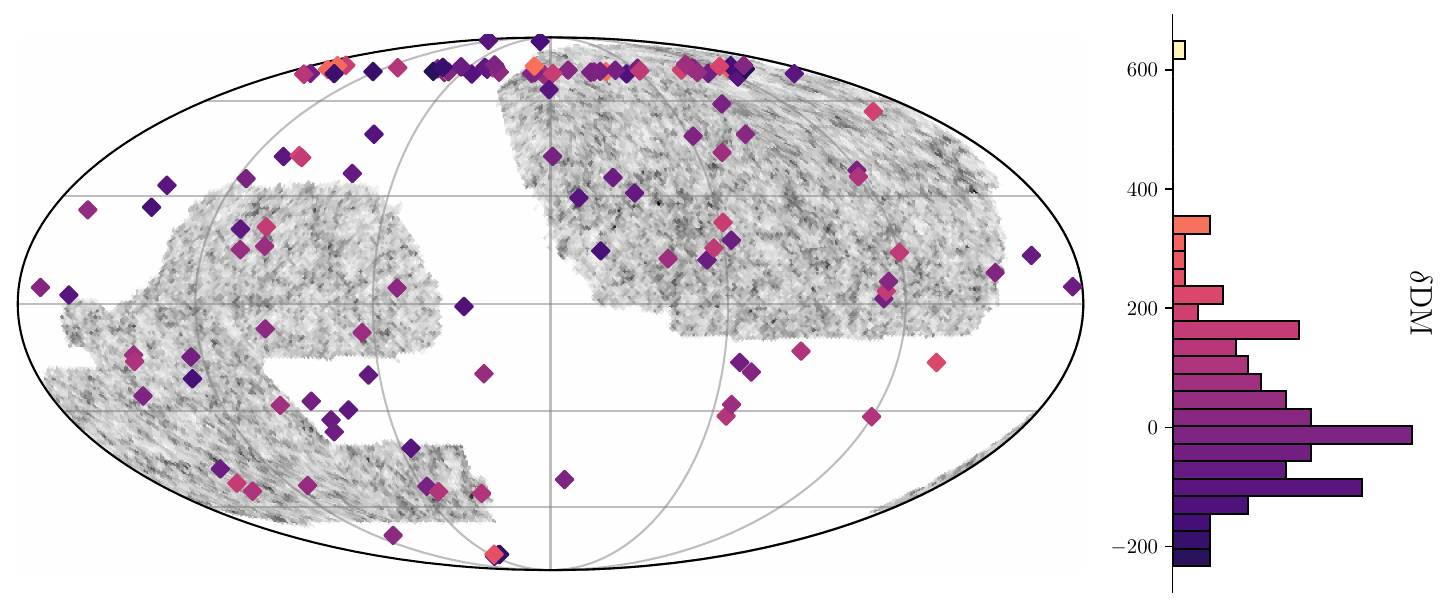}
    \caption{The sky distribution of FRBs and galaxies in our sample. In grey is a healipix map of galaxy number counts and overlaid are the positions of the FRBs, color coded by their $\delta \mathrm{DM}$. }
    \label{fig:sky}
\end{figure*}

However, DM one-point statistics do not directly measure the spatial distribution of baryons and are subject to biases from host galaxy DM. There is significant interest in DM two-point statistics \citep{Rafiei_Ravandi_2021,connor2022observedimpactgalaxyhalo, Wu_2023, wang2025measurementdispersionunicodex2013galaxycrosspowerspectrum, leung2025nullingbaryonicfeedbackweak, sharma2025probingbaryonicfeedbackcosmology, hussaini2025correlationfrbdispersionmeasure, takahashi2025measurementangularcrosscorrelationcosmological, sharma2026backlightingcosmicwebfast, shirasaki2026crosscorrelatinggalaxiescosmicdispersion, Wayland_2026}, which can map baryons as a function of scale or separation from other tracers. In particular, the correlation between galaxies and DM reveals the radial profile of the gas, which depends sensitively on feedback. Further, this statistic allows us to study feedback as a function of mass, redshift and other galaxy properties, while the one-point statistic intrinsically integrates over them. However, the dearth of known FRBs has limited useful measurements of this correlation until recently. 

In this work, we use a large sample of localized FRBs to measure the DM-galaxy correlation at the highest significance to date. In contrast to armcimute-scale ``unlocalized" FRBs (e.g. \cite{wang2025measurementdispersionunicodex2013galaxycrosspowerspectrum, sharma2026backlightingcosmicwebfast}), arcsecond-scale localized FRBs with host galaxy identifications are desirable because knowledge of the FRB redshift allows the mean DM-$z$ relation to be subtracted (Section \ref{sec:theory}), significantly reducing noise. Further, precise localization enables measurements of small scales and the host redshift provides a clean identification of foreground galaxies. 

We further explore, for the first time with data, the synergy between FRBs and the kSZ effect. kSZ studies have been extremely successful at measuring the baryons in the past decade. However, these measurements are fundamentally limited by the fact that the kSZ signal measures the product of electron density and velocity. When studying electrons, the velocity field must be reconstructed by surrounding large scale structure, which is becoming a dominant source of uncertainty \citep{hadzhiyska2023velocityreconstructioneradesi,Ried_Guachalla_2024}. Using the kSZ as a probe of velocities is similarly limited by this so-called ``optical depth degeneracy". It was first proposed by \cite{Madhavacheril_2019} that FRBs, which measure only the electron density, could be used to break this degeneracy and allow for unbiased measurements of cosmological velocities. Further, while the kSZ signal is excellent at measuring small scales, it cannot measure large scales ($>$ a few Mpc) due to primary CMB fluctuations. FRB measurements, on the other hand, have better constraining power on large scales, as we'll show. Therefore, a combination of kSZ and FRB signals is a promising path towards a complete picture of cosmic baryons.

In Section \ref{sec:theory} and the Appendix, we introduce the basic principles and modeling framework used throughout the paper. Section \ref{sec:data} describes the data used in our measurement, Section \ref{sec:measurement} outlines our measurement procedure, and Section \ref{sec:fit} details how we fit the models to our measuremnt. Section \ref{sec:results} presents our results and a discussion of their significance. Finally, in Section \ref{sec:conclusion} we provide a brief conclusion. Throughout this work we assume the best-fit Planck 2018+BAO cosmology \citep{Planck18}.

\section{Theory \& Background}
\label{sec:theory}

The theoretical framework underpinning our measurement and analysis is detailed in Sections \ref{sec:dmstats}, \ref{sec:hm}, \ref{sec:ksztheory} of the Appendix. Here we provide an overview of the core concepts. 

The FRB DM gives an electron column along the line of sight. We can define a per FRB DM contrast, $\delta \mathrm{DM}$, by subtracting the mean DM to the FRB redshift. Then, $\delta \mathrm{DM}$ traces the clustering of electrons:

\begin{equation}
    \mathrm{\delta DM}(\mathbf{\hat{n}})  = \int_0^{\chi_f} \frac{\overline{n_e}(\chi)}{(1+z)^2} \delta_e(\mathbf{\hat{n}},\chi)d\chi.
\end{equation}

\noindent $\chi_f$ is the comoving radial distance of the FRB, $z$ is the redshift at comoving distance $\chi$, and $n_e(\chi)$ is the electron density at a distance $\chi$ in physical (not comoving) units. $\delta_e(\mathbf{\hat{n}},\chi)$ is the electron overdensity at line-of-sight direction $\mathbf{\hat{n}}$ and distance $\chi$, which contains the cosmological and astrophysical information we wish to extract. 

By cross correlating the $\mathrm{\delta DM}$ field with the positions of galaxies we can isolate the clustering of electrons around those galaxies, which measures gas in halos and surrounding large scale structure. This is a projected configuration space count-scalar correlation, similar to weak lensing. The resulting correlation function is an integral of the galaxy-electron correlation function:

\begin{equation}
\label{eqn:xitodm}
    \xi_{g,\delta\mathrm{DM}}(r) =  \int_0^{\chi_f} d\chi \frac{\overline{n_e}(\chi)}{(1+z)^2} \xi_{g,e}\left(\sqrt{(\chi-\chi_g)^2+r^2}, z\right),
\end{equation}

\noindent where $r$ is the transverse distance between the galaxy and the FRB line of sight. 

To model the correlation function we use a halo model. The premise is to regard all fields as being associated to dark matter halos. A one-halo term accounts for correlations within an individual halo, and a two-halo term accounts for correlations between the fields in distinct halos. In our case, the two fields are galaxies and electrons. The galaxy field is determined by a Halo Occupation Distribution (HOD), which connects galaxies to their host dark matter halos. For studies of cosmic baryons, the interesting ingredient is the halo electron profile. By parameterizing this profile and fitting to our measurements, we can learn about feedback physics. See Section \ref{sec:hm} for details.

Finally, we wish to compare our measurements to the kSZ effect. The kSZ effect arises from the scattering of CMB photons off halo electrons, which are moving with a bulk velocity towards overdense regions of the Universe. The kSZ signal reduces to (Section \ref{sec:ksztheory}):

\begin{equation}
    \frac{\Delta T_\mathrm{kSZ}}{T_\mathrm{CMB}} \approx - \tau \frac{v_\mathrm{halo}}{c},
\end{equation}

\noindent where $\Delta T_\mathrm{kSZ}/T_\mathrm{CMB}$ is the fractional change in CMB temperature, $\tau$ is the optical depth of the halo, and $v_\mathrm{halo}$ is the halo velocity. Dispersion measure can be converted to optical depth trivially (Equation \ref{eqn:dtau}), allowing us to compare the two signals. 

The kSZ optical depth degeneracy can be formulated in two ways (Section \ref{sec:ksztheory}). When the kSZ signal is used to study electrons by estimating the velocities from surrounding large scale structure, the $\tau$ signal is degenerate in $r/\sigma_v$. $r$ is a correlation coefficient quantifying how well the velocities are reconstructed. $\sigma_v$ is the characteristic velocity dispersion of the sample, which by linear theory is proportional to $f\sigma_8$ (i.e. the growth rate of large-scale structure). On the other hand, when the goal of the study is to measure the velocities to infer cosmology, the electrons become a nuisance parameter encoded by the kSZ velocity bias factor $b_v$. $b_v$ is essentially an integral over the electron-galaxy power spectrum. We will use the kSZ effect and our FRB measurement to constrain both $f\sigma_8$ (or $r$) and $b_v$. 

\section{Data}
\label{sec:data}

\subsection{FRB Sample}
\label{sec:frbsample}

We start with a list of 153 FRBs with reported host redshifts from the literature. We then apply the following cuts:
First, we exclude all FRBs with galactic latitude $|b|<5$\textdegree, because we expect residual errors from subtracting $\mathrm{DM_{MW}}$ to be large near the plane. We exclude FRBs with known dwarf galaxy hosts, which tend to have high excess host DM (FRBs 20121102A \citep{Tendulhar2017}, 20190520B \citep{Niu_2022}, and 20210117A \citep{Shannon_2025}), as well as FRB 20190417A, which has a very large host DM \citep{moroianu2025milliarcsecondlocalizationassociatesfrb}. Following \cite{leung2025stellarmassdispersionmeasurecorrelations}, we exclude the so-called ``low-DM" localizations, which could bias the measurement because their identification depended explicitly on their DM: FRBs 20181030A, 20180814A, 20181220A, 20181223C, 20190110C, 20190303A, 20190418A, 20190425A, 20191106C, and 20200223B \citep{Bhardwaj_2021,Michilli_2023,bhardwaj2023hostgalaxiesnearbychimefrb,Ibik2024}. From \cite{frbcollaboration2025cataloglocaluniversefast} we exclude host galaxy associations that are less than 90\% confident. 

\begin{figure}
    \centering
    \includegraphics[width=0.95\linewidth]{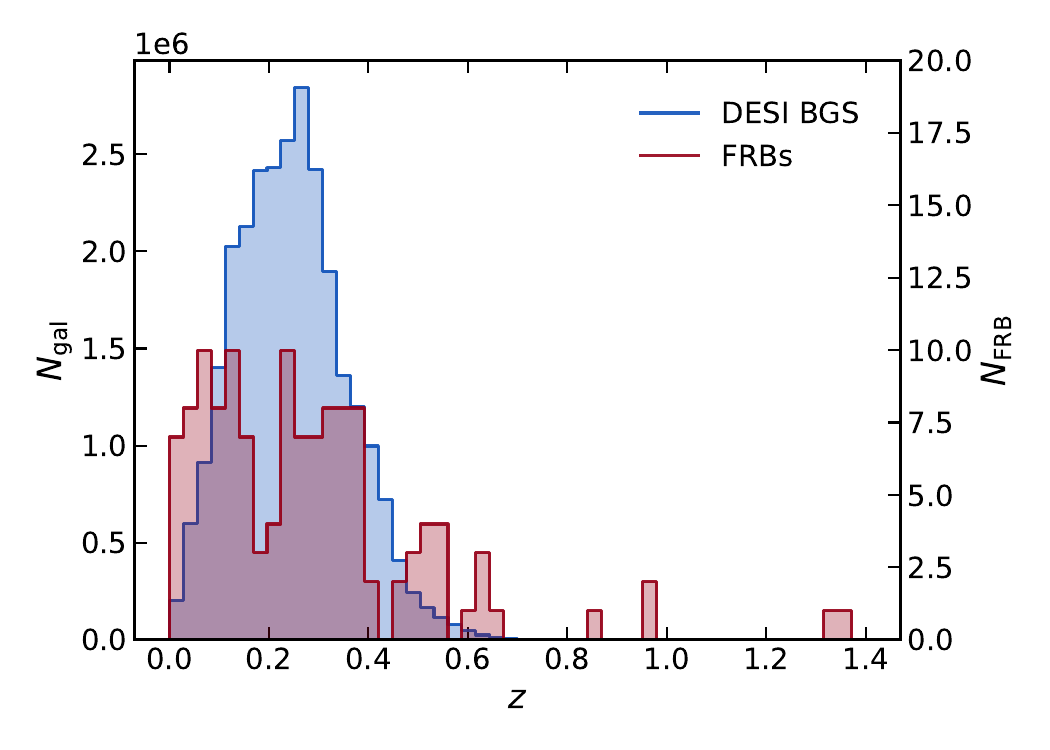}
    \caption{The redshift distributions of the 130 FRBs and 27.3 million DESI BGS galaxies in our sample.}
    \label{fig:zdist}
\end{figure}

\begin{figure}
    \centering
    \includegraphics[width=0.95\linewidth]{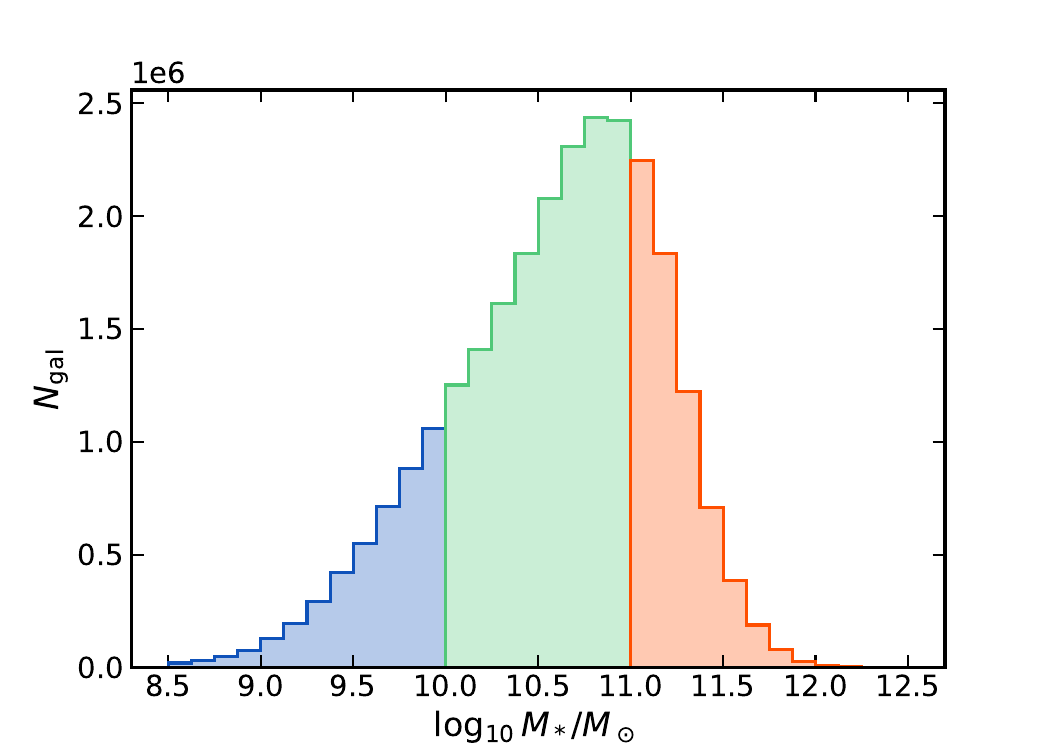}
    \caption{DESI BGS stellar mass distribution. Highlighted in color are the three mass bins we consider: all BGS galaxies, $\mathrm{log}_{10}M_*/M_\odot>10$, and $\mathrm{log}_{10}M_*/M_\odot>11$.}
    \label{fig:mdist}
\end{figure}

During our analysis, we also make two ad hoc cuts that we justify here. FRB 20240304B is excluded because its very high redshift $z\approx2$ means that it has an outsized impact on the signal \citep{caleb2025fastradioburst3}. 
Further, we find that the average $\mathrm{\delta DM}$ of FRBs in the southern portion of DESI (see next section) is quite high ($>100$ pc/cm$^{-3}$), which increases the noise in our measurement. For this reason, we cut FRBs with $\mathrm{\delta DM}>300$ in the south: FRBs 20220610A, 20240310A, 20230222A, 20251130A, 20220224C, 20230907D and 20231020B \citep{Shannon_2025,frbcollaboration2025cataloglocaluniversefast,2026ATel17619....1G,pastormarazuela2025localisationhostgalaxyidentification}. We do not expect this to bias our measurement significantly (Section \ref{sec:issues}).

In addition to localized FRBs from the literature, we include 11 new FRBs discovered by the DSA-110 and so far unpublished. These are detailed in Section \ref{sec:newdsa110}. The total number of FRBs in our sample after cuts is 130. Note that the actual number of FRBs incorporated into a given measurement will be smaller. We do not explicitly cut FRBs that are not in the footprint of DESI because those near the edge still contribute to the signal, but those far from the edge will not. Further, we apply various redshift cuts to the galaxies (Section \ref{sec:measurement}), which sets a minimum FRB redshift due to the requirement of foreground galaxies. The FRB sample and basic properties are listed in Table \ref{tab:sample}.

\subsection{Galaxy Sample}
\label{sec:gals}

\begin{figure*}
    \centering
    \includegraphics[width=0.9\linewidth]{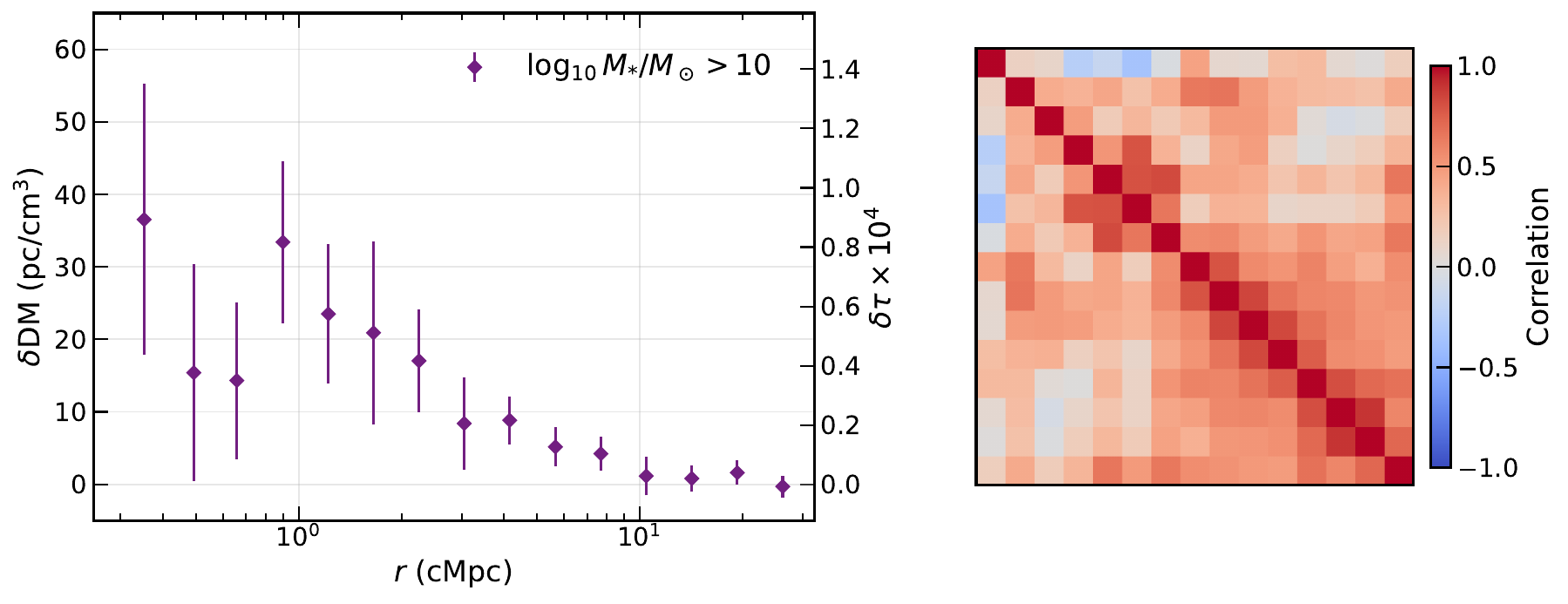}
    \caption{\textit{Left:} the measured correlation as a function of separation in comoving Mpc for BGS galaxies with $M_*>10^{10}\,M_\odot$. We show the corresponding optical depth $\delta \tau$ using the conversion in Equation \ref{eqn:dtau}. \textit{Right:} the correlation matrix.}
    \label{fig:biggtr10}
\end{figure*}

The Dark Energy Spectroscopic Instrument is a robotic fiber spectrograph operating on the Mayall 4-meter telescope at Kitt Peak National Observatory \citep{DESIoverview}. DESI will survey approximately 17,000 deg$^2$ of the sky, obtaining the largest spectroscopic galaxy sample to date. The target selection for DESI relies on the DESI Legacy Imaging Surveys (LS), which combines data from the Mayall z-band Legacy Survey, the Dark Energy Camera Legacy Survey, and the Beijing-Arizona Sky Survey with a resulting photometric catalog of roughly one billion objects \citep{DESILS}. The targets are divided into several tracers, with rigorous selection criteria that produce a nearly uniform density of sources across the survey footprint and redshift range \citep{Adame2025}. In this work we use the Bright Galaxy Sample (BGS), which is the low redshift ($z\lesssim0.6$) sample \citep{BGS}. BGS is the obvious choice considering the redshift distribution of the FRBs. 

Starting with the BGS sample from LS Data Release 9\footnote{\url{https://data.desi.lbl.gov/public/ets/target/catalogs/dr9/1.1.1/targets/main/resolve/bright/}}, we cross-match with the photometric redshift and stellar mass catalogs of \citep{Zhou_2023pz} and \citep{Zhou_2023sm} respectively.  Both the photometric redshifts and stellar masses are estimated with a Random Forest algorithm, reaching statistical uncertainties of $\sigma_z\approx2\%$ and $<0.2\,$dex for the BGS. We use both the BGS Bright and BGS Faint samples. Following \cite{sharma2026backlightingcosmicwebfast}, we exclude the region DEC $< -10.5^\circ$ and $120^\circ<$RA$ < 260^\circ$ that is poorly connected to the rest of the footprint. We cut galaxies with $\Delta z/(1+z) > 0.1$, where $\Delta z$ is the redshift uncertainty estimate. While a previous study restricted their analysis to the north of the DESI LS \citep{wang2025measurementdispersionunicodex2013galaxycrosspowerspectrum}, here we use both the north and south regions. Despite being observed with different instruments, the sample selection ensures that the sample is nearly uniform across the regions and we expect any residual systematics to be well below the current statistical uncertainty. 

 In Figure \ref{fig:sky} we show the sky distribution of the FRBs and galaxies. After cuts, we have a sample of 27.3 million galaxies. The redshift distribution of the galaxies and the FRBs is shown in Figure \ref{fig:zdist}, and the stellar mass distribution is shown in Figure \ref{fig:mdist}. We consider three galaxy stellar mass cuts: all BGS galaxies, log$_{10}M_*/M_\odot>10$, and log$_{10}M_*/M_\odot>11$. The average $(M_*,\,z)$ is ($8.0\times10^{10}M_\odot,\,0.25$), ($9.1\times10^{10}M_\odot,\,0.27$) and ($1.8\times10^{11}M_\odot,\,0.30$) respectively. We note that because the FRBs and galaxies overlap in redshift, the correlation will select a lower redshift subset of the galaxies, which is discussed later in this work. 

We also make use of a random catalog to control for systematics and subtract any DC offset in the signal. The random catalog\footnote{\url{https://data.desi.lbl.gov/public/ets/target/catalogs/dr9/0.48.0/randoms/resolve/}} is generated by sampling pixel locations during observations for the LS. Each random is given a redshift drawn from the redshift distribution of the real galaxies for a given mass bin. 

\subsection{Simulations}
\label{sec:sims}

To aid interpretation of our measurements, we compare to the predictions of hydrodynamical simulations. Specifically, we use the DM and subhalo lightcone outputs of the FLAMINGO simulation suite \citep{FLAM1,FLAM2,FLAM3}\footnote{\url{https://dataweb.cosma.dur.ac.uk:8443/flamingo/lightcones/index.html}}. The DM lightcones are available as HEALPIX maps in redshift shells. We use both the L1\_m9 and L1\_m9\_fgas-8sigma FLAMINGO simulations, corresponding to the fiducial model and an extreme feedback model both at medium resolution in a 1 Gpc box. To test a no-feedback scenario, we further use the HEALPIX lightcone dark matter maps to simulate FRBs whose DM traces the dark matter. In general, matching a real galaxy population to simulation subhalos is difficult without a full forward model of the survey, we will use an approach that is akin to abundance matching and rely on external calibration of the BGS sample. Our procedure is as follows:

\begin{itemize}
    \item To simulate FRBs, we create cumulative $\delta \mathrm{DM}$ maps by subtracting the mean DM of each redshift shell and then summing the shells. We sample random pixels with the same redshift distribution as our observed FRBs. To create ``no-feedback" FRBs, we turn the dark matter maps into overdensity maps and then multiply by the mean DM in the same redshift shell, before repeating the same sampling process as the normal FRBs. 
    \item For the galaxies, we first measure the observed redshift distribution in several redshift bins. We match this distribution in the simulations by selecting the same number of lightcone subhalos in each bin, keeping the most massive subhalos by stellar mass. After correcting for the observed sky fraction, this is equivalent to matching the number density of galaxies over redshift, assuming the sample is stellar mass limited. 
    \item Then, for each stellar mass cut in our measurements we apply a stellar mass cut on the simulated galaxies that leaves a population with approximately the expected mean halo mass of the observed galaxies. The expected mean halo mass is taken from the CMB lensing fits by \cite{hadzhiyska2026precisionkinematicsunyaevzeldovichmeasurements} on the BGS galaxies.  Because the stellar mass in a simulated subhalo does not necessarily correspond to the measured stellar mass of a real galaxy, the stellar mass cuts applied in the simulation are not the same as those applied to the data. 
    \item  Given a simulated set of FRBs and galaxies, we apply the same measurement pipeline as our real data, which is described in the next section.
\end{itemize}

We expect inevitable differences between our simulated galaxy population and the BGS, which we mitigate by fitting the simulation correlation function to our data on large scales, see Section \ref{sec:fit}. We note that, in the one-halo regime, changing feedback strength will be degenerate with halo mass, which is relevant if the halo mass of our simulated sample is biased. 

\section{Measurement}
\label{sec:measurement}

In this section we describe our measurement procedure, including covariance estimates and how we quantify detection significance. Section \ref{sec:estimator} deals with the FRB DM-galaxy correlation in real (configuration) space. Section \ref{sec:clestimator} details the harmonic space angular power spectrum measurement. The kSZ optical depth signal is estimated using a Compensated Aperture Photometry (CAP) fiter (Section \ref{sec:ksztheory}); we consider an estimator for $\tau^\mathrm{CAP}_\mathrm{FRB}$ to enable direct comparison with kSZ studies in Section \ref{sec:cap}.

\begin{figure}
    \centering
    \includegraphics[width=1.0\linewidth]{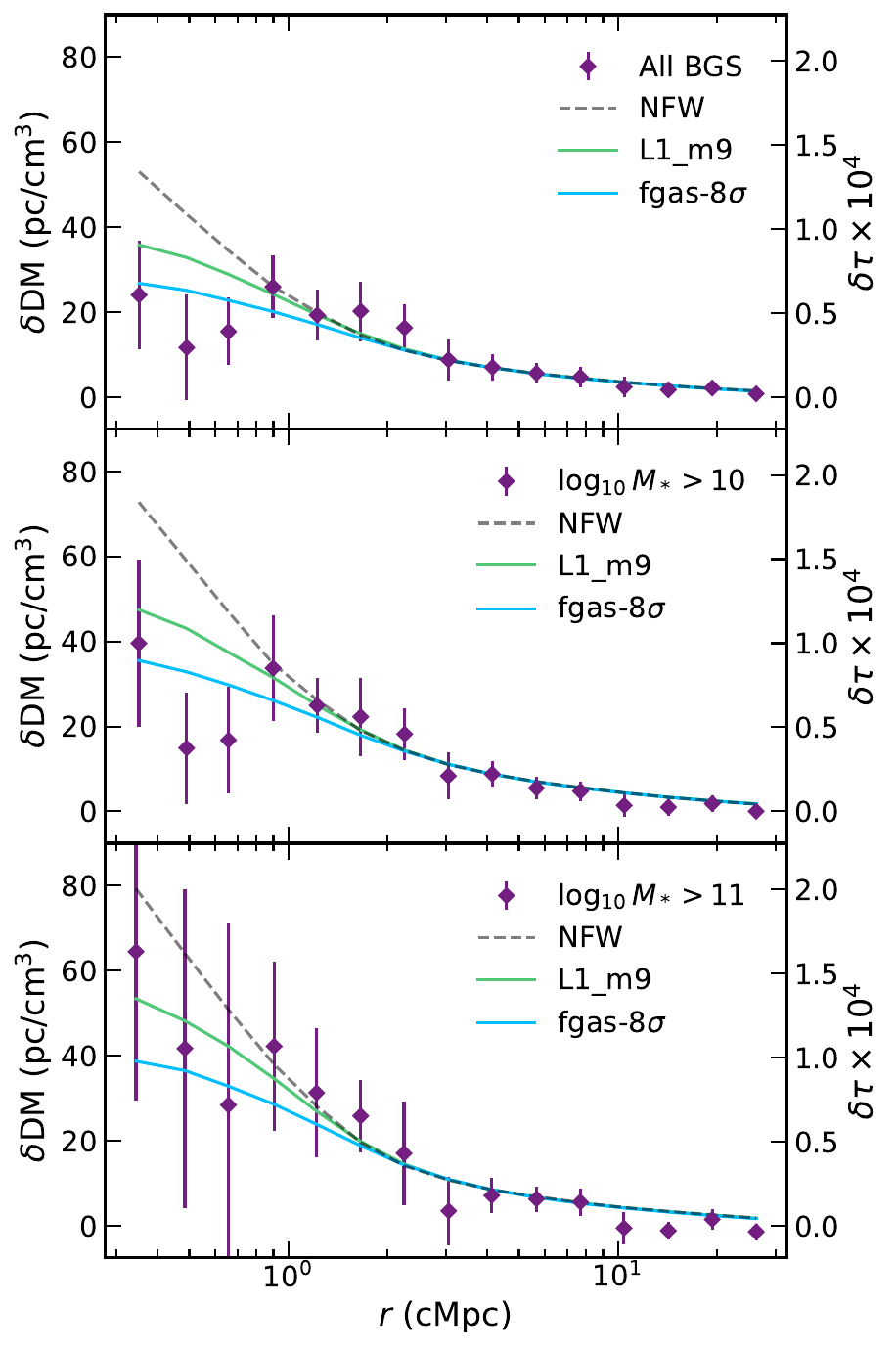}
    \caption{Comparison of our measurement to the FLAMINGO simulations. L1\_m9 is the fiducial FLAMINGO model, and fgas-8$\sigma$ is a strong feedback model. NFW is the no-feedback scenario in which gas traces the underlying dark matter.}
    \label{fig:flam}
\end{figure}

\subsection{Real Space Estimator} 
\label{sec:estimator}

For a given radial bin $R_i$, the estimator of the correlation function is an average over FRB-galaxy pairs weighted by $\delta \mathrm{DM}$ minus the same signal computed for randoms:

\begin{equation}
    \begin{split}
        \hat\xi_{g,\delta\mathrm{DM}}(R_i) =& \frac{\sum_{g,f\in R_i} w_gw_f \,\delta\mathrm{DM}_f }{ \sum_{g,f\in R_i} w_gw_f } \\ &- \frac{\sum_{r,f\in R_i} w_rw_f \,\delta\mathrm{DM}_f }{ \sum_{r,f\in R_i} w_rw_f } 
    \end{split}
\end{equation}

\noindent where $w_g$, $w_f$, and $w_r$ are any weights on each galaxy, FRB, and random respectively (all set to one in this work). We estimate the covariance of our measurement using jackknife resampling. The sky is divided into patches, labeled $i$, then the estimator of the  covariance between radial bins $j$ and $k$ is: 

\begin{equation}
\hat{C}_{jk} = \frac{N_\mathrm{patch}-1}{N_\mathrm{patch}} \sum\limits_{i=1}^{N_\mathrm{patch}} \left( \hat{\xi}_j^i - \bar{\xi}_j \right) \left( \hat{\xi}_k^i - \bar{\xi}_k \right),
\label{eqn:jackknife_covariance}
\end{equation}
where $\hat{\xi}_j^i$ is the measurement in separation bin $j$ with patch $i$ omitted, and $\bar{\xi}_j$ is the mean over all jackknife realizations. While the jackknife estimate of the covariance is unbiased, its inverse is not. We use the correction proposed by \cite{Hartlap2007} to account for this. For visualization purposes we will use the correlation matrix, which is a normalized version of the covariance matrix:

\begin{equation}
    r_{jk} = \frac{\hat{C}_{jk}}{\sqrt{\hat{C}_{jj}\hat{C}_{kk}}}.
\end{equation}

We quantify the statistical power of our measurements in the following way. For a given data vector $\mathbf{d}$, and model vector $\mathbf{m}$, the $\chi^2$ is defined as

\begin{equation}
    \chi^2\equiv(\mathbf{d}-\mathbf{m})^\mathrm{T}\,\mathrm{C}^{-1}\,(\mathbf{d}-\mathbf{m}).
    \label{eqn:chi2}
\end{equation}

\noindent The corresponding signal-to-noise ratio (SNR) is defined as:

\begin{equation}
    \mathrm{SNR} \equiv \sqrt{\chi^2_\mathrm{null}-\chi^2_\mathrm{bf}}
    \label{eqn:snr}
\end{equation}

\noindent where $\chi^2_\mathrm{null}$ is the $\chi^2$ under the null hypothesis, i.e. $\mathbf{m}=\mathbf{0}$, and $\chi^2_\mathrm{bf}$ is the $\chi^2$ for the best-fitting model. To quantify the disagreement between our measurement and a different model (e.g., an NFW profile), we replace $\chi^2_\mathrm{null}$ in the above with $\chi^2_\mathrm{model}$. 

We use the \texttt{treecorr} package for computing the correlation function and jackknife resampling \citep{treecorr}. We use 15 logarithmically spaced bins between 0.3-30 comoving Mpc. This covers the scales of interest: the one-halo regime at $\lesssim$ a few Mpc and well into the linear two-halo regime at large separation. Because we are measuring in transverse distance, we restrict the galaxies to $z>0.1$ such that the largest angular scales of the measurement are reasonable. Within \texttt{treecorr}, the sky is partitioned into jackknife patches using the \texttt{k-means} algorithm. In essence, \texttt{k-means} divides the catalog into $k$ optimal clusters that minimize the distance between sources and their corresponding cluster centers. We set $N_\mathrm{patch} = 150$, which gives many more patches than data bins but the patches are still larger in angular size than the largest scales we intend to measure. As an example of how $N_\mathrm{patch}$ affects our resulting SNR, the fiducial SNR of the log$_{10}M_*/M_\odot>10$ mass bin changes by approximately $\sim1\sigma$ in either direction as $N_\mathrm{patch}$ changes by a factor of 2. We enforce, at the FRB-galaxy pair level, that the minimum distance between foreground galaxies and background FRBs is 100 comoving Mpc to eliminate any correlation with the FRB host and its environment. This requirement eliminates approximately half of the possible FRB galaxy pairs.

\subsection{Harmonic Space Estimator}
\label{sec:clestimator}

While interpretation of the real space signal $\xi_{g,\delta\mathrm{DM}}$ is more intuitive, the signal in harmonic space $C_\ell^{g,\delta\mathrm{DM}}$ is useful because in general the powers at different multipoles $\ell$ are uncorrelated. We will use $C_\ell^{g,\delta\mathrm{DM}}$ to validate the detection significance of our measurement and to compare with other studies. 

The angular cross power spectrum is defined by decomposing the galaxy overdensity and FRB DM fields into spherical harmonics:

\begin{equation}
    \delta_g^{2D}(\mathbf{\hat{n}}) = \sum_{\ell m} a_{\ell m}^{\delta_g} Y_{\ell m} (\mathbf{\hat{n}}), \quad
    \delta\mathrm{DM}(\mathbf{\hat{n}}) = \sum_{\ell m} a_{\ell m}^{\delta\mathrm{DM}} Y_{\ell m} (\mathbf{\hat{n}}).
\end{equation}

\noindent Then the cross-spectrum is 

\begin{equation}
    \tilde{C_\ell}^{g,\delta\mathrm{DM}} = \frac{1}{2\ell+1} \sum_{m=-\ell}^{\ell} a_{\ell m}^{\delta_g }  (a_{\ell m}^{\delta\mathrm{DM}})^*.
    \label{eqn:pcl}
\end{equation}

For gaussian fields the covariance of the power spectrum is diagonal in $\ell$. For a given bandpower $b$, and using Wick's theorem, the diagonal terms are defined analytically as 

\begin{equation}
    \mathrm{Var}(C_\ell^{g,\delta\mathrm{DM}})=\frac{1}{(2\ell_b+1) \Delta \ell f_\mathrm{sky}} \left( \hat{C}_\ell^{g,g} \hat{C}_\ell^{\delta\mathrm{DM},\delta\mathrm{DM}} + (C_\ell^{g,\delta\mathrm{DM}})^2 \right),
    \label{eqn:varcl}
\end{equation}

\noindent where $\ell_b$ is the multipole of the bandpower, $\Delta \ell$ is the width of the bandpower, and $f_\mathrm{sky}$ is the fraction of sky covered by the data. Here $\hat{C}_\ell^{g,g}$ and $\hat{C}_\ell^{\delta\mathrm{DM},\delta\mathrm{DM}}$ are the measured (including noise) auto power spectra of the two fields.

Equation \ref{eqn:pcl} is referred to as a \textit{pseudo}-$C_\ell$ because it has not been corrected for the mask of the two fields. The key result of the pseudo-$C_\ell$ formalism is that Equation \ref{eqn:pcl} can be related to the underlying true power spectrum through a mode coupling matrix that depends only on the masks. We use the \texttt{NaMaster} package \citep{Alonso_2019} to compute the mode coupling matrix and generalize Equations \ref{eqn:pcl} and \ref{eqn:varcl}. We need a model for each of the power spectra in Equation \ref{eqn:varcl} to compute the uncertainty of our measurement. We use our halo model, with the details described in Section \ref{sec:cltheory}.

A main advantage of using localized FRBs is that we can restrict our analysis to galaxies foreground to the FRBs. However, the separation is not as clean in harmonic space as in the real space measurement. Following the procedure of \cite{wang2025measurementdispersionunicodex2013galaxycrosspowerspectrum}, we divide the galaxies into 5 redshift slices with edges 0.05, 0.1, 0.2, 0.3, 0.4, and 0.5. Each slice is correlated with all FRBs background to it. To compute the power spectrum  we use the \texttt{NaMaster} \texttt{NmtFieldCatalogClustering} and \texttt{NmtFieldCatalog} classes for the galaxies and FRBs respectively, which avoids the need to pixelize. We use the same random catalog for each redshift slice. As in \cite{wang2025measurementdispersionunicodex2013galaxycrosspowerspectrum}, to calculate the SNR we use ten log spaced bins between $40\le\ell\le8000$ for each slice. When plotting we choose a smaller number of $\ell$ bins. We choose to  make this measurement only for the log$_{10}M_*/M_\odot>10$ mass bin for convenience.

It is important to consider the covariance between redshift bins as they share FRBs. We neglect this covariance under the following two assumptions: the uncertainty is dominated by noise in the FRB field $N_\ell^{\delta\mathrm{DM}}$ (which shows up in the autocorrelation in Equation \ref{eqn:varcl}), and the galaxy fields between redshift slices are uncorrelated. Given the former, the covariance between two slices $i,j$ is dominated by a term like $\hat{C}_\ell^{g_i,g_j} \hat{C}_\ell^{\delta\mathrm{DM}_i,\delta\mathrm{DM}_j}$ (Equation \ref{eqn:varcl}). Under the second assumption, $\hat{C}_\ell^{g_i,g_j}$ is negligible. We indeed find that the covariance is dominated by $N_\ell^{\delta\mathrm{DM}}$, and given wide redshift bins the galaxies between bins should be uncorrelated. Note that galaxy photometric redshift uncertainties may complicate this picture.

\subsection{CAP Filter Estimator}
\label{sec:cap}

\begin{figure}
    \centering
    \includegraphics[width=1.0\linewidth]{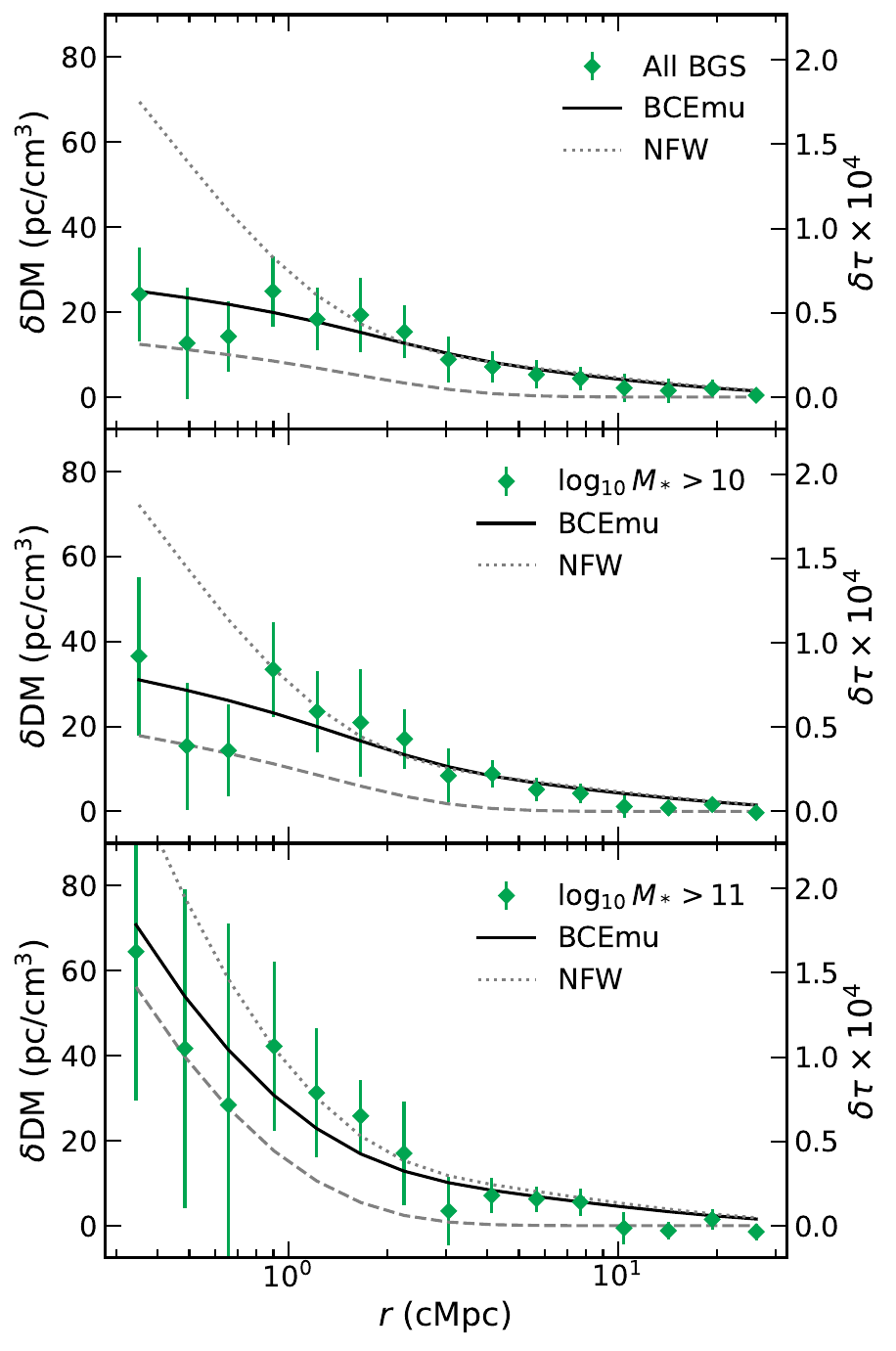}
    \caption{Halo model fit to each mass bin. In black is the best fit model,  with the one-halo term shown as a dashed line. NFW is the no-feedback scenario in which gas traces the underlying dark matter.}
    \label{fig:bcemu}
\end{figure}

Following the definition of the CAP filter (Equation \ref{eqn:cap}), the desired quantity is a combination of 2D integrals over sky area of the FRB signal. In contrast to the kSZ signal, which is naturally a 2D cutout, we can think of each FRB-galaxy pair as a noisy sample of the full 2D optical depth function around the stacked galaxies. Because the positions of these samples will be distributed randomly in the integration area, we can define an estimator of the integral that is akin to Monte Carlo integration. The (not yet CAP filtered) optical depth is:

\begin{equation}
  \begin{split}
    \hat{\tau}_\mathrm{FRB} (<\theta)
      = {}& \sigma_T \pi \theta^2
      \left(
        \frac{\sum_{g,f \le \theta} w_g (1+z_g)\,\delta\mathrm{DM}_f}
             {\sum_{g,f \le \theta} w_g}
      \right. \\
      &\left.
        - \frac{\sum_{r,f \le \theta} w_r (1+z_r)\,\delta\mathrm{DM}_f}
               {\sum_{r,f \le \theta} w_r}
      \right)
  \end{split}
\end{equation}

\noindent That is, an average of the random samples in the integration bounds multiplied by the area of integration (minus the same signal for randoms). Here we explicitly include a factor of $1+z$ to account for the same factor in Equation \ref{eqn:dtau}. Then the CAP filtered signal is:

\begin{equation}
    \hat{\tau}^\mathrm{CAP}_\mathrm{FRB} (<\theta_i) = 2\hat{\tau}_\mathrm{FRB} (<\theta_i) - \hat{\tau}_\mathrm{FRB} (<\sqrt{2}\theta_i).
\end{equation}

\noindent This signal is directly comparable to the standard kSZ signal. 

Because the CAP filter measurement is made in angle instead of transverse distance, we do not restrict the redshift range of the galaxy sample in this measurement. Apart from the different estimator, the code and covariance estimation for this measurement is the same as the real space correlation. We choose bins in $\theta$ that correspond to matching comoving distances with the kSZ signals we wish to compare to \citep{hadzhiyska2026precisionkinematicsunyaevzeldovichmeasurements}. In each case we keep $N_\mathrm{patch}=10 \times N_\mathrm{data}$ for the jackknife covariance.

\section{Fitting}
\label{sec:fit}

\begin{table}[]
    \centering
    \begin{tabular}{lccc}
         & All BGS & log$_{10}M_*>10$ & log$_{10}M_*>11$ \\
         \hline
         $\overline{z}$ & 0.17 & 0.19 & 0.22 \\
         $\mathrm{log}_{10}\,\overline{M}_*$ & 10.65 & 10.78 & 11.25 \\
         \bf{SNR} & \bf{5.0} & \bf{6.5} & \bf{4.9} \\
         \bf{SNR$_\mathrm{\mathbf{DM}}$} & \bf{8.8} & \bf{4.9} & \bf{2.4} \\
         SNR$_\mathrm{flamingo}$ & 3.9 & 6.2 & 5.0 \\
         SNR$_\mathrm{flamingo,DM}$ & 7.6 & 3.2 & -- \\
         SNR$_{C_\ell}$ & -- & 6.2 & -- \\
         SNR$_{C_\ell,\mathrm{DM}}$ & -- & 8.0 & -- \\
         \hline
         $\mathrm{log}_{10}\,M_c$ & $14.3^{+0.5}_{-0.5}$ & $13.4^{+0.9}_{-1.4}$ & $12.0^{+0.8}_{-0.6}$\\
         $\theta_{ej}$ & $4.4^{+2.1}_{-1.4}$ & $4.9^{+2.1}_{-1.9}$ & $3.7^{+2.4}_{-1.3}$\\
         $\eta$ & $0.24^{+0.11}_{-0.12}$ & $0.24^{+0.11}_{-0.12}$ & $0.19^{+0.14}_{-0.10}$\\
         $A_\mathrm{flamingo}$ & 1.1 & 1.2 & 1.2 \\
         \hline
    \end{tabular}
    \caption{The mean redshift, mean stellar mass, SNR, and best-fit params for each mass bin. SNR$_\mathrm{DM}$ quantifies how much the measurement disfavors the NFW/no-feedback scenario. Our fiducial SNR (SNR/SNR$_\mathrm{DM}$ above) use the best-fit halo model in real space. The FLAMINGO SNR uses the best-fitting of the three FLAMINGO models, and SNR$_{C_\ell}$ refers to the harmonic space measurement again assuming the best-fit halo model from the real space measurement. The mean redshift and stellar mass are for the real space measurement. The best-fit BCEmu parameters are reported as median and $1\sigma$ confidence intervals of the posteriors, while $A_\mathrm{flamingo}$ is the maximum likelihood amplitude of the FLAMINGO default model on large scales. }
    \label{tab:params}
\end{table}

The following explains how we fit the halo model (Sections \ref{sec:theory}, \ref{sec:dmstats}, \ref{sec:hm}) and simulations (Section \ref{sec:sims}) to our data (Section \ref{sec:measurement}). We restrict model fitting to the real space measurements and do not attempt to fit either the harmonic space or CAP filter measurements. However, comparing the halo model fit by the real space measurement to the harmonic space measurement provides a useful cross check. 

\subsection{Simulations}

The simulated galaxy sample in FLAMINGO will not match the observed galaxy population exactly. For example, we have approximated the sample as being stellar mass limited, while the real DESI sample selection is a more complex function of other properties such as color. This would affect the galaxy bias and the mean halo mass. Further, less significant issues such as differences in assumed cosmology and simulation finite box size effects will cause further discrepancies. 

We can approximate the net result of these discrepancies as an overall amplitude difference $A$ between the simulation and data. This would include, for example, differences in the galaxy bias. Note that in reality a changing galaxy population does not affect the one- and two-halo terms identically, but because we find $A$ to be nearly unity (see below) we take this as a decent approximation. To compare the kSZ signal with simulations, \cite{hadzhiyska2025evidencelargebaryonicfeedback} utilize an analogous approach in which the largest aperture is matched between the data and the simulations, resulting in an overall amplitude shift.

Our measurement includes data on large scales, which contain information about the  amplitude of the signal while containing little information about feedback physics. For this reason, we choose to fit the FLAMINGO predictions to the last 7 bins of our measurement, corresponding to scales $\gtrsim4\,$Mpc. To perform the fit, we take the data vector, FLAMINGO model vector, and covariance matrix, restrict each to the last 7 bins, and calculate analytically the amplitude $A$ that minimizes the $\chi^2$ when multiplied into the model vector. For each mass bin we find that $A$ is close to unity (Table \ref{tab:params}). This lends confidence to our simulation procedure and our conclusions. After the FLAMINGO curves have been corrected for $A$, we can then compare between the different feedback models. We note again, however, that residual differences between the real data and simulated data, for example the mean halo mass of the sample, would be degenerate with changing feedback strength.

\subsection{Halo Model}

When fitting the halo model, we choose to free three baryonification parameters: $\mathrm{log}_{10}\,M_c$, $\theta_{ej}$, and $\eta$. $\mathrm{log}_{10}\,M_c$ sets the mass scale below which feedback becomes important, represented as a change in the power law slope of the profile. $\theta_{ej}$ controls how far baryons are moved out of the halo, and $\eta$ controls the halo gas fractions. We fix the other parameters  $(\mu, \gamma, \delta)$ as $(1.0,2.5,7.0)$. Our purpose in fitting the halo model here is to demonstrate that the measurement can constrain feedback, so we choose not to fit a free amplitude on the two-halo term, which is common in the literature. We find that allowing a free amplitude tends to produce unphysical fits at the current statistical power of the measurement. Further, the two-halo term is set primarily by the galaxy bias, which is well known for the BGS. The default halo model prediction of the signal in the two-halo regime already matches our measurement on large scales well. The halo model curves are close to the best-fitting FLAMINGO curves, and the galaxy bias predicted by the halo model matches the expectation for the BGS \citep{desicollaboration2016desiexperimentisciencetargeting}. 

We use uniform priors on the parameters $( \mathrm{log}_{10}\,M_c, \theta_{ej},\eta)$ with ranges $[11,15]$, $[2,8]$, and $[0.05,0.4]$ respectively. We will refer to this model as BCEmu, following other authors. 
For model fitting, we use the affine invariant Markov Chain Monte Carlo (MCMC) sampler \texttt{emcee} \citep{Foreman_Mackey_2013}. The likelihood is defined by the $\chi^2$ (Equation \ref{eqn:chi2}) given a prediction of the model. For a given halo model and set of parameters, we evaluate the model at the mean $r$ within each $r$ bin and the mean redshift of the FRB-galaxy pairs in the measurement. 

\section{Results \& Discussion} 
\label{sec:results}

\begin{figure}
    \centering
    \includegraphics[width=0.8\linewidth]{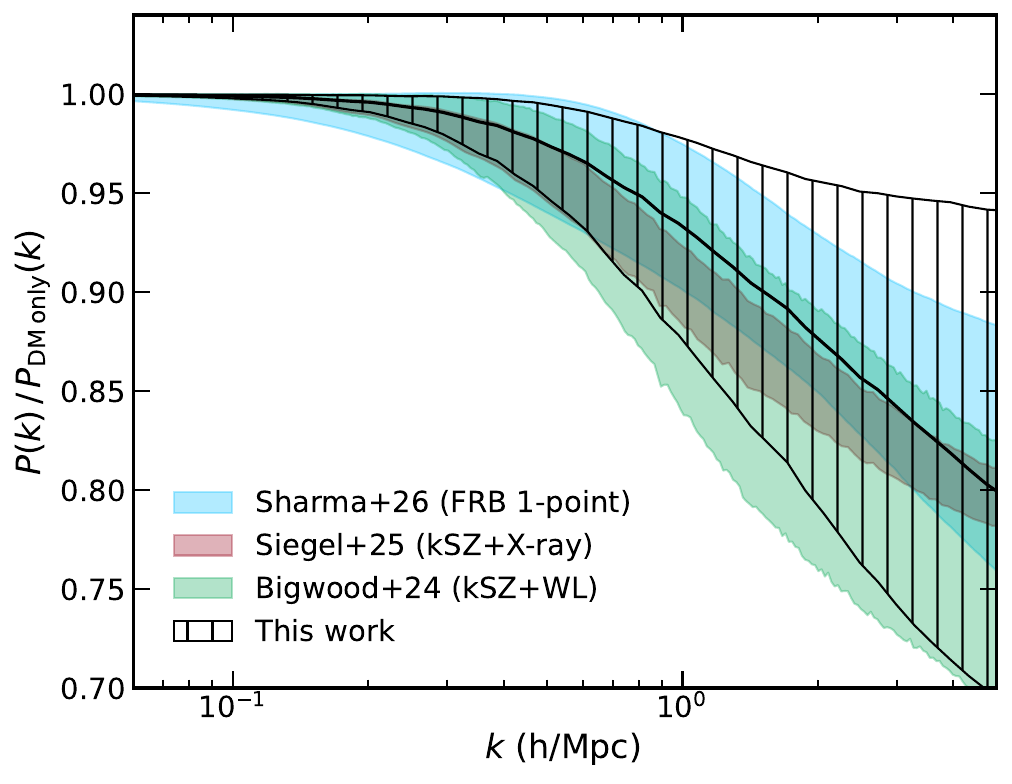}
    \caption{The suppression of the total matter power spectrum relative to the dark-matter only power spectrum due to baryonic feedback. The hatched curve is determined from the baryonification halo model fit to our log$_{10}M_*/M_\odot>10$ measurement. }
    \label{fig:spk}
\end{figure}

\begin{figure*}
    \centering
    \includegraphics[width=1.0\linewidth]{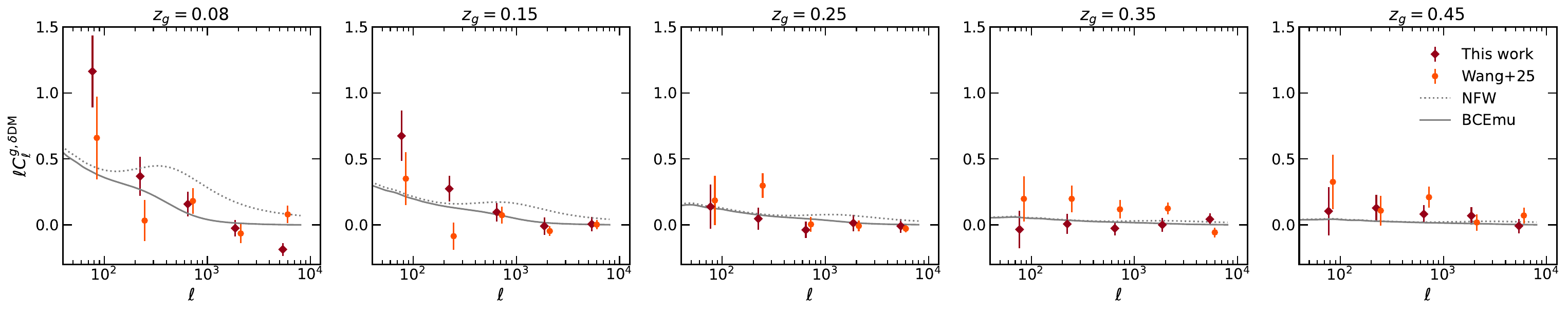}
    \caption{The results of our harmonic space measurement for BGS galaxies with $M_*>10^{10}\,M_\odot$ in each of the five redshift bins, compared to Figure 2 of \cite{wang2025measurementdispersionunicodex2013galaxycrosspowerspectrum}. The localized and unlocalized FRB measurements agree broadly. In this figure, we have multiplied our measurement and model by the fraction of FRBs that are background to each redshift slice to facilitate comparison, see text. Note that the two measurements will still differ in the exact redshift distribution of the FRBs and galaxy selection. In each plot, the solid grey line is the halo model fit to the real space measurement and the dotted line is the no-feedback scenario in which gas traces the underlying dark matter. }
    \label{fig:clz}
\end{figure*}

\begin{figure*}
    \centering
    \includegraphics[width=1.0\linewidth]{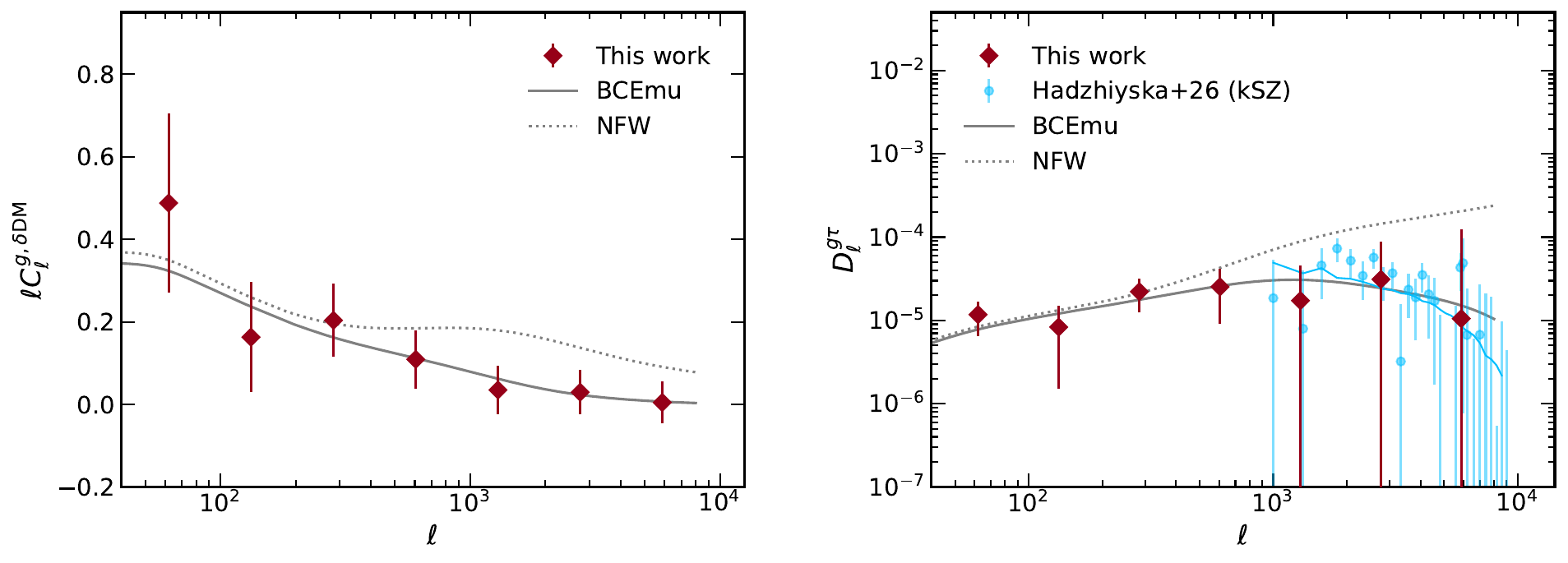}
    \caption{\textit{Left:} The harmonic space measurement for all redshift bins combined, weighted by the fraction of galaxies in each bin. Here we use 7 log spaced bins between $40< \ell <8000$. The solid grey line is the halo model fit to the real space measurement and the dotted line is the no-feedback scenario in which gas traces the underlying dark matter. \textit{Right:} We convert to optical depth ($D_\ell^{g \tau} = C_\ell^{g,\delta\mathrm{DM}} \ell(\ell+1) \sigma_\mathrm{T}(1+z) /2\pi$) and compare to the kSZ measurement by \cite{hadzhiyska2026precisionkinematicsunyaevzeldovichmeasurements}. The blue line is the best-fit model of \cite{hadzhiyska2026precisionkinematicsunyaevzeldovichmeasurements}. }
    \label{fig:cl}
\end{figure*}

This section presents the results and discussion of our measurements and analysis.

\subsection{The real space measurement}

In Figure \ref{fig:biggtr10}, we show the measured correlation function and corresponding correlation matrix for galaxies with log$_{10}M_*/M_\odot>10$. The mean redshift, stellar mass, and SNR for this measurement and others are listed in Table \ref{tab:params}. As expected, FRBs do well at measuring large scales, but the measurement becomes uncertain at low separation due to the small number of FRB-galaxy pairs in these bins. The measurement is in good agreement with the results of \cite{hussaini2025correlationfrbdispersionmeasure}, which measured the correlation on a similar selection of galaxies. In configuration space, the signal is highly correlated between bins because many scales $k$ are mixed together in each bin. This is exacerbated by the small sample of FRBs, which are shared between bins. The noticeable dip in the signal at the second and third bins from the left in Figure \ref{fig:biggtr10} may be due to such correlated noise. 

In Figure \ref{fig:flam}, we compare our measurement in each mass bin to FLAMINGO. This is the first time that this statistic has been measured in mass bins; we see a trend of increasing $\delta \mathrm{DM}$ with increasing $M_*$, as expected. In each case the best-fit amplitude of the FLAMINGO curves is close to unity, demonstrating good agreement between the simulations and our measurement. We can calculate an SNR of the measurement with the best-fitting of the three FLAMINGO curves, and quantify the significance at which the NFW curve is disfavored (Table \ref{tab:params}). The measurements in the lower two mass bins strongly disfavor the NFW (no-feedback) scenario, and in both cases the strong feedback scenario (fgas-8$\sigma$) is the preferred model. The log$_{10}M_*/M_\odot>11$ bin is very uncertain in the one-halo regime due to a lack of FRB-galaxy pairs, and does not significantly favor any model. 

Figure \ref{fig:bcemu} is the same as Figure \ref{fig:flam} but now comparing to the halo model. There is good agreement between the halo model and FLAMINGO predictions. For the halo model, the disagreement with NFW in the lower two mass bins is even more significant. These bins prefer strong feedback scenarios, which can be seen by the posterior values of the BCEmu parameters (Table \ref{tab:params}, Figure \ref{fig:corner}), and the implied suppression of the matter power spectrum (Figure \ref{fig:spk}). Note that the total suppression of the matter power spectrum depends on the gas profiles over a large range of halo masses ($\sim10^{12}-10^{15}\,M_\odot$). Here we have measured the gas profile for our specific galaxy sample and used the baryonification framework to generalize \citep{Giri_2021,schneider2025baryonificationalternativehydrodynamicalsimulations}. On the other hand, the log$_{10}M_*/M_\odot>11$ bin favors a weak feedback scenario, although the uncertainty is large. We await further data to confirm what kind of feedback scenario is preferred by the DM-galaxy statistic, but take this as a promising demonstration that FRB two-point statistics can constrain baryonic physics. At the least, we have detected the effects of baryonic processes and ruled out a no feedback scenario in which gas traces the underlying dark matter. 

\subsection{The harmonic space measurement}

Our harmonic space measurement, presented in Figures \ref{fig:clz} and \ref{fig:cl}, provides a useful cross-check of our results. In Figure \ref{fig:clz} we show the measurement in each redshift bin and compare to the CHIME measurement by \cite{wang2025measurementdispersionunicodex2013galaxycrosspowerspectrum}. Because the CHIME FRBs in \cite{wang2025measurementdispersionunicodex2013galaxycrosspowerspectrum} are unlocalized, each galaxy redshift bin is correlated with the full FRB sample. Therefore the correlation will be suppressed by a factor equal to the fraction of FRBs that are actually background to the bin: $f_f(z)=\int_z^\infty dz \,n_\mathrm{FRB}(z)$. For the purpose of comparison, in Figure \ref{fig:clz} we artificially multiply our measurement and model by the same factor computed for our FRB sample in each bin. While the two measurements still differ in the details of the FRB redshift distribution and galaxy selection, we find broad agreement between them. Assuming the degrees of freedom is the number of bins, the two measurements are consistent at $2.2\sigma$, which is reasonable considering the discrepancies. Consistency across FRB studies is a promising sign for the field. We determine the SNR of this measurement using the halo model fit to the real space measurement. Both the halo model and the resulting SNR agree well between real and harmonic space (Table \ref{tab:params}). In harmonic space, the disagreement with NFW at small scales is also clear, with the significance of the disagreement even higher than in real space. In the left of Figure \ref{fig:cl} we combine the redshift bins, weighting by the number of galaxies in each and without the $f_f(z)$ factor. 

Collectively, we have measured the DM-galaxy correlation at a maximum SNR of $6.5\sigma$. This is the strongest detection of this signal to date in both real and harmonic space. To the authors' knowledge, this represents one of the first measurements of the clustering of electrons around galaxies on large scales. Electrons are expected to be unbiased tracers on large scales \citep{andrew2026fastradioburstdispersion}, meaning this correlation contains cosmological information that could be utilized by future studies. Further, we have demonstrated for the first time that this measurement can constrain the strength of baryonic feedback on small scales. While the constraint is less competitive currently than other probes, future FRB samples such as from the Deep Synoptic Array \citep{Hallinan,DSAforecast} will provide precision measurements.

Our measurement of the DM-galaxy correlation with localized FRBs is less constraining for feedback than recent DM one-point studies with similar samples (Figure \ref{fig:spk}, \cite{sharma2026signaturessuppressedmatterclustering}). The $p(\mathrm{DM}|z)$ PDF is sensitive to a wider range of mass and redshifts (but the contributions from different halos are largely degenerate), while here we isolate feedback from the target sample (BGS in this case), allowing us to study specific samples in detail. The two statistics probe distinct halo masses, redshifts, and physical scales \citep{sharma2026signaturessuppressedmatterclustering}, with complementary strengths and weaknesses. 

\subsection{CAP filter measurement \& comparison to kSZ}

\begin{figure*}
    \centering
    \includegraphics[width=0.95\linewidth]{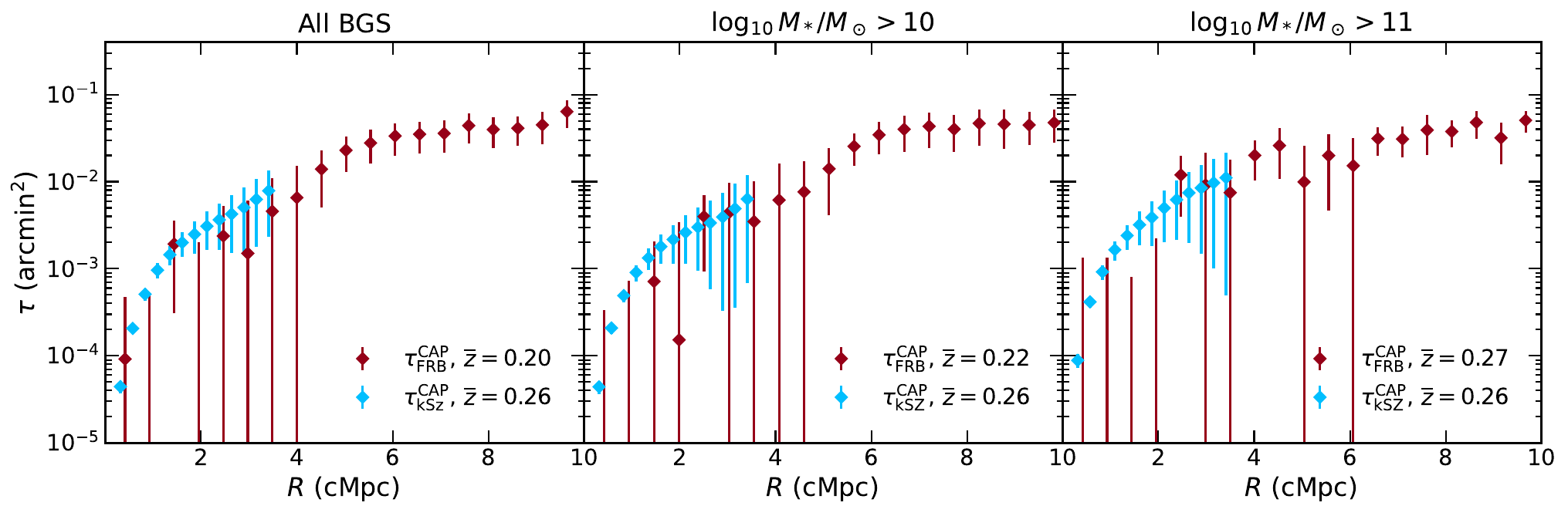}
    \caption{The CAP filtered $\tau$ signal measured by FRBs in each mass bin is compared to the results of the kSZ study by \cite{hadzhiyska2026precisionkinematicsunyaevzeldovichmeasurements}. Despite the different probes and sample selection, each mass bin is consistent across the two measurements at $\sim1\sigma$.}
    \label{fig:tau}
\end{figure*}

\begin{figure*}
    \centering
    \includegraphics[width=0.95\linewidth]{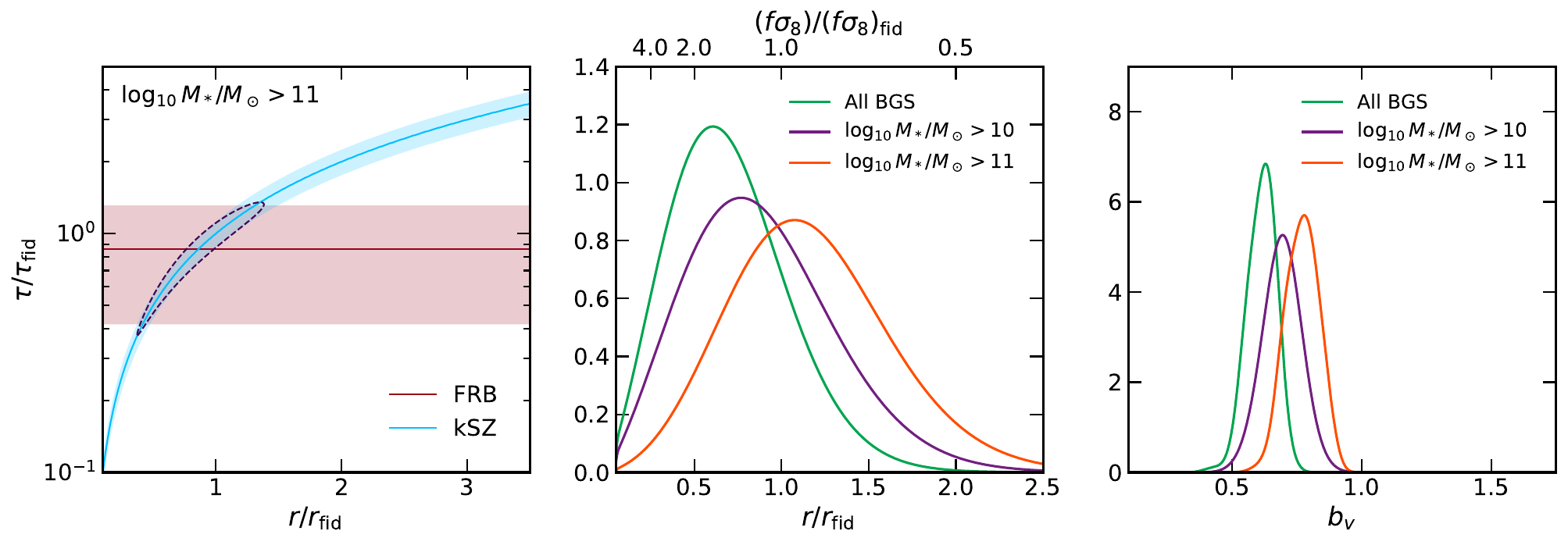}
    \caption{A first attempt at breaking the kSZ optical depth degeneracy with FRBs. \textit{Left:} the kSZ measurements are degenerate in $\tau$ and $r/\sigma_v$, while the FRB measurement depends only on $\tau$. We show the 1$\sigma$ constraint from each and the combined posterior as the dashed contour, for the log$_{10}M_*/M_\odot>11$ bin. \textit{Middle:} the 1D posteriors on $r/r_\mathrm{fid}$ or $(f\sigma_8)/(f\sigma_8)_\mathrm{fid}$. \textit{Right:} constraints on the kSZ velocity bias factor $b_v$. }
    \label{fig:degeneracy}
\end{figure*}

Given our measurement, we would like to make a direct comparison to the kSZ effect. kSZ measurements have been invaluable for studying the diffuse gas over the last $\sim5\,$years. Each probe (kSZ, FRB) is sensitive to a distinct range of spatial scales, so there is potential for them to be highly complementary. kSZ studies are bounded on small scales by the beam of CMB experiments, and on large scales by the overwhelming primary CMB fluctuations. Current kSZ measurements are strongest on scales of order $1\,$Mpc. On the other hand, we have shown that FRBs are best at measuring large scales greater than a few Mpc. It is worth noting, however, that there is no fundamental limit on the scales that FRBs can access. Given enough data, FRBs can measure the entire range of spatial scales down to the localization uncertainty, which will be sub-arcsecond in near-term surveys. 

We will compare our measurements to the recent work by \cite{hadzhiyska2026precisionkinematicsunyaevzeldovichmeasurements}, which studies the kSZ signal around the BGS galaxies. Note a few caveats in this comparison: \cite{hadzhiyska2026precisionkinematicsunyaevzeldovichmeasurements} use the DESI DR2 spectroscopic sample, while we use the photometric LS DR9 sample. The spectroscopic sample is known to undersample overdense regions of the sky, which would reduce the signal relative to the photometric sample. Further, \cite{hadzhiyska2026precisionkinematicsunyaevzeldovichmeasurements} use only the BGS Bright sample, while here we use both the Bright and Faint samples, which works in the opposite direction in terms of the relative amplitudes of the two measurements. Note also that the stellar masses in the two samples are determined differently. The photometric sample relies on the Random Forest algorithm of \cite{Zhou_2023sm}, while the spectroscopic sample uses SED fitting including the galaxy spectra. Finally, as noted previously, the overlapping FRB redshift kernel will preferentially select a lower redshift subset of the galaxies (Table \ref{tab:params}). 

In the right panel of Figure \ref{fig:cl} we have converted our harmonic space measurement to an optical depth and compared it to the kSZ measurement by \cite{hadzhiyska2026precisionkinematicsunyaevzeldovichmeasurements}. By weighting each redshift slice by the fraction of galaxies in the slice, we have effectively removed the effect of the FRB redshift kernel selecting a lower redshift subsample of the galaxies (only for this harmonic space measurement).  Note that this combination of redshift bins is not SNR optimal, as the statistical power of the measurement is dominated by the lower redshift bins, which have more background FRBs. Qualitatively, the FRB and kSZ measurements agree well despite the differences in sample selection and observational methods. Over the overlapping scales, the measurements are statistically consistent. The FRB measurement agrees with the best-fit kSZ model well below the $1\sigma$ level. We see clearly the complementary nature of the two signals: FRBs constrain large scales well but uncertainties blow up at small scales, and vice versa for kSZ.

We can further compare the kSZ and FRB signals by converting DM to optical depth $\tau$ (Eqn. \ref{eqn:dtau}). The results of our measurement with the CAP-filtered $\tau$ estimator (Section \ref{sec:cap}) are shown in Figure \ref{fig:tau}. We compare to \cite{hadzhiyska2026precisionkinematicsunyaevzeldovichmeasurements}, but because the FRBs select a lower redshift subsample of the galaxies we make the comparison in comoving transverse distance instead of angle. Again, we see clearly the complementary nature of the two signals in terms of the scales they constrain. The kSZ and FRB signals appear to agree in this space as well as harmonic space. Calculating the $\chi^2$ between the two measurements and assuming the number of degrees of freedom equals the number of overlapping bins, we find that the measurements are consistent at $0.4$, $0.2$, and $1.2\sigma$ for each of the mass cuts.

We show how FRBs can break the kSZ optical depth degeneracy in the following. Because the kSZ $\tau$ signal is degenerate in $\tau$ and $r/\sigma_v$ (Section \ref{sec:ksztheory}), an overall amplitude difference between the FRB and kSZ signals may be caused by a misestimate of $r/\sigma_v$ (assuming no other factors affect the amplitude), which requires detailed mocks to measure. We can cross-check the amplitude of the kSZ estimator by requiring that the optical depth measured by the two probes is the same. Assuming a LCDM model with known parameters, we can break the degeneracy with $\tau$ and solve for $r/r_\mathrm{fid}$, shown in Fig \ref{fig:degeneracy}. Alternatively, assuming that the kSZ analyses properly calibrated $r$, we can get a constraint on the growth of structure, $(f\sigma_8)/(f\sigma_8)_\mathrm{fid}$, by a change of variables. 

Note that this formulation ignores the fact that uncertainty in $r$ is already believed to be at the 10-15\% level or below \citep{hadzhiyska2023velocityreconstructioneradesi,Ried_Guachalla_2024}, but serves to demonstrate how the FRB and kSZ measurements could be combined. The results of this analysis are shown in the left two panels of Figure \ref{fig:degeneracy}. We evaluate the likelihood using the overlapping FRB and kSZ bins. For each mass bin, the amplitude of the FRB and kSZ signals again agree within a factor of $\sim2$ and within the uncertainty. 

In another formulation of the optical depth degeneracy, the parameter of interest is the kSZ velocity bias factor $b_v$, which is an integral over the galaxy-electron power spectrum (Section \ref{sec:ksztheory}, \cite{Madhavacheril_2019}). We demonstrate a constraint on $b_v$ by sampling from the posteriors of our halo model fit. We take $P_{ge}^\mathrm{fid}$ as the default Battaglia profile (Section \ref{sec:ksztheory}), which is the same as that used in \cite{chaussidon2026measurementgalaxyvelocitypowerspectrum}, and use the window function $F(l)$ from the same study. Again, we note that the exact value of $b_v$ that we arrive at is unimportant because it is specific to a given analysis (for example, it depends not only on the galaxy sample, but also on the noise of the CMB maps being used); we wish only to demonstrate that a constraint can be made. The result is shown in the rightmost plot of Figure \ref{fig:degeneracy}.

\subsection{Measurement Systematics \& Limitations}\
\label{sec:issues}

It is worth considering what kind of systematics might impact the measurement. Because we are using FRBs as backlights, this analysis is inherently robust to many systematics that might affect the FRB one-point statistic, for example. \cite{sharma2025quantifyingimpactselectioneffects} analyze several of the most important systematics for the one-point statistic, including the DM selection function and evolution of the FRB population with redshift. They find that these effects can bias the one-point statistic significantly, but that this bias is well below the current statistical uncertainty. \cite{Cheng_2026} investigate how properties of the host galaxy and DM or scattering dependent selection affect an angular DM-galaxy cross correlation for a CHIME-like sample. We would not expect host galaxy properties to impact the two-point statistic because they will not correlate with foreground galaxies. Indeed, \cite{Cheng_2026} finds that these systematics are negligible.

A more concerning issue for the two-point statistic is a DM dependent cut on the FRB sample. DM affects the detectability of an FRB in a given search pipeline, and FRB searches are carried out over a limited range of DM. \cite{Cheng_2026} find that for a realistic CHIME DM selection function, the bias in the cross-correlation is negligible. However, stronger DM selection functions, including a sharp cutoff at 600 pc/cm$^3$, can bias the measurement by a factor of 2 or more. Modeling the selection function for our sample is not possible because the sample is very heterogeneous. However, because the bias becomes significant only in relatively extreme cases \citep{Cheng_2026}, we take this as a promising sign for our measurement. Another point to consider is whether the high-DM tail of the FRB sample is dominated by correlated signal or uncorrelated host DM, the latter case reducing the bias. In our sample selection we cut out 7 FRBs in the southern portion of DESI with $\delta \mathrm{DM}>300\,$pc/cm$^3$ (Section \ref{sec:frbsample}). However, qualitatively we find that many of the highest $\delta \mathrm{DM}$ FRBs are those with known extreme host DM. Further, we have verified that including the 7 cut FRBs does not change the amplitude of the measurement significantly but does increase the noise. More extensive modeling and large homogeneous samples from forthcoming surveys will be able to characterize the DM selection function and verify our assumption of negligible bias.

The two-point statistic is further expected to be robust to host galaxy DM and residuals from the MW. However, several additional issues may persist that will need to be accounted for in future analyses. Misidentification of FRB host galaxies and thus redshift could bias the measurement. Of course, the standard galaxy related systematics are present; survey completeness, masks, depth variations, etc. need careful accounting. The photometric redshift uncertainties and catastrophic outliers, which are poorly constrained currently, will need to be corrected, if relevant. Photo-z errors should convolve the measured signal by changing the inferred $r$, at the level of a few percent. 
One issue relevant to the small sample in our measurement is that at small $r$, where there are few FRB-galaxy pairs, we may be missing rare systems. Over repeated realizations of the measurement the mean will be unbiased, but for our given sample if we simply miss the rarest and most massive systems at small $r$, and these dominate the signal, the measurement could be biased low. This could plausibly explain the dip seen in bins two and three of Figure \ref{fig:biggtr10}. 

\subsection{Outlook}

We have demonstrated that $\mathcal{O}(100)$ FRBs can already provide interesting constraints on the spatial distribution of cosmic baryons through the DM-galaxy cross correlation. While we are limited by this small sample size, the number of known FRBs is expected to increase dramatically in coming years. Upcoming surveys by the Deep Synoptic Array (DSA) \citep{Hallinan,DSAforecast} and Canadian Hydrogen Observatory and Radio-transient Detector (CHORD) \citep{chord} should produce a sample of localized FRBs up to three orders of magnitude larger than present. Such a sample will reveal the Universe's diffuse gas in unprecedented detail, with the DM-galaxy cross correlation likely becoming a pillar measurement of the field. This statistic could be used to benchmark astrophysical models in simulations across a wide range of halo masses, similarly to the use of X-ray cluster gas fractions today. 

Achieving a deeper understanding of cosmic baryons and feedback will probably require a multi-probe approach. Each probe has limitations and systematics to contend with, and we can expect those of FRBs to be revealed as statistical power increases. These challenges may be overcome by identifying synergies between observational tools and validating results across them. In this work we have taken a small step in that direction, but much work remains. The upcoming Simon's Observatory will deliver precise measurements of the kSZ and tSZ effects \citep{thesimonsobservatorycollaboration2025simonsobservatorysciencegoals}. Combining these measurements with those of a large sample of localized FRBs is a promising avenue forward. 

\section{Conclusion} 
\label{sec:conclusion}

In this work we have measured the FRB DM-galaxy cross correlation with localized FRBs and the DESI BGS sample. This statistic measures circumgalactic gas and the large-scale baryon distribution. The DM-galaxy cross correlation is a promising tool for constraining baryonic feedback. However, only recently has the sample of FRBs become sufficient to make a useful measurement. We choose to use localized FRBs with host galaxy redshifts for this work because the mean cosmological DM can be subtracted, significantly reducing the largest source of sightline variance. We make the measurement in both real and harmonic space, detecting the signal at the highest significance to date. The measurements are in good agreement with previous FRB studies. Across several stellar mass cuts we detect for the first time the expected trend of increasing DM with increasing halo mass in this statistic. Our results show that FRBs can constrain the clustering of electrons over orders of magnitude in scale, which is in contrast to other baryon probes like tSZ, kSZ, and X-rays.

By analyzing the measurement in the context of simulations and a halo model, we show that the DM-galaxy correlation can constrain the strength of baryonic feedback. With only $\sim100$ FRBs, we find that the electron clustering is suppressed relative to the dark matter at high significance ($\sim$\,9$\sigma$). We find tentative evidence that FRBs prefer a strong feedback scenario. Apart from small scales, the measurement is broadly in agreement with simulations and theory in the two-halo regime. 

We further present a novel estimator of the kSZ integrated optical depth signal using FRBs, and compare to a recent kSZ study of a similar galaxy sample. The two probes agree at the factor of $\sim2$ level and within the uncertainty. This is remarkable given the difference in observational techniques. Finally, we show how to calibrate the kSZ optical depth degeneracy with FRBs, unlocking access to unbiased measurements of velocities and growth of structure. Although the uncertainty is large on the FRB side, this is a promising direction for future studies. 

This work demonstrates methodology by which future FRB studies can make precise constraints on cosmic baryons and be combined with important probes such as the kSZ effect. With near term radio surveys, we expect the DM-galaxy statistic to become a leading cosmological and astrophysical tool. 

\begin{acknowledgments}

S.M. and L.C. acknowledge support from the U.S. National Science Foundation Astronomy and Astrophysics Research Grants program under award AST-2508734. We acknowledge the Virgo Consortium for making their simulation data available. The FLAMINGO simulations were performed using the Durham Memory Intensive system managed by the Institute for Computational Cosmology on behalf of the STFC DiRAC facility (www.dirac.ac.uk). We thank Jared Siegel for helpful discussions. 
The authors also thank staff members of the Owens Valley Radio Observatory and the Caltech radio group for building and supporting the DSA-110.

\end{acknowledgments}

\appendix

\section{Dispersion Measure Statistics}
\label{sec:dmstats}

Because of its additive nature, the observed DM of an FRB can be divided into parts representing contributions from different bodies of gas:

\begin{equation}
    \mathrm{DM_{obs}} = \mathrm{DM_{MW}} + \mathrm{DM_{cosmic}} + \mathrm{DM_{host}}.
\end{equation}

\noindent $\mathrm{DM_{obs}}$ is the observed DM, $\mathrm{DM_{MW}}$ represents the contributions from the interstellar medium (ISM) and CGM of the MW, and $\mathrm{DM_{cosmic}}$ is the contribution from the IGM and intervening halos (the cosmic web). $\mathrm{DM_{host}}$ is the same as $\mathrm{DM_{MW}}$ but for the FRB host galaxy, and includes any gas in the local environment of the FRB. The interesting cosmological signal is contained in $\mathrm{DM_{cosmic}}$, which can be written:

\begin{equation}
    \mathrm{DM_{cosmic}}(z)=\int_0^{\chi_f} \frac{n_e(\chi)}{(1+z)^2}d\chi,
    \label{eqn:DM_z}
\end{equation}

\noindent where $\chi_f$ is the comoving radial distance of the FRB, $z$ is the redshift at comoving distance $\chi$, and $n_e(\chi)$ is the electron density at a distance $\chi$ in physical (not comoving) units. Here and in the following the dependence on sky position $\mathbf{\hat{n}}$ is implicit. One factor of $1/(1+z)$ arises from the conversion between comoving and proper units, and the other is due to the redshifting of DM. 

We can define a DM contrast as:

\begin{equation}
    \mathrm{\delta DM} = \mathrm{DM_{obs}} -  \mathrm{DM_{MW}} - \langle\mathrm{DM_{cosmic}}\rangle - \langle\mathrm{DM_{host}}\rangle,
    \label{eqn:ddm}
\end{equation}

\noindent which quantifies the measured cosmological DM excess over the mean. If we expand $n_e(\chi)=\overline{n_e}(\chi)(1+\delta_e(\chi))$, where $\delta_e(\chi)$ is the electron overdensity, and $\overline{n_e}(\chi)$ is the average electron density of the universe, then:

\begin{equation}
    \mathrm{DM_{cosmic}} = \int_0^{\chi_f} \frac{\overline{n_e}(\chi)}{(1+z)^2} (1+\delta_e(\chi))d\chi,
\end{equation}

\begin{equation}
    \langle\mathrm{DM_{cosmic}}\rangle  = \int_0^{\chi_f} \frac{\overline{n_e}(\chi)}{(1+z)^2} d\chi,
\end{equation}

\noindent and finally

\begin{equation}
    \mathrm{\delta DM}  = \int_0^{\chi_f} \frac{\overline{n_e}(\chi)}{(1+z)^2} \delta_e(\chi)d\chi.
\end{equation}

\noindent Therefore, $\mathrm{\delta DM}$ allows us to study the clustering of electrons.

The mean electron density can be written

\begin{equation}
    \overline{n_e}(z) = f_d (z)f_e(z) \rho_b(z) m^{-1}_p
\end{equation}

\noindent where $\rho_b(z)$ is the cosmological baryon density, $f_e$ is the number of electrons per baryon, and $f_d$ is the fraction of electrons in the diffuse ionized state. For computing $\langle\mathrm{DM_{cosmic}}\rangle$, we assume constant $f_d=0.94$ \citep{Connor2025}, and $f_e=0.88$ (from the cosmological abundance of hydrogen and helium). This is a good approximation as we expect these values to be only weakly redshift dependent at moderate $z$. Further, the essential function of subtracting the mean components in Equation \ref{eqn:ddm} is to decrease the noise in the measurement. Incorrect subtraction will not bias the measurement because it will not correlate with foreground galaxies. We assume $\langle\mathrm{DM_{host}}\rangle=150/(1+z)$ pc/cm$^{-3}$ \citep{Connor2025}, where the $(1+z)$ factor again arises from the redshifting of DM. For $\mathrm{DM_{MW}}$, we use the ne2001 model \citep{cordes2003ne2001inewmodelgalactic} evaluated at the sky position of each FRB, implemented in the \texttt{pygedm} package \citep{pygedm}. Although newer galactic electron models exist, we choose ne2001 for convenience and because we only expect this choice to affect the noise properties of the measurement. We ignore the contribution to the DM from the MW CGM due to the lack of data on this component. 

In this work, we are interested in the 2-point correlation between $\delta \mathrm{DM}$ and galaxies. In a flat-sky approximation, given a galaxy at redshift $\chi_g$, the correlation can be written as: 

\begin{equation}
    \xi_{g,\delta\mathrm{DM}}(r) =  \int_0^{\chi_f} d\chi \frac{\overline{n_e}(\chi)}{(1+z)^2} \xi_{g,e}\left(\sqrt{(\chi-\chi_g)^2+r^2}, z\right).
\end{equation}

\noindent $r$ is the comoving distance between the galaxy and the FRB line of sight defined at the position of the galaxy, i.e. $r=\chi_g\theta$ where $\theta$ is the angular separation between the FRB and the galaxy. Because $\xi_{g,e}$ becomes negligible at distances $\gtrsim100$ Mpc from the galaxy, and we do not expect it to be a strong function of $z$, we make the approximation that $z$ is constant over the integration range when modeling the correlation function.

\section{Halo Model}
\label{sec:hm}

To model the observed signal, we follow the halo model formalism and separate the correlation function into one- and two-halo terms:

\begin{equation}
    P^\mathrm{1h}_{ge}(k,z) = \frac{1}{\overline{n}_g(z) \overline{n}_e(z)}\int dM \frac{dn}{dM}(M,z) \, \tilde{u}_e(k|M,z) \, \tilde{u}_g(k|M,z)
\end{equation}

\begin{equation}
    \begin{split}
        P^\mathrm{2h}_{ge}(k,z) =& \frac{P_\mathrm{lin}(k)}{\overline{n}_g(z) \overline{n}_e(z)} \int dM \frac{dn}{dM}(M,z) \, b(M,z) \, \tilde{u}_g(k|M,z) \\
        & \times \int dM \frac{dn}{dM}(M,z) \, b(M,z) \, \tilde{u}_e(k|M,z).
    \end{split}
\end{equation}

\noindent The former captures the correlation within a halo, i.e. the galaxy electron profile, and the second captures the correlation between distinct halos. $\frac{dn}{dM}$ is the halo mass function, $P_\mathrm{lin}(k)$ is the linear matter power spectrum, $\overline{n}_g(z)$ is the mean galaxy number density, and $\overline{n}_e(z)$ is the mean electron number density. $\tilde{u}_e(k|M,z)$ and $\tilde{u}_g(k|M,z)$ are the fourier transforms of the electron and galaxy profiles within a halo of mass $M$, normalized to the total electron content and number of galaxies within the halo. $\xi_{g,e}$ can be obtained from $P_{ge}$ by the usual fourier transform. 

We use the \texttt{hmvec}\footnote{\url{https://github.com/simonsobs/hmvec}} code to compute the halo model prediction, which follows the formalism in the appendix of \cite{smith2018ksztomographybispectrum}. $\tilde{u}_g(k|M,z)$ is a sum of a central and satellite galaxy profile, where the satellite distribution is assumed to follow the NFW profile. In \texttt{hmvec}, the HOD is specified by a single parameter $m_*^\mathrm{thresh}$, which determines the minimum halo mass to host a central galaxy. The other HOD parameters depend on $m_*^\mathrm{thresh}$ and are calibrated to external data. $m_*^\mathrm{thresh}$ is determined iteratively by matching the number density of observed galaxies at a given redshift to the number density predicted by the halo model. 

For $\tilde{u}_e(k|M,z)$, we use the baryonification framework \citep{Schneider_2015,Schneider_2019,Giri_2021,schneider2025baryonificationalternativehydrodynamicalsimulations,kovac2025baryonificationiiconstrainingfeedback}. The basic premise of the baryonification method is that the effects of baryonic processes on the matter distribution can be approximated in N-body simulations by displacing particles around halos, which is computationally more efficient than full hydrodynamic simulations. The displacement is governed by a set of flexible analytical profiles, which incorporate feedback by moving baryons beyond the halo virial radius. An attractive feature of the baronification framework is that it can efficiently predict the suppression of the matter power spectrum, which is of interest for weak lensing. We will use the emulator trained in \cite{Giri_2021} to translate contraints on the baryonification parameters into the matter power spectrum suppression. 

In the baryonification framework, the halo electron profile is given by:

\begin{equation}
    n^\mathrm{1h}_{e}(r|M)\propto\frac{\Omega_b/\Omega_m-f_\mathrm{star}(M)}{\left[ 1+\left( \frac{r}{r_\mathrm{core}} \right) \right]^{\beta (M)} \left[1 + \left( \frac{r}{r_{ej}} \right)^\gamma \right]^{\frac{\delta-\beta(M)}{\gamma}}}.
\end{equation}

\noindent Here $M$ refers to $M_{200}$, defined w.r.t. the critical density of the Universe. $f_\mathrm{star}$ is the fraction of the halo mass in stars, including the central galaxy, satellites, and halo stars. This profile has a core of radius $r_\mathrm{core}=\theta_\mathrm{co}R_{200}$, followed by a power law of slope $\beta(M)$. $\theta_\mathrm{core}$ is fixed to 0.1, and  

\begin{equation}
    \beta(M)= \frac{3(M/M_c)^\mu}{1+(M/M_c)^\mu}.
\end{equation}

\noindent At high mass $\beta$ approaches 3, while at low mass the gas has been ejected farther out so the slope is more shallow. $M_c$ controls where this transition happens and $\mu$ controls the abruptness. The profile is truncated at the ejection radius $r_{ej}=\theta_{ej}R_{200}$, beyond which the profile has slope $\delta$ and $\gamma$ controls the abruptness of this transition. 

The profile is normalized by requiring that the total gas mass integrated to infinity is $f_\mathrm{gas}M$, with $f_\mathrm{gas}=\Omega_b/\Omega_m-f_\mathrm{star}(M)$. $f_\mathrm{star}(M)$ is the fraction of the halo mass in stars, given by 

\begin{equation}
    f_\mathrm{star}(M)=A\left(\left(\frac{M}{M_s}\right)^{\eta}+\left(\frac{M}{M_s}\right)^{-\tau}\right)^{-1}.
\end{equation} 

\noindent Following standard practice, we fix the normalization $A=0.055$, turnover mass $M_s =2.5\times10^{11}\,M_\odot/h$, and low mass slope $\tau=1.376$ \citep{Moster_2012}. 

For dark matter halo and cosmological calculations we use the \texttt{COLOSSUS} package \citep{Colossus}. For computing the linear matter power spectrum we use the \texttt{CAMB} package \citep{CAMB}.

\subsection{Harmonic space}
\label{sec:cltheory}

With our halo model we can calculate the signal in harmonic space, which is relevant for our harmonic space measurement. Under the Limber approximation, we can model the angular power spectra as

\begin{equation}
    C_\ell^{A,B} = \int_0^\chi d\chi \frac{W_A W_B}{\chi^2} P_{AB}\left(k=\frac{\ell+1/2}{\chi}, z\right),
    \label{eqn:theoryCl}
\end{equation}

\noindent where $W_A$ and $W_B$ are radial weight functions and $P_{AB}$ is the 3D cross power spectrum between $A$ and $B$. 

For the two fields of interest:

\begin{equation}
    \begin{split}
        & W_{\delta\mathrm{DM}}(\chi) = \frac{\overline{n_e}(\chi)}{(1+z)^2} \int_\chi^{\infty}d\chi'n_f(\chi')\\
        & W_g(\chi) = n_g(\chi),
    \end{split}
\end{equation}

\noindent where $n_g(\chi)$, $n_f(\chi)$ are the normalized radial distributions of galaxies and FRBs respectively. For the noise contribution to the galaxy auto power spectrum we have $N_\ell^g = 1/n_g^\mathrm{2D}$, where $n_g^\mathrm{2D}$ is the 2D galaxy number density. For the DM auto correlation we have $N_\ell^{\delta\mathrm{DM}} = \sigma_D^2/n_f^\mathrm{2D}$, where $\sigma_D\approx100$ pc/cm$^3$ is the variance in host DM and $n_f^\mathrm{2D}$ is the areal FRB density.  
We use \texttt{hmvec} to compute the three relevant power spectra $P_{gg}$, $P_{ge}$, and $P_{ee}$. For the halo electron profile, we use the best-fitting BCEmu model from the real space measurement. 

\begin{figure}
    \centering
    \includegraphics[width=0.55\linewidth]{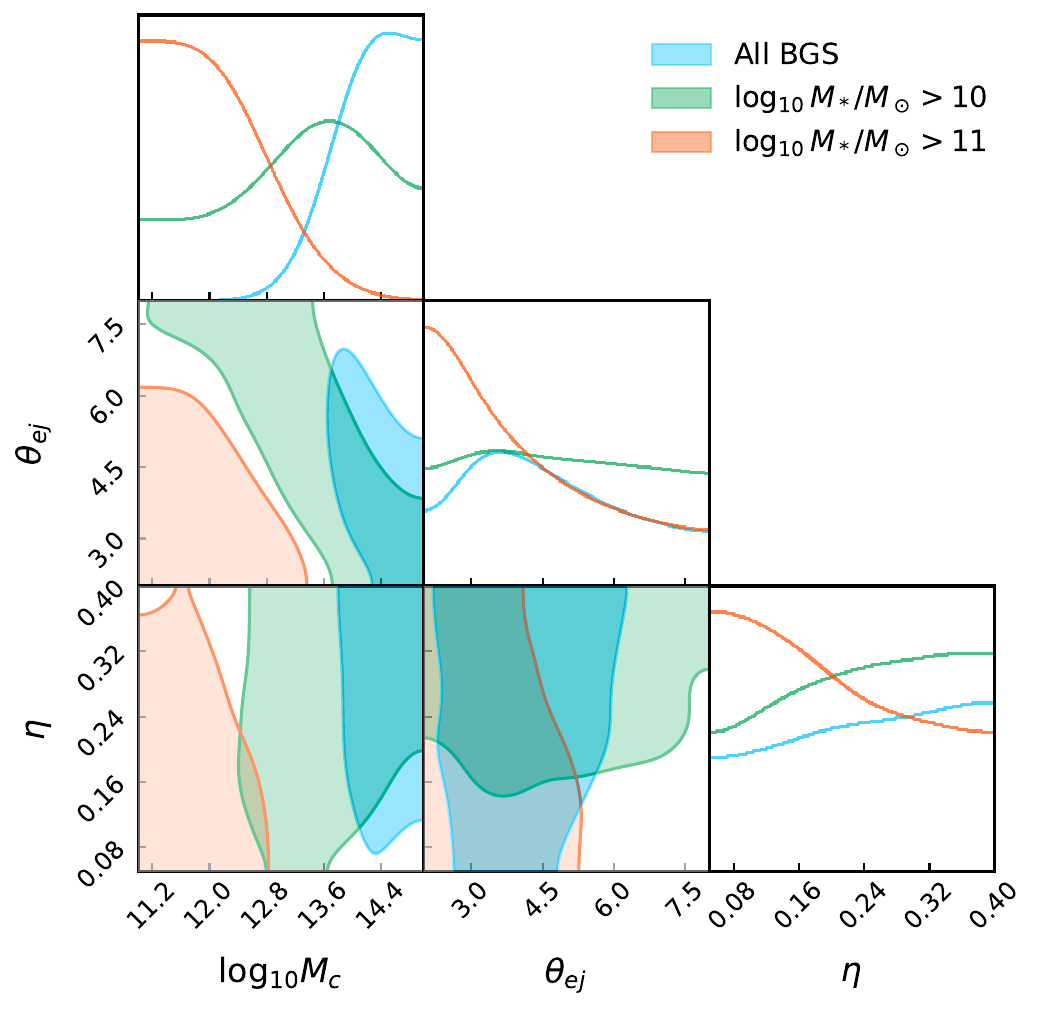}
    \caption{Posterior corner plot for parameters of the BCEmu model fit in Figure \ref{fig:bcemu}. Contours are $1\sigma$ confidence intervals.}
    \label{fig:corner}
\end{figure}

\section{The kSZ Effect}
\label{sec:ksztheory}

We are interested in comparing the DM signal with measurements of the kSZ effect. The fractional change in the CMB temperature due to the kSZ effect is

\begin{equation}
    \frac{\Delta T_\mathrm{kSZ}}{T_\mathrm{CMB}} = - \int d\chi \frac{\sigma_\mathrm{T}\, n_e(\chi)}{(1+z)} \frac{v_r(\chi)}{c} e^{-\tau(\chi)}.
    \label{eqn:ksz}
\end{equation}

\noindent $\sigma_\mathrm{T}$ is the Thompson cross section, $v_r$ is the bulk line-of-sight velocity of the electrons, and $\tau$ is the optical depth. The similarity between the DM and kSZ signals is apparent, differing mainly in the velocity weighting. Following convention, the optical depth is assumed to be small such that the $e^{-\tau(\chi)}$ term can be dropped. We also assume the velocity is due to the motion of a single halo, leading to:

\begin{equation}
    \frac{\Delta T_\mathrm{kSZ}}{T_\mathrm{CMB}} \approx - \tau \frac{v_\mathrm{halo}}{c}.
\end{equation}

\noindent DM can be turned into an optical depth as 

\begin{equation}
    \delta\tau = \sigma_\mathrm{T} (1+z) \delta\mathrm{DM},
    \label{eqn:dtau}
\end{equation}

\noindent so optical depth is a natural space in which to compare the two signals. 

The kSZ signal is typically measured as an integral over sky area with compensated aperture photometry (CAP). To maintain consistency with kSZ literature, we will use $\tau$ to refer to the integrated quantity, and $\tau^\mathrm{CAP}$ if it has been CAP filtered. We will use $\delta\tau$ to refer to the unintegrated quantity obtained from Equation \ref{eqn:dtau}. For an aperature of size $\theta_i$, the signal is then

\begin{equation}
    \tau^\mathrm{CAP}_\mathrm{kSZ} (\theta_i)= \frac{c}{v_\mathrm{halo}} \int d\theta^2 \frac{\Delta T_\mathrm{kSZ}}{T_\mathrm{CMB}}(\theta)\,   W_{\theta_i}(\theta),
\end{equation}

\noindent with $W_{\theta_i}(\theta)$ the CAP filter:

\begin{equation}
    \label{eqn:cap}
    W_{\theta_i}(\theta) =
        \begin{cases}
          1      & \text{if } 0\le\theta < \theta_i,\\
          -1       & \text{if } \theta_i \le\theta < 
          \sqrt{2}\theta_i,\\
          0        & \text{otherwise.}
        \end{cases}
\end{equation}

The kSZ effect is used to study both the electrons (e.g. \cite{hadzhiyska2026precisionkinematicsunyaevzeldovichmeasurements}) and cosmological velocity field (e.g. \cite{chaussidon2026measurementgalaxyvelocitypowerspectrum}). In either case, a main challenge is handling the so-called kSZ optical depth degeneracy between these two elements of the signal (Equation \ref{eqn:ksz}). FRBs, which are sensitive only to the electrons, are a promising approach to breaking this degeneracy \citep{Madhavacheril_2019}. The optical depth degeneracy is typically quantified in two ways, depending on the goal of the study. 

First, when using the kSZ effect to study electrons, the halo velocities are ``reconstructed" from the surrounding large scale structure and linear theory. The measured signal is degenerate in $\tau$ and the combination $r/\sigma_v$, where $r$ quantifies how well the reconstruction correlates with the true velocities and $\sigma_v$ is the characteristic velocity dispersion of the sample (the raw kSZ signal $T_\mathrm{kSZ}$ is proportional to $\sigma_v/r$, when converted to an optical depth $\tau$ it becomes inversely proportional to that quantity) \citep{hadzhiyska2026precisionkinematicsunyaevzeldovichmeasurements}. According to linear theory:

\begin{equation}
    v_r(\mu,k) \propto f \delta_m(k),
\end{equation}

\noindent where $f$ is the logarithmic growth rate of structure and $\delta_m$ is the matter overdensity field ($f \equiv \frac{d\ln D}{d\ln a}$ with $D$ the growth factor). 
Then the velocity dispersion $\sigma_v$ is proportional to the product $f\sigma_8$. We can take one of two perspectives: 1) Cosmology is fixed, and the ratio of the kSZ and FRB signals can be used to check and calibrate the correlation coefficient $r$, or 2) We believe the velocity reconstruction and therefore the ratio of the kSZ and FRB signals can be used to constrain the growth rate $f\sigma_8$ \citep{Madhavacheril_2019}.

Second, when using kSZ to study the cosmological velocity field, the electrons become a nuisance parameter. The measured signal is degenerate with the velocity bias factor $b_v$. $b_v$ is an integral over the electron-galaxy power spectrum $P_{ge}$ on small scales \citep{Madhavacheril_2019}:

\begin{equation}
    \label{eqn:bv}
    b_v = \frac{\int dk F(k) P_{ge}^\mathrm{true} (k)}{\int dk F(k) P_{ge}^\mathrm{fid} (k)},
\end{equation}

\begin{equation}
    F(k) = k \frac{P_{ge}^\mathrm{fid} (k)}{P_{gg}^\mathrm{tot} (k)} \left( \frac{1}{C^{TT,\mathrm{tot}}_l} \right)_{l=k\chi_g}.
\end{equation}

\noindent Here $F(k)$ is a window function, $P_{ge}^\mathrm{fid}$ is the fiducial electron-galaxy power spectrum assumed in the measurement, $P_{gg}^\mathrm{tot} (k)$ is the galaxy auto power spectrum, $C^{TT,\mathrm{tot}}_l$ is the total temperature power spectrum of the CMB, and the integration runs over the range $0.1\lesssim k/\mathrm{Mpc}^{-1}\lesssim 10$. Note that $b_v$ is specific to each analysis through the details of the CMB map, galaxy sample, and assumed $P_{ge}^\mathrm{fid}$. We would like to demonstrate that our FRB measurement can constrain $b_v$, but the exact value is unimportant. To calculate $b_v$ we use the window function $F(k)$ from \cite{chaussidon2026measurementgalaxyvelocitypowerspectrum}, keeping in mind that this is computed for a different galaxy sample. We take $P_{ge}^\mathrm{fid}$ as the default Battaglia profile \citep{Battaglia_2016} in \texttt{hmvec}, which is the same used in \cite{chaussidon2026measurementgalaxyvelocitypowerspectrum}

\section{New DSA-110 FRB discoveries}
\label{sec:newdsa110}

\begin{figure}[h!]
  \centering
  \includegraphics[width=0.9\linewidth]{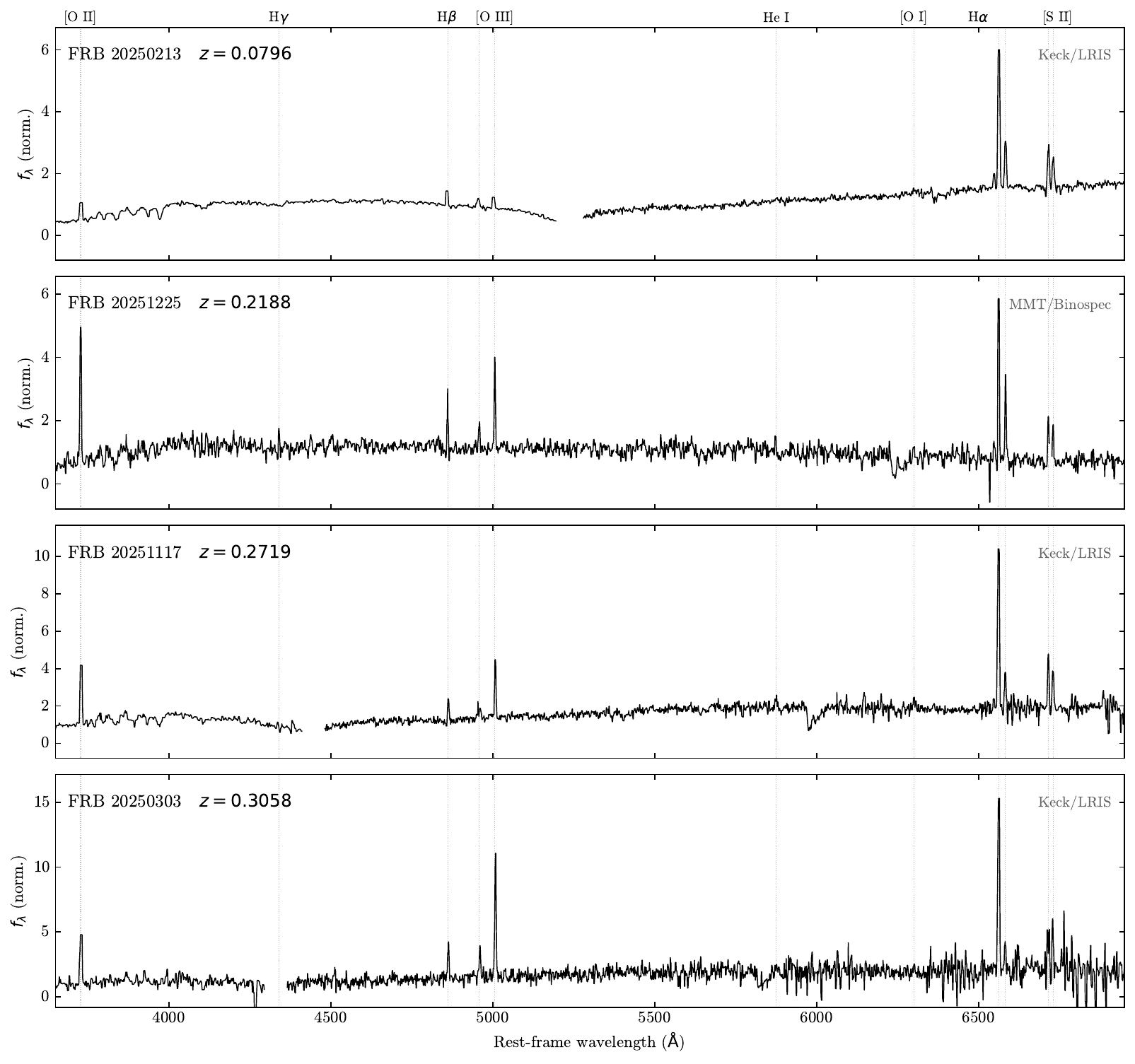}
  \caption{Host galaxy spectra for four new FRBs, obtained with either Keck/LRIS or MMT/Binospec.}
  \label{fig:spectra}
\end{figure}

Included in our analysis are 11 unpublished FRBs discovered by the The Deep Synoptic Array 110 (DSA-110). 
The DSA-110 is a radio interferometer built for detecting and 
localizing FRBs to their host galaxy \citep{ravidsa}. The DSA-110 operates at 1.28-1.54 MHz 
at Caltech's Owens Valley Radio Observatory (OVRO). Its localization precision is 1-2'' 
for most FRBs. 
The localizations and host galaxy properties of seven of 11 are detailed here; the remaining four will be reported in \citet{verdi26}. All new FRBs are non-repeaters thus far. Figure~\ref{fig:spectra} shows spectra for four host galaxies. The host of FRB\,20260407E is 
in the DESI DR1 spectroscopic galaxy catalog and is 
not presented here. Two sources, FRB\,20251125A and FRB\,20251225B, will be presented in host galaxy papers that are in preparation.  


\vspace{3mm}

\noindent \textit{FRB\,20250213C (Harmony)} The optimal total DM 
of this source was 173.3\,pc\,cm$^{-3}$ at $\mathrm{R.A.} = 07^{\rm h}06^{\rm m}27\fs7$ and $\mathrm{Decl.} = +70\degr11\arcmin44\farcs0$. Detection S/N was 16.9 at MJD 60719.2296360. The host galaxy position is 07:06:27.93 +70:11:44.7, detected in DESI Legacy DR10 and in PanSTARRS. It is a bright ($g = 18.49$, $r = 17.95$, $z = 17.60$), nearby star forming galaxy, consistent with its low DM. Spectroscopic follow up with Keck/LRIS confirmed that $z=0.0796\pm0.0003$. 

\vspace{3mm}

\noindent \textit{FRB\,20250303A (Irving)} The radio localization of this source is
$\mathrm{R.A.} = 06^{\rm h}59^{\rm m}11\fs1$ and
$\mathrm{Decl.} = +13\degr17\arcmin00\farcs8$. It was detected at MJD 60737.1705390 with S/N=18.1. The host 
galaxy position is $\mathrm{R.A.} = 06^{\rm h}59^{\rm m}11.20\fs1$ and
$\mathrm{Decl.} = +13\degr17\arcmin01\farcs7$. Our Keck/LRIS spectroscopy of the host galaxy of FRB\,20250303 reveals a suite
of nebular emission lines at a common redshift of $z = 0.3058 \pm 0.0004$: [O\,\textsc{ii}]\,$\lambda3727$, H$\beta$,
[O\,\textsc{iii}]\,$\lambda\lambda4959,5007$, H$\alpha$,
[N\,\textsc{ii}]\,$\lambda6584$, and the
[S\,\textsc{ii}]\,$\lambda\lambda6716,6731$ doublet. The line ratios
([N\,\textsc{ii}]/H$\alpha \approx 0.1$,
[O\,\textsc{iii}]\,$\lambda5007$/H$\beta \approx 3$) place the galaxy in the
star-forming region of the BPT diagram, indicating a moderately
metal-poor, high-excitation star-forming galaxy with no evidence for AGN activity.

\vspace{3mm}

\noindent \textit{FRB\,20251117A (Mira):} The optimal total DM of this source is DM=473.9\,pc\,cm$^{-3}$. It was detected by the real-time system with S/N=8.7 at MJD 60996.2198775 and localized to 
$\mathrm{R.A.} = 01^{\rm h}08^{\rm m}05\fs2$\,(1.1'') and $\mathrm{Decl.} = +14\degr51\arcmin44\farcs2$\,(0.7''). The most likely host galaxy is located at $\mathrm{R.A.} = 01^{\rm h}08^{\rm m}05\fs4$ and $\mathrm{Decl.} = +14\degr51\arcmin43\farcs9$. It was detected in DESI Legacy Surveys DR10 with $g = 22.10$, $r = 21.36$, and $z = 20.90$. 
We obtained Keck/LRIS longslit spectroscopy of the host galaxy of FRB on UT 2025 December 17, using the 1\arcsec\ slit with the 400/3400
grism (D560 dichroic) on the blue arm ($5\times1200$\,s) and the 400/8500
grating on the red arm ($10\times560$\,s). The spectrum exhibits nebular emission lines at a common redshift of $z = 0.2719 \pm 0.0003$:
[O\,\textsc{ii}]\,$\lambda3727$, H$\beta$,
[O\,\textsc{iii}]\,$\lambda\lambda4959,5007$, He\,\textsc{i}\,$\lambda5876$,
[O\,\textsc{i}]\,$\lambda6300$, H$\alpha$, [N\,\textsc{ii}]\,$\lambda6584$,
and the [S\,\textsc{ii}]\,$\lambda\lambda6716,6731$ doublet, yielding a
secure spectroscopic redshift. The line ratios
([N\,\textsc{ii}]/H$\alpha \approx 0.2$,
[O\,\textsc{iii}]\,$\lambda5007$/H$\beta \approx 1$) are consistent with a
star-forming galaxy of roughly solar metallicity, with no indication of AGN activity.

\vspace{3mm}

\noindent \textit{FRB\,20251225B (Grinch)}: The FRB was detected with a total DM of 
300.3\,pc\,cm$^{-3}$ at S/N of 12.7 at MJD 61034.6433580. 
Its radio localization was 
$\mathrm{R.A.} = 13^{\rm h}48^{\rm m}22\fs8$\,(1'') and
$\mathrm{Decl.} = +17\degr30\arcmin28\farcs2$\,(0.75''). We obtained MMT/Binospec longslit spectroscopy of the host galaxy of FRB\,20251225 at $\mathrm{R.A.} = 13^{\rm h}48^{\rm m}22\fs8$ and
$\mathrm{Decl.} = +17\degr30\arcmin27\farcs9$. The spectrum shows 
nebular emission lines at  $z = 0.2188$:
H$\alpha$, H$\beta$, [O\,\textsc{ii}]\,$\lambda3727$,
[O\,\textsc{iii}]\,$\lambda\lambda4959,5007$,
[N\,\textsc{ii}]\,$\lambda6584$, and the
[S\,\textsc{ii}]\,$\lambda\lambda6716,6731$ doublet. The emission is spatially extended along
the slit, consistent with ongoing star formation distributed across the
galaxy.

\vspace{3mm}

\noindent \textit{FRB\,20251125A (Copernicus):} The real-time detection significance was S/N=24.4 with an optimal DM of 610.5\,pc\,cm$^{-3}$ at MJD 61004.7393690. The radio localization and host-galaxy position are $\mathrm{R.A.} = 14^{\rm h}14^{\rm m}13\fs9$\,(1.1''), $\mathrm{Decl.} = +15\degr19\arcmin17\farcs4$\,(0.6''), and $\mathrm{R.A.} = 14^{\rm h}14^{\rm m}14\fs00$, $\mathrm{Decl.} = +15\degr19\arcmin17\farcs0$, respectively. We obtained Keck/LRIS longslit spectroscopy of the host galaxy of FRB, finding an unresolved emission line on the blue side identified as
[O\,\textsc{ii}]\,$\lambda3727$ and a red spectrum with a $\sim$$10\sigma$ detection of
H$\alpha$ at $\lambda_{\rm obs} \approx 8709$\,\AA, along with marginal
[N\,\textsc{ii}]\,$\lambda6583$ and [S\,\textsc{ii}]\,$\lambda\lambda6716,6731$
emission at the same redshift. The redshift of this galaxy is $z = 0.327$.

\vspace{3mm}

\vspace{3mm}

\noindent \textit{FRB\,20220801A (Augustine)}: This source was localized to $\mathrm{R.A.} = 03^{\rm h}39^{\rm m}51\fs23$, $\mathrm{Decl.} = +70\degr10\arcmin53\farcs7$. It was detected at MJD 59792.6177637 with an optimal DM of $413.0\ {\rm pc\ cm^{-3}}$ and S/N$\approx10$. The FRB position sits at Galactic latitude +11.8 behind a significant dust screen with nearly 2 mag of extinction in $r$. The FRB is coincident with a compact, reddened galaxy in a tight pair with another galaxy at $z = 0.3675$. Host imaging at this FRB position required long integrations with P200/WIRC in $J$, $H$, and $K$ bands, as no host candidates were found in public imaging data. We obtained MMT/Binospec longslit spectroscopy of the two galaxies, and a Keck/LRIS spectrum of the apparent host. We will detail the host galaxy and its companion in a paper currently in preparation, as the host system is unusual. We note that removing FRB\,20220801A from 
this work's cross-correlation does not noticeably impact 
the results.

\vspace{3mm}

\noindent \textit{FRB\,20260407E (Aviva):} The source was detected with a total DM of 
494.1\,pc\,cm$^{-3}$ at S/N of 7.6. Its radio localization is at $\mathrm{R.A.} = 16^{\rm h}50^{\rm m}40\fs4$ and $\mathrm{Decl.} = +14\degr31\arcmin57\farcs6$. The host galaxy redshift is in the DESI DR1 spectroscopic galaxy catalog with $z=0.487767\pm0.000024$ \citep{DESIDR1}, obviating the need for targeted follow up by the DSA-110 collaboration. There is a foreground $G=18.6$ star in Gaia DR3 roughly 3'' from the FRB position and galaxy.

\vspace{0.5cm}

The new DSA-110 FRBs are included at the bottom of Table \ref{tab:sample}.

\begin{longtable}{lcccccc}
\caption{Localized FRB sample.} \label{tab:sample} \\
\toprule
Name & RA & DEC & DM$_\mathrm{obs}$ & z & Telescope & References \\
\midrule
\endfirsthead
\caption[]{Localized FRB sample continued.} \\
\toprule
Name & RA & DEC & DMobs & z & Telescope & References \\
\midrule
\endhead
\midrule
\midrule
\endfoot
\bottomrule
\endlastfoot
FRB20220307B & 350.87450 & 72.19239 & 499.1 & 0.25 & DSA & \cite{law2024deepsynopticarrayscience}, \cite{Sherman_2024} \\
FRB20220920A & 240.25708 & 70.91880 & 315.0 & 0.16 & DSA & \cite{law2024deepsynopticarrayscience}, \cite{Sherman_2024} \\
FRB20201124A & 77.01462 & 26.06070 & 411.0 & 0.10 & ASKAP & \cite{Kumar2021} \\
FRB20221029A & 141.96342 & 72.45232 & 1391.0 & 0.97 & DSA & \cite{Sherman_2024}, \cite{sharma2024preferentialoccurrencefastradio} \\
FRB20220330D & 163.75125 & 70.35075 & 468.1 & 0.37 & DSA & \cite{Sherman_2024}, \cite{sharma2024preferentialoccurrencefastradio} \\
FRB20220509G & 282.67000 & 70.24383 & 269.5 & 0.09 & DSA & \cite{Connor_2023} \\
FRB20190102C & 322.41567 & -79.47569 & 364.5 & 0.29 & ASKAP & \cite{Bhandari2020}, \cite{Day2020} \\
FRB20220506D & 318.04483 & 72.82728 & 396.9 & 0.30 & DSA & \cite{law2024deepsynopticarrayscience}, \cite{Sherman_2024} \\
FRB20221101B & 342.21579 & 70.68150 & 490.7 & 0.24 & DSA & \cite{Sherman_2024}, \cite{sharma2024preferentialoccurrencefastradio} \\
FRB20200120E & 149.47783 & 68.81889 & 87.8 & 0.00 & CHIME & \cite{Bhardwaj2021}, \cite{Kirsten_2022} \\
FRB20220208A & 322.57513 & 70.04104 & 437.0 & 0.35 & DSA & \cite{Sherman_2024}, \cite{sharma2024preferentialoccurrencefastradio} \\
FRB20230521B & 351.03600 & 71.13803 & 1345.7 & 1.35 & DSA & \cite{Connor2025} \\
FRB20230814B & 335.97480 & 73.02590 & 696.4 & 0.55 & DSA & \cite{Connor2025} \\
FRB20220204A & 274.22625 & 69.72250 & 612.2 & 0.40 & DSA & \cite{Sherman_2024}, \cite{sharma2024preferentialoccurrencefastradio} \\
FRB20220418A & 219.10557 & 70.09594 & 623.5 & 0.62 & DSA & \cite{law2024deepsynopticarrayscience}, \cite{Sherman_2024} \\
FRB20190711A & 329.41724 & -80.35803 & 587.9 & 0.52 & ASKAP & \cite{Macquart_2020}, \cite{Day2020} \\
FRB20181112A & 327.34846 & -52.97094 & 589.3 & 0.48 & ASKAP & \cite{Prochaska_2019}, \cite{Day2020} \\
FRB20220310F & 134.72050 & 73.49083 & 462.1 & 0.48 & DSA & \cite{law2024deepsynopticarrayscience}, \cite{Sherman_2024} \\
FRB20190611B & 320.74558 & -79.39758 & 332.6 & 0.38 & ASKAP & \cite{Macquart_2020}, \cite{Day2020} \\
FRB20180924A & 326.10523 & -40.90003 & 362.2 & 0.32 & ASKAP & \cite{Bannister2019}, \cite{Bhandari2020} \\
FRB20191001A & 323.35168 & -54.74831 & 507.0 & 0.23 & ASKAP & \cite{Bhandari_2020b}, \cite{Day2020} \\
FRB20220319D & 32.17792 & 71.03526 & 111.0 & 0.01 & DSA & \cite{Ravi2025}, \cite{law2024deepsynopticarrayscience} \\
FRB20220207C & 310.19952 & 72.88233 & 262.3 & 0.04 & DSA & \cite{law2024deepsynopticarrayscience}, \cite{Sherman_2024} \\
FRB20221012A & 280.79871 & 70.52421 & 442.2 & 0.28 & DSA & \cite{law2024deepsynopticarrayscience}, \cite{Sherman_2024} \\
FRB20190608B & 334.01988 & -7.89825 & 340.1 & 0.12 & ASKAP & \cite{Bhandari2020}, \cite{Day2020} \\
FRB20220726A & 73.94554 & 69.92953 & 686.5 & 0.36 & DSA & \cite{Sherman_2024}, \cite{sharma2024preferentialoccurrencefastradio} \\
FRB20220825A & 311.98145 & 72.58497 & 651.2 & 0.24 & DSA & \cite{Sherman_2024}, \cite{sharma2024preferentialoccurrencefastradio} \\
FRB20220831A & 338.69554 & 70.53844 & 1146.2 & 0.26 & DSA & \cite{Sherman_2024}, \cite{Connor2025} \\
FRB20171020A & 333.82729 & -19.66979 & 114.1 & 0.01 & ASKAP & \cite{Mahony_2018} \\
FRB20180301A & 93.22683 & 4.67106 & 536.0 & 0.33 & ASKAP & \cite{Bhandari_2022} \\
FRB20190523A & 207.06500 & 72.46972 & 760.8 & 0.66 & DSA & \cite{ravi2019} \\
FRB20190614D & 65.07552 & 73.70674 & 959.0 & 0.60 & VLA & \cite{Law_2020} \\
FRB20190714A & 183.97971 & -13.02100 & 504.1 & 0.24 & ASKAP & \cite{Heintz2020}, \cite{Day2020} \\
FRB20191228A & 344.43054 & -29.59412 & 298.0 & 0.24 & ASKAP & \cite{Bhandari_2022}, \cite{Day2020} \\
FRB20200430A & 229.70645 & 12.37633 & 380.0 & 0.16 & ASKAP & \cite{Bhandari_2022}, \cite{Day2020} \\
FRB20200906A & 53.49554 & -14.08301 & 577.8 & 0.37 & ASKAP & \cite{Bhandari_2022}, \cite{Day2020} \\
FRB20201123A & 263.67000 & -50.76000 & 433.6 & 0.05 & MeerKAT & \cite{Rajwade_2022} \\
FRB20210320C & 204.45877 & -16.12267 & 384.6 & 0.28 & ASKAP & \cite{Shannon_2025}, \cite{gordon2025mappingspatialdistributionfast} \\
FRB20210410D & 326.08625 & -79.31819 & 575.0 & 0.14 & MeerKAT & \cite{Caleb_2023} \\
FRB20210807D & 299.22143 & -0.76236 & 251.3 & 0.13 & ASKAP & \cite{Shannon_2025},  \cite{gordon2025mappingspatialdistributionfast} \\
FRB20211127I & 199.80881 & -18.83791 & 235.0 & 0.05 & ASKAP & \cite{Shannon_2025},  \cite{gordon2025mappingspatialdistributionfast} \\
FRB20211203C & 204.56250 & -31.38027 & 635.0 & 0.34 & ASKAP & \cite{Shannon_2025},  \cite{gordon2025mappingspatialdistributionfast} \\
FRB20211212A & 157.35078 & 1.36043 & 209.0 & 0.07 & ASKAP & \cite{Shannon_2025},  \cite{gordon2025mappingspatialdistributionfast} \\
FRB20220105A & 208.80336 & 22.46623 & 580.0 & 0.28 & ASKAP & \cite{Shannon_2025}, \cite{gordon2025mappingspatialdistributionfast} \\
FRB20220501C & 352.37915 & -32.49072 & 449.5 & 0.38 & ASKAP & \cite{Shannon_2025} \\
FRB20220725A & 353.31522 & -35.99026 & 290.4 & 0.19 & ASKAP & \cite{Shannon_2025} \\
FRB20220914A & 282.05679 & 73.33691 & 631.0 & 0.11 & DSA & \cite{Connor_2023} \\
FRB20220918A & 17.59214 & -70.81138 & 656.8 & 0.49 & ASKAP & \cite{Shannon_2025} \\
FRB20221027A & 130.87200 & 72.10094 & 452.5 & 0.54 & DSA & \cite{Sherman_2024}, \cite{sharma2024preferentialoccurrencefastradio} \\
FRB20221106A & 56.70481 & -25.56981 & 343.8 & 0.20 & ASKAP & \cite{Shannon_2025} \\
FRB20221113A & 71.41100 & 70.30739 & 411.4 & 0.25 & DSA & \cite{sharma2024preferentialoccurrencefastradio} \\
FRB20221116A & 21.21096 & 72.65375 & 640.6 & 0.28 & DSA & \cite{sharma2024preferentialoccurrencefastradio} \\
FRB20221203A & 315.12954 & 72.03756 & 602.2 & 0.51 & DSA & \cite{Connor2025} \\
FRB20221219A & 257.62979 & 71.62684 & 706.7 & 0.55 & DSA & \cite{sharma2024preferentialoccurrencefastradio} \\
FRB20230124A & 231.91700 & 70.96809 & 590.0 & 0.09 & DSA & \cite{sharma2024preferentialoccurrencefastradio} \\
FRB20230216A & 156.47217 & 3.43682 & 828.3 & 0.53 & DSA & \cite{sharma2024preferentialoccurrencefastradio} \\
FRB20230307A & 177.78158 & 71.69513 & 610.1 & 0.27 & DSA & \cite{sharma2024preferentialoccurrencefastradio} \\
FRB20230501A & 340.02708 & 70.92214 & 533.7 & 0.30 & DSA & \cite{sharma2024preferentialoccurrencefastradio} \\
FRB20230526A & 22.23262 & -52.71733 & 361.4 & 0.16 & ASKAP & \cite{Shannon_2025} \\
FRB20230626A & 235.62958 & 71.13355 & 451.2 & 0.33 & DSA & \cite{sharma2024preferentialoccurrencefastradio} \\
FRB20230628A & 166.78671 & 72.28184 & 345.1 & 0.13 & DSA & \cite{sharma2024preferentialoccurrencefastradio} \\
FRB20230708A & 303.11554 & -55.35627 & 411.5 & 0.10 & ASKAP & \cite{Shannon_2025} \\
FRB20230712A & 167.35853 & 72.55778 & 587.0 & 0.45 & DSA & \cite{sharma2024preferentialoccurrencefastradio} \\
FRB20230814A & 335.97475 & 73.02591 & 696.4 & 0.55 & DSA & \cite{Connor2025} \\
FRB20230902A & 52.13977 & -47.33350 & 440.1 & 0.36 & ASKAP & \cite{Shannon_2025} \\
FRB20231120A & 143.98396 & 73.28467 & 438.9 & 0.04 & DSA & \cite{sharma2024preferentialoccurrencefastradio} \\
FRB20231123B & 242.53817 & 70.78506 & 396.7 & 0.26 & DSA & \cite{sharma2024preferentialoccurrencefastradio} \\
FRB20231220A & 123.90871 & 73.65992 & 491.2 & 0.34 & DSA & \cite{Connor2025} \\
FRB20231226A & 155.36374 & 6.11026 & 329.9 & 0.16 & ASKAP & \cite{Shannon_2025} \\
FRB20240119A & 224.46717 & 71.61176 & 483.1 & 0.37 & DSA & \cite{Connor2025} \\
FRB20240123A & 68.26250 & 71.94528 & 1462.0 & 0.97 & DSA & \cite{Connor2025} \\
FRB20240201A & 149.90558 & 14.08803 & 374.5 & 0.04 & ASKAP & \cite{Shannon_2025} \\
FRB20240210A & 8.77958 & -28.27075 & 283.7 & 0.02 & ASKAP & \cite{Shannon_2025} \\
FRB20240213A & 166.16829 & 74.07539 & 357.4 & 0.12 & DSA & \cite{Connor2025} \\
FRB20240215A & 268.44125 & 70.23236 & 549.5 & 0.21 & DSA & \cite{Connor2025} \\
FRB20240229A & 169.98354 & 70.67622 & 491.1 & 0.29 & DSA & \cite{Connor2025} \\
FRB20240304A & 136.33083 & -16.16663 & 652.6 & 0.24 & ASKAP & \cite{Shannon_2025} \\
FRB20240318A & 150.39316 & 37.61636 & 256.4 & 0.11 & ASKAP & \cite{Shannon_2025} \\
FRB20230203A & 151.66159 & 35.69410 & 417.3 & 0.15 & CHIME & \cite{frbcollaboration2025cataloglocaluniversefast}  \\
FRB20230222B & 238.73912 & 30.89870 & 188.0 & 0.11 & CHIME & \cite{frbcollaboration2025cataloglocaluniversefast}  \\
FRB20230703A & 184.62445 & 48.72993 & 290.7 & 0.12 & CHIME & \cite{frbcollaboration2025cataloglocaluniversefast}  \\
FRB20230730A & 54.66456 & 33.15930 & 312.6 & 0.21 & CHIME & \cite{frbcollaboration2025cataloglocaluniversefast}  \\
FRB20230926A & 269.12488 & 41.81430 & 222.8 & 0.06 & CHIME & \cite{frbcollaboration2025cataloglocaluniversefast}  \\
FRB20231005A & 246.02800 & 35.44871 & 188.8 & 0.07 & CHIME & \cite{frbcollaboration2025cataloglocaluniversefast}  \\
FRB20231011A & 18.24110 & 41.74910 & 186.4 & 0.08 & CHIME & \cite{frbcollaboration2025cataloglocaluniversefast}  \\
FRB20231017A & 346.75429 & 36.65268 & 344.1 & 0.24 & CHIME & \cite{frbcollaboration2025cataloglocaluniversefast}  \\
FRB20231025B & 270.78807 & 63.98908 & 368.4 & 0.32 & CHIME & \cite{frbcollaboration2025cataloglocaluniversefast}  \\
FRB20231123A & 82.62325 & 4.47554 & 302.1 & 0.07 & CHIME & \cite{frbcollaboration2025cataloglocaluniversefast}  \\
FRB20231128A & 199.57820 & 42.99271 & 332.4 & 0.11 & CHIME & \cite{frbcollaboration2025cataloglocaluniversefast}  \\
FRB20231201A & 54.58929 & 26.81767 & 169.4 & 0.12 & CHIME & \cite{frbcollaboration2025cataloglocaluniversefast}  \\
FRB20231204A & 207.99903 & 48.11600 & 222.0 & 0.06 & CHIME & \cite{frbcollaboration2025cataloglocaluniversefast}  \\
FRB20231206A & 112.44285 & 56.25627 & 457.8 & 0.07 & CHIME & \cite{frbcollaboration2025cataloglocaluniversefast}  \\
FRB20231223C & 259.54465 & 29.49794 & 179.5 & 0.11 & CHIME & \cite{frbcollaboration2025cataloglocaluniversefast}  \\
FRB20231229A & 26.46783 & 35.11292 & 198.5 & 0.02 & CHIME & \cite{frbcollaboration2025cataloglocaluniversefast}  \\
FRB20231230A & 72.79761 & 2.39398 & 131.4 & 0.03 & CHIME & \cite{frbcollaboration2025cataloglocaluniversefast}  \\
FRB20231230D & 118.76079 & 8.51410 & 676.5 & 0.51 & ASKAP & \cite{gordon2025mappingspatialdistributionfast} \\
FRB20240117B & 53.87750 & -15.85231 & 881.0 & 0.64 & ASKAP & \cite{gordon2025mappingspatialdistributionfast} \\
FRB20220222C & 203.90450 & -28.02690 & 1071.2 & 0.85 & MeerKAT & \cite{pastormarazuela2025localisationhostgalaxyidentification} \\
FRB20230125D & 150.20500 & -31.54470 & 640.1 & 0.33 & MeerKAT & \cite{pastormarazuela2025localisationhostgalaxyidentification} \\
FRB20230613A & 356.85270 & -27.05280 & 483.5 & 0.39 & MeerKAT & \cite{pastormarazuela2025localisationhostgalaxyidentification} \\
FRB20240224A & 70.11490 & 73.51280 & 881.2 & 0.37 & DSA & \cite{Connor2025} \\
FRB20240114A & 321.91600 & 4.32900 & 330.0 & 0.13 & CHIME & \cite{2024ATel16613....1B} \\
FRB20210603A & 10.27400 & 21.22600 & 500.1 & 0.18 & CHIME & \cite{Cassanelli_2024}\\
FRB20220529 & 19.10400 & 20.63200 & 250.2 & 0.18 & CHIME & \cite{Li_2026} \\
FRB20220717A & 293.30400 & -19.28800 & 637.3 & 0.36 & NaN & \cite{Rajwade2024} \\
FRB20220912A & 347.27000 & 48.70700 & 219.5 & 0.08 & DSA & \cite{ravi2023} \\
FRB20221022A & 48.62900 & 86.87200 & 116.8 & 0.01 & NaN & \cite{mckinven2024pulsarlikeswingpolarisationposition} \\
FRB20230506C & 12.10000 & 42.00600 & 772.0 & 0.39 & VLA & \cite{Anna_Thomas_2025} \\
FRB20230808F & 53.30400 & -51.93500 & 653.2 & 0.35 & MeerKAT & \cite{Hanmer_2025} \\
FRB20230930A & 10.50700 & 41.41700 & 456.0 & 0.09 & VLA & \cite{Anna_Thomas_2025} \\
FRB20240209A & 289.85000 & 86.06000 & 176.6 & 0.14 & CHIME & \cite{shah2024repeatingfastradioburst}, \cite{eftekhari2024massivequiescentellipticalhost} \\
FRB20241228A & 216.38600 & 12.02500 & 246.5 & 0.16 & CHIME & \cite{curtin2025discoverylocalizationswiftobservedfrb} \\
FRB20250316A & 182.43500 & 58.84900 & 161.8 & 0.01 & CHIME & \cite{frbcollaboration2025frb20250316abrilliantnearby} \\
FRB20240203C & 8.89754 & 15.84086 & 454.0 & 0.24 & ASKAP & \cite{gordon2025mappingspatialdistributionfast} \\
FRB20240312D & 51.62338 & -54.60097 & 330.0 & 0.05 & ASKAP & \cite{gordon2025mappingspatialdistributionfast} \\
FRB20240525A & 6.78287 & -6.88961 & 491.6 & 0.33 & ASKAP & \cite{gordon2025mappingspatialdistributionfast} \\
FRB20240615B & 33.96363 & -14.62450 & 201.0 & 0.07 & ASKAP & \cite{gordon2025mappingspatialdistributionfast} \\
FRB20241027B & 36.02900 & -20.70353 & 320.0 & 0.34 & ASKAP & \cite{gordon2025mappingspatialdistributionfast} \\

\midrule[\heavyrulewidth]             
\multicolumn{7}{c}{\textbf{New DSA-110 FRBs}} \\[0.5ex]  
\midrule                              

FRB\,20230913G & 305.03717 & 70.79277 & 518.7 & 0.30 & DSA & Verdi+2026 \\
FRB\,20240104A & 348.87400 & 72.82058 & 1351.0 & 1.33 & DSA & Verdi+2026 \\
FRB\,20240203D & 312.61912 & 73.90000 & 272.6 & 0.07 & DSA & Verdi+2026 \\
FRB\,20250213C & 106.61542 & 70.19556 & 173.3 & 0.08 & DSA & This work \\
FRB\,20250303A & 104.79625 & 13.28356 & 489.5 & 0.30 & DSA & This work \\
FRB\,20250518  & 207.99179 & 71.28235 & 921.2 & 0.64 & DSA & Verdi+2026 \\
FRB\,20251117A & 17.02171 & 14.86229 & 473.9 & 0.27 & DSA & This work \\
FRB\,20251225A & 213.55792 & 15.32150 & 610.5 & 0.33 & DSA & This work \\
FRB\,20251225B & 207.09512 & 17.50784 & 300.3 & 0.22 & DSA & This work \\
FRB\,20220801A & 54.96346 & 70.18158 & 413.0 & 0.37 & DSA & This work \\
FRB\,20260407E & 252.66833 & 14.53267 & 494.1 & 0.49 & DSA & This work \\

\end{longtable}

\bibliography{sample701}{}

@misc{kovac2025baryonificationiiconstrainingfeedback,
      title={Baryonification II: Constraining feedback with X-ray and kinematic Sunyaev-Zel'dovich observations}, 
      author={Michael Kovač and Andrina Nicola and Jozef Bucko and Aurel Schneider and Robert Reischke and Sambit K. Giri and Romain Teyssier and Matthieu Schaller and Joop Schaye},
      year={2025},
      eprint={2507.07991},
      archivePrefix={arXiv},
      primaryClass={astro-ph.CO},
      url={https://arxiv.org/abs/2507.07991}, 
}

@ARTICLE{Tendulhar2017,
       author = {{Tendulkar}, S.~P. and {Bassa}, C.~G. and {Cordes}, J.~M. and {Bower}, G.~C. and {Law}, C.~J. and {Chatterjee}, S. and {Adams}, E.~A.~K. and {Bogdanov}, S. and {Burke-Spolaor}, S. and {Butler}, B.~J. and {Demorest}, P. and {Hessels}, J.~W.~T. and {Kaspi}, V.~M. and {Lazio}, T.~J.~W. and {Maddox}, N. and {Marcote}, B. and {McLaughlin}, M.~A. and {Paragi}, Z. and {Ransom}, S.~M. and {Scholz}, P. and {Seymour}, A. and {Spitler}, L.~G. and {van Langevelde}, H.~J. and {Wharton}, R.~S.},
        title = "{The Host Galaxy and Redshift of the Repeating Fast Radio Burst FRB 121102}",
      journal = {\apjl},
         year = 2017,
        month = jan,
       volume = {834},
       number = {2},
          eid = {L7},
        pages = {L7},
          doi = {10.3847/2041-8213/834/2/L7},
archivePrefix = {arXiv},
       eprint = {1701.01100},
 primaryClass = {astro-ph.HE},
       adsurl = {https://ui.adsabs.harvard.edu/abs/2017ApJ...834L...7T}
}

@article{Niu_2022,
   title={A repeating fast radio burst associated with a persistent radio source},
   volume={606},
   ISSN={1476-4687},
   url={http://dx.doi.org/10.1038/s41586-022-04755-5},
   DOI={10.1038/s41586-022-04755-5},
   number={7916},
   journal={Nature},
   publisher={Springer Science and Business Media LLC},
   author={Niu, C.-H. and Aggarwal, K. and Li, D. and Zhang, X. and Chatterjee, S. and Tsai, C.-W. and Yu, W. and Law, C. J. and Burke-Spolaor, S. and Cordes, J. M. and Zhang, Y.-K. and Ocker, S. K. and Yao, J.-M. and Wang, P. and Feng, Y. and Niino, Y. and Bochenek, C. and Cruces, M. and Connor, L. and Jiang, J.-A. and Dai, S. and Luo, R. and Li, G.-D. and Miao, C.-C. and Niu, J.-R. and Anna-Thomas, R. and Sydnor, J. and Stern, D. and Wang, W.-Y. and Yuan, M. and Yue, Y.-L. and Zhou, D.-J. and Yan, Z. and Zhu, W.-W. and Zhang, B.},
   year={2022},
   month={June}, pages={873–877} }

@article{Shannon_2025,
   title={The commensal real-time ASKAP fast transient incoherent-sum survey},
   volume={42},
   ISSN={1448-6083},
   url={http://dx.doi.org/10.1017/pasa.2025.8},
   DOI={10.1017/pasa.2025.8},
   journal={Publications of the Astronomical Society of Australia},
   publisher={Cambridge University Press (CUP)},
   author={Shannon, Ryan M. and Bannister, Keith W. and Bera, Apurba and Bhandari, Shivani and Day, Cherie K. and Deller, Adam T. and Dial, Tyson and Dobie, Dougal and Ekers, Ron D. and Fong, Wen-fai and Glowacki, Marcin and Gordon, Alexa C. and Gourdji, Kelly and Jaini, Akhil and James, Clancy W. and Kumar, Pravir and Mahony, Elizabeth K. and Marnoch, Lachlan and Muller, August R. and Prochaska, Xavier and Qiu, Hao and Ryder, Stuart D. and Sadler, Elaine M. and Scott, Danica R. and Tejos, N. and Uttarkar, Pavan A. and Wang, Yuanming},
   year={2025} }

@article{Bhardwaj_2021,
   title={A Local Universe Host for the Repeating Fast Radio Burst FRB 20181030A},
   volume={919},
   ISSN={2041-8213},
   url={http://dx.doi.org/10.3847/2041-8213/ac223b},
   DOI={10.3847/2041-8213/ac223b},
   number={2},
   journal={The Astrophysical Journal Letters},
   publisher={American Astronomical Society},
   author={Bhardwaj, M. and Kirichenko, A. Yu. and Michilli, D. and Mayya, Y. D. and Kaspi, V. M. and Gaensler, B. M. and Rahman, M. and Tendulkar, S. P. and Fonseca, E. and Josephy, Alexander and Leung, C. and Merryfield, Marcus and Petroff, Emily and Pleunis, Z. and Sanghavi, Pranav and Scholz, P. and Shin, K. and Smith, Kendrick M. and Stairs, I. H.},
   year={2021},
   month={Sept}, pages={L24} }

@article{Michilli_2023,
   title={Subarcminute Localization of 13 Repeating Fast Radio Bursts Detected by CHIME/FRB},
   volume={950},
   ISSN={1538-4357},
   url={http://dx.doi.org/10.3847/1538-4357/accf89},
   DOI={10.3847/1538-4357/accf89},
   number={2},
   journal={The Astrophysical Journal},
   publisher={American Astronomical Society},
   author={Michilli, Daniele and Bhardwaj, Mohit and Brar, Charanjot and Gaensler, B. M. and Kaspi, Victoria M. and Kirichenko, Aida and Masui, Kiyoshi W. and Mckinven, Ryan and Ng, Cherry and Patel, Chitrang and Sand, Ketan R. and Scholz, Paul and Shin, Kaitlyn and Siegel, Seth R. and Stairs, Ingrid and Cassanelli, Tomas and Cook, Amanda M. and Dobbs, Matt and Dong, Fengqiu Adam and Fonseca, Emmanuel and Ibik, Adaeze and Kaczmarek, Jane and Leung, Calvin and Pearlman, Aaron B. and Petroff, Emily and Pleunis, Ziggy and Rafiei-Ravandi, Masoud and Sanghavi, Pranav and Shaw, J. Richard and Tendulkar, Shriharsh P.},
   year={2023},
   month={June}, pages={134} }

@misc{leung2025stellarmassdispersionmeasurecorrelations,
      title={Stellar Mass-Dispersion Measure Correlations Constrain Baryonic Feedback in Fast Radio Burst Host Galaxies}, 
      author={Calvin Leung and Sunil Simha and Isabel Medlock and Daisuke Nagai and Kiyoshi W. Masui and Lordrick A. Kahinga and Adam E. Lanman and Shion Andrew and Kevin Bandura and Alice P. Curtin and B. M. Gaensler and Nina Gusinskaia and Ronniy C. Joseph and Mattias Lazda and Lluis Mas-Ribas and Bradley W. Meyers and Kenzie Nimmo and Aaron B. Pearlman and J. Xavier Prochaska and Mawson W. Sammons and Kaitlyn Shin and Kendrick Smith and Haochen Wang},
      year={2025},
      eprint={2507.16816},
      archivePrefix={arXiv},
      primaryClass={astro-ph.GA},
      url={https://arxiv.org/abs/2507.16816}, 
}

@misc{bhardwaj2023hostgalaxiesnearbychimefrb,
      title={Host Galaxies for Four Nearby CHIME/FRB Sources and the Local Universe FRB Host Galaxy Population}, 
      author={Mohit Bhardwaj and Daniele Michilli and Aida Yu. Kirichenko and Obinna Modilim and Kaitlyn Shin and Victoria M. Kaspi and Bridget C. Andersen and Tomas Cassanelli and Charanjot Brar and Shami Chatterjee and Amanda M. Cook and Fengqiu Adam Dong and Emmanuel Fonseca and B. M. Gaensler and Adaeze L. Ibik and J. F. Kaczmarek and Adam E. Lanman and Calvin Leung and K. W. Masui and Ayush Pandhi and Aaron B. Pearlman and Ziggy Pleunis and J. Xavier Prochaska and Masoud Rafiei-Ravandi and Ketan R. Sand and Paul Scholz and Kendrick M. Smith},
      year={2023},
      eprint={2310.10018},
      archivePrefix={arXiv},
      primaryClass={astro-ph.HE},
      url={https://arxiv.org/abs/2310.10018}, 
}

@ARTICLE{Ibik2024,
       author = {{Ibik}, Adaeze L. and {Drout}, Maria R. and {Gaensler}, B.~M. and {Scholz}, Paul and {Michilli}, Daniele and {Bhardwaj}, Mohit and {Kaspi}, Victoria M. and {Pleunis}, Ziggy and {Cassanelli}, Tomas and {Cook}, Amanda M. and {Dong}, Fengqiu A. and {Kaczmarek}, Jane F. and {Leung}, Calvin and {Lu}, Katherine J. and {Masui}, Kiyoshi W. and {Pearlman}, Aaron B. and {Rafiei-Ravandi}, Masoud and {Sand}, Ketan R. and {Shin}, Kaitlyn and {Smith}, Kendrick M. and {Stairs}, Ingrid H.},
        title = "{Proposed Host Galaxies of Repeating Fast Radio Burst Sources Detected by CHIME/FRB}",
      journal = {\apj},
         year = 2024,
        month = jan,
       volume = {961},
       number = {1},
          eid = {99},
        pages = {99},
          doi = {10.3847/1538-4357/ad0893},
archivePrefix = {arXiv},
       eprint = {2304.02638},
 primaryClass = {astro-ph.HE},
       adsurl = {https://ui.adsabs.harvard.edu/abs/2024ApJ...961...99I}
}

@misc{moroianu2025milliarcsecondlocalizationassociatesfrb,
      title={A milliarcsecond localization associates FRB 20190417A with a compact persistent radio source and an extreme magneto-ionic environment}, 
      author={Alexandra M. Moroianu and Shivani Bhandari and Maria R. Drout and Jason W. T. Hessels and Danté M. Hewitt and Franz Kirsten and Benito Marcote and Ziggy Pleunis and Mark P. Snelders and Navin Sridhar and Uwe Bach and Emmanuel K. Bempong-Manful and Vladislavs Bezrukovs and Richard Blaauw and Justin D. Bray and Salvatore Buttaccio and Shami Chatterjee and Alessandro Corongiu and Roman Feiler and B. M. Gaensler and Marcin P. Gawroński and Marcello Giroletti and Adaeze L. Ibik and Ramesh Karuppusamy and Mattias Lazda and Calvin Leung and Michael Lindqvist and Kiyoshi W. Masui and Daniele Michilli and Kenzie Nimmo and Omar S. Ould-Boukattine and Ayush Pandhi and Zsolt Paragi and Aaron B. Pearlman and Weronika Puchalska and Paul Scholz and Kaitlyn Shin and Jurjen J. Sluman and Matteo Trudu and David Williams-Baldwin and Jun Yang},
      year={2025},
      eprint={2509.05174},
      archivePrefix={arXiv},
      primaryClass={astro-ph.HE},
      url={https://arxiv.org/abs/2509.05174}, 
}

@misc{frbcollaboration2025cataloglocaluniversefast,
      title={A Catalog of Local Universe Fast Radio Bursts from CHIME/FRB and the KKO Outrigger}, 
      author={Mandana Amiri and Daniel Amouyal and Bridget C. Andersen and Shion Andrew and Kevin Bandura and Mohit Bhardwaj and P. J. Boyle and Charanjot Brar and Alyssa Cassity and Shami Chatterjee and Alice P. Curtin and Matt Dobbs and Fengqiu Adam Dong and Yuxin Dong and Gwendolyn M. Eadie and Tarraneh Eftekhari and Wen-fai Fong and Emmanuel Fonseca and B. M. Gaensler and Mark Halpern and Jason W. T. Hessels and Hans Hopkins and Adaeze L. Ibik and Ronniy C. Joseph and Jane Kaczmarek and Lordrick Kahinga and Victoria Kaspi and Kholoud Khairy and Charles D. Kilpatrick and Adam E. Lanman and Mattias Lazda and Calvin Leung and Robert Main and Lluis Mas-Ribas and Kiyoshi W. Masui and Ryan Mckinven and Juan Mena-Parra and Bradley W. Meyers and Daniele Michilli and Nikola Milutinovic and Kenzie Nimmo and Gavin Noble and Ayush Pandhi and Swarali Shivraj Patil and Aaron B. Pearlman and Emily Petroff and Ziggy Pleunis and J. Xavier Prochaska and Masoud Rafiei-Ravandi and Mubdi Rahman and Andre Renard and Mawson W. Sammons and Ketan R. Sand and Paul Scholz and Vishwangi Shah and Kaitlyn Shin and Seth R. Siegel and Sunil Simha and Kendrick Smith and Ingrid Stairs and Keith Vanderlinde and Haochen Wang and Dallas Wulf and Tarik J. Zegmott},
      year={2025},
      eprint={2502.11217},
      archivePrefix={arXiv},
      primaryClass={astro-ph.HE},
      url={https://arxiv.org/abs/2502.11217}, 
}

@misc{caleb2025fastradioburst3,
      title={A fast radio burst from the first 3 billion years of the Universe}, 
      author={Manisha Caleb and Themiya Nanayakkara and Benjamin Stappers and Inés Pastor-Marazuela and Ilya S. Khrykin and Karl Glazebrook and Nicolas Tejos and J. Xavier Prochaska and Kaustubh Rajwade and Lluis Mas-Ribas and Laura N. Driessen and Wen-fai Fong and Alexa C. Gordon and Jordan Hoffmann and Clancy W. James and Fabian Jankowski and Lordrick Kahinga and Michael Kramer and Sunil Simha and Ewan D. Barr and Mechiel Christiaan Bezuidenhout and Xihan Deng and Zeren Lin and Lachlan Marnoch and Christopher D. Martin and Anya Nugent and Kavya Shaji and Jun Tian},
      year={2025},
      eprint={2508.01648},
      archivePrefix={arXiv},
      primaryClass={astro-ph.HE},
      url={https://arxiv.org/abs/2508.01648}, 
}

@article{Prochaska_2019,
   title={The low density and magnetization of a massive galaxy halo exposed by a fast radio burst},
   volume={366},
   ISSN={1095-9203},
   url={http://dx.doi.org/10.1126/science.aay0073},
   DOI={10.1126/science.aay0073},
   number={6462},
   journal={Science},
   publisher={American Association for the Advancement of Science (AAAS)},
   author={Prochaska, J. Xavier and Macquart, Jean-Pierre and McQuinn, Matthew and Simha, Sunil and Shannon, Ryan M. and Day, Cherie K. and Marnoch, Lachlan and Ryder, Stuart and Deller, Adam and Bannister, Keith W. and Bhandari, Shivani and Bordoloi, Rongmon and Bunton, John and Cho, Hyerin and Flynn, Chris and Mahony, Elizabeth K. and Phillips, Chris and Qiu, Hao and Tejos, Nicolas},
   year={2019},
   month=Oct, pages={231–234} }

@ARTICLE{2026ATel17619....1G,
       author = {{Gordon}, A.~C. and {Fong}, W. and {Caleb}, M. and {Pastor-Marazuela}, I. and {Shaji}, K. and {Stappers}, B.~W. and {Surnis}, Mayuresh and {Tian}, J. and {Jankowski}, Fabian},
        title = "{A redshift for the host galaxy of the repeating FRB20251130A}",
      journal = {The Astronomer's Telegram},
         year = 2026,
        month = jan,
       volume = {17619},
        pages = {1},
       adsurl = {https://ui.adsabs.harvard.edu/abs/2026ATel17619....1G}
}

@misc{pastormarazuela2025localisationhostgalaxyidentification,
      title={Localisation and host galaxy identification of new Fast Radio Bursts with MeerKAT}, 
      author={Inés Pastor-Marazuela and Alexa C. Gordon and Ben Stappers and Ilya S. Khrykin and Nicolas Tejos and Kaustubh Rajwade and Manisha Caleb and Mayuresh P. Surnis and Laura N. Driessen and Sunil Simha and Jun Tian and J. Xavier Prochaska and Ewan Barr and Sarah Buchner and Wen-Fai Fong and Fabian Jankowski and Lordrick Kahinga and Charles D. Kilpatrick and Michael Kramer and Lluis Mas-Ribas and Joseph Hennawi},
      year={2025},
      eprint={2507.05982},
      archivePrefix={arXiv},
      primaryClass={astro-ph.HE},
      doi={https://doi.org/10.1093/mnras/staf2144},
      url={https://arxiv.org/abs/2507.05982}, 
}

@misc{sharma2026backlightingcosmicwebfast,
      title={Backlighting the Cosmic Web with Fast Radio Bursts: An Anthology of Dispersion Measure Cross-Correlations with Large-Scale Structure and Baryon Tracers}, 
      author={Kritti Sharma and Elisabeth Krause and Vikram Ravi and Dhayaa Anbajagane and Liam Connor and W. L. Kimmy Wu and Simone Ferraro and Sebastian Grandis and David Alonso and Yi-Kuan Chiang and Casey J. Law and Pranjal R. S. and Samuel McCarty and Shivam Pandey},
      year={2026},
      eprint={2604.22105},
      archivePrefix={arXiv},
      primaryClass={astro-ph.CO},
      url={https://arxiv.org/abs/2604.22105}, 
}

@ARTICLE{Colossus,
       author = {{Diemer}, Benedikt},
        title = "{COLOSSUS: A Python Toolkit for Cosmology, Large-scale Structure, and Dark Matter Halos}",
      journal = {\apjs},
         year = 2018,
        month = dec,
       volume = {239},
       number = {2},
          eid = {35},
        pages = {35},
          doi = {10.3847/1538-4365/aaee8c},
archivePrefix = {arXiv},
       eprint = {1712.04512},
 primaryClass = {astro-ph.CO},
       adsurl = {https://ui.adsabs.harvard.edu/abs/2018ApJS..239...35D}
}

@misc{CAMB,
       author = {{Lewis}, Antony and {Challinor}, Anthony},
        title = "{CAMB: Code for Anisotropies in the Microwave Background}",
 howpublished = {Astrophysics Source Code Library, record ascl:1102.026},
         year = 2011,
        month = feb,
          eid = {ascl:1102.026},
archivePrefix = {ascl},
       eprint = {1102.026},
       adsurl = {https://ui.adsabs.harvard.edu/abs/2011ascl.soft02026L}
}

@article{Connor2025,
  author    = {Connor, Liam and Ravi, Vikram and Sharma, Kritti and Ocker, Stella Koch and Faber, Jakob
               and Hallinan, Gregg and Harnach, Charlie and Hellbourg, Greg and Hobbs, Rick
               and Hodge, David and Hodges, Mark and Kosogorov, Nikita and Lamb, James
               and Law, Casey and Rasmussen, Paul and Sherman, Myles and Somalwar, Jean
               and Weinreb, Sander and Woody, David and Konietzka, Ralf M.},
  title     = {A gas-rich cosmic web revealed by the partitioning of the missing baryons},
  journal   = {Nature Astronomy},
  year      = {2025},
  volume    = {9},
  number    = {8},
  pages     = {1226--1239},
  doi       = {10.1038/s41550-025-02566-y},
  url       = {https://doi.org/10.1038/s41550-025-02566-y},
  issn      = {2397-3366}
}

@misc{cordes2003ne2001inewmodelgalactic,
      title={NE2001.I. A New Model for the Galactic Distribution of Free Electrons and its Fluctuations}, 
      author={J. M. Cordes and T. J. W. Lazio},
      year={2003},
      eprint={astro-ph/0207156},
      archivePrefix={arXiv},
      primaryClass={astro-ph},
      url={https://arxiv.org/abs/astro-ph/0207156}, 
}

@ARTICLE{pygedm,
       author = {{Price}, D.~C. and {Flynn}, C. and {Deller}, A.},
        title = "{A comparison of Galactic electron density models using PyGEDM}",
      journal = {\pasa},
         year = 2021,
        month = aug,
       volume = {38},
          eid = {e038},
        pages = {e038},
          doi = {10.1017/pasa.2021.33},
archivePrefix = {arXiv},
       eprint = {2106.15816},
 primaryClass = {astro-ph.GA},
       adsurl = {https://ui.adsabs.harvard.edu/abs/2021PASA...38...38P}
}

@article{Giri_2021,
   title={Emulation of baryonic effects on the matter power spectrum and constraints from galaxy cluster data},
   volume={2021},
   ISSN={1475-7516},
   url={http://dx.doi.org/10.1088/1475-7516/2021/12/046},
   DOI={10.1088/1475-7516/2021/12/046},
   number={12},
   journal={Journal of Cosmology and Astroparticle Physics},
   publisher={IOP Publishing},
   author={Giri, Sambit K. and Schneider, Aurel},
   year={2021},
   month=Dec, pages={046} }

@article{Madhavacheril_2019,
   title={Cosmology with the kinematic Sunyaev-Zeldovich effect: Breaking the optical depth degeneracy with fast radio bursts},
   volume={100},
   ISSN={2470-0029},
   url={http://dx.doi.org/10.1103/PhysRevD.100.103532},
   DOI={10.1103/physrevd.100.103532},
   number={10},
   journal={Physical Review D},
   publisher={American Physical Society (APS)},
   author={Madhavacheril, Mathew S. and Battaglia, Nicholas and Smith, Kendrick M. and Sievers, Jonathan L.},
   year={2019},
   month=Nov }

@ARTICLE{verdi26,
  author  = {{Verdi et al. in prep}},
  journal = {\apj},
  year    = 2026,
  note    = {in preparation}
}

@ARTICLE{ravidsa,
       author = {{Ravi}, Vikram and {Catha}, Morgan and {Chen}, Ge and {Connor}, Liam and {Faber}, Jakob T. and {Lamb}, James W. and {Hallinan}, Gregg and {Harnach}, Charlie and {Hellbourg}, Greg and {Hobbs}, Rick and {Hodge}, David and {Hodges}, Mark and {Law}, Casey and {Rasmussen}, Paul and {Sharma}, Kritti and {Sherman}, Myles B. and {Shi}, Jun and {Simard}, Dana and {Squillace}, Reynier and {Weinreb}, Sander and {Woody}, David P. and {Yadlapalli}, Nitika and {Ahumada}, Tomas and {Dong}, Dillon and {Fremling}, Christoffer and {Huang}, Yuping and {Karambelkar}, Viraj and {Miller}, Jessie M.},
        title = "{Deep Synoptic Array Science: Discovery of the Host Galaxy of FRB 20220912A}",
      journal = {\apjl},
         year = 2023,
        month = may,
       volume = {949},
       number = {1},
          eid = {L3},
        pages = {L3},
          doi = {10.3847/2041-8213/acc4b6},
archivePrefix = {arXiv},
       eprint = {2211.09049},
 primaryClass = {astro-ph.HE},
       adsurl = {https://ui.adsabs.harvard.edu/abs/2023ApJ...949L...3R}
}

@misc{hadzhiyska2026precisionkinematicsunyaevzeldovichmeasurements,
      title={Precision Kinematic Sunyaev--Zel'dovich Measurements Across Halo Mass and Redshift with DESI DR2 and ACT DR6: Part II. Bright Galaxy Survey and Emission-Line Galaxies}, 
      author={B. Hadzhiyska and S. Ferraro and F. J. Qu and B. Ried Guachalla and E. Schaan and J. Aguilar and S. Ahlen and D. Bianchi and D. Brooks and F. J. Castander and E. Chaussidon and T. Claybaugh and A. de la Macorra and Arjun Dey and Biprateep Dey and P. Doel and J. E. Forero-Romero and E. Gaztañaga and S. Gontcho A Gontcho and G. Gutierrez and J. Guy and K. Honscheid and C. Howlett and D. Huterer and M. Ishak and R. Joyce and R. Kehoe and T. Kisner and A. Kremin and O. Lahav and M. Landriau and L. Le Guillou and A. Leauthaud and M. Manera and P. Martini and A. Meisner and R. Miquel and S. Nadathur and N. Palanque-Delabrouille and W. J. Percival and F. Prada and I. Pérez-Ràfols and G. Rossi and L. Samushia and E. Sanchez and E. F. Schlafly and D. Schlegel and J. Silber and D. Sprayberry and G. Tarlé and B. A. Weaver and R. Zhou and H. Zou},
      year={2026},
      eprint={2604.19745},
      archivePrefix={arXiv},
      primaryClass={astro-ph.CO},
      url={https://arxiv.org/abs/2604.19745}, 
}

@ARTICLE{DESIoverview,
       author = {{DESI Collaboration} and {Abareshi}, B. and {Aguilar}, J. and {Ahlen}, S. and {Alam}, Shadab and {Alexander}, David M. and {Alfarsy}, R. and {Allen}, L. and {Allende Prieto}, C. and {Alves}, O. and {Ameel}, J. and {Armengaud}, E. and {Asorey}, J. and {Aviles}, Alejandro and {Bailey}, S. and {Balaguera-Antol{\'\i}nez}, A. and {Ballester}, O. and {Baltay}, C. and {Bault}, A. and {Beltran}, S.~F. and {Benavides}, B. and {BenZvi}, S. and {Berti}, A. and {Besuner}, R. and {Beutler}, Florian and {Bianchi}, D. and {Blake}, C. and {Blanc}, P. and {Blum}, R. and {Bolton}, A. and {Bose}, S. and {Bramall}, D. and {Brieden}, S. and {Brodzeller}, A. and {Brooks}, D. and {Brownewell}, C. and {Buckley-Geer}, E. and {Cahn}, R.~N. and {Cai}, Z. and {Canning}, R. and {Capasso}, R. and {Carnero Rosell}, A. and {Carton}, P. and {Casas}, R. and {Castander}, F.~J. and {Cervantes-Cota}, J.~L. and {Chabanier}, S. and {Chaussidon}, E. and {Chuang}, C. and {Circosta}, C. and {Cole}, S. and {Cooper}, A.~P. and {da Costa}, L. and {Cousinou}, M.-C. and {Cuceu}, A. and {Davis}, T.~M. and {Dawson}, K. and {de la Cruz-Noriega}, R. and {de la Macorra}, A. and {de Mattia}, A. and {Della Costa}, J. and {Demmer}, P. and {Derwent}, M. and {Dey}, A. and {Dey}, B. and {Dhungana}, G. and {Ding}, Z. and {Dobson}, C. and {Doel}, P. and {Donald-McCann}, J. and {Donaldson}, J. and {Douglass}, K. and {Duan}, Y. and {Dunlop}, P. and {Edelstein}, J. and {Eftekharzadeh}, S. and {Eisenstein}, D.~J. and {Enriquez-Vargas}, M. and {Escoffier}, S. and {Evatt}, M. and {Fagrelius}, P. and {Fan}, X. and {Fanning}, K. and {Fawcett}, V.~A. and {Ferraro}, S. and {Ereza}, J. and {Flaugher}, B. and {Font-Ribera}, A. and {Forero-Romero}, J.~E. and {Frenk}, C.~S. and {Fromenteau}, S. and {G{\"a}nsicke}, B.~T. and {Garcia-Quintero}, C. and {Garrison}, L. and {Gazta{\~n}aga}, E. and {Gerardi}, F. and {Gil-Mar{\'\i}n}, H. and {Gontcho A Gontcho}, S. and {Gonzalez-Morales}, Alma X. and {Gonzalez-de-Rivera}, G. and {Gonzalez-Perez}, V. and {Gordon}, C. and {Graur}, O. and {Green}, D. and {Grove}, C. and {Gruen}, D. and {Gutierrez}, G. and {Guy}, J. and {Hahn}, C. and {Harris}, S. and {Herrera}, D. and {Herrera-Alcantar}, Hiram K. and {Honscheid}, K. and {Howlett}, C. and {Huterer}, D. and {Ir{\v{s}}i{\v{c}}}, V. and {Ishak}, M. and {Jelinsky}, P. and {Jiang}, L. and {Jimenez}, J. and {Jing}, Y.~P. and {Joyce}, R. and {Jullo}, E. and {Juneau}, S. and {Kara{\c{c}}ayl{\i}}, N.~G. and {Karamanis}, M. and {Karcher}, A. and {Karim}, T. and {Kehoe}, R. and {Kent}, S. and {Kirkby}, D. and {Kisner}, T. and {Kitaura}, F. and {Koposov}, S.~E. and {Kov{\'a}cs}, A. and {Kremin}, A. and {Krolewski}, Alex and {L'Huillier}, B. and {Lahav}, O. and {Lambert}, A. and {Lamman}, C. and {Lan}, Ting-Wen and {Landriau}, M. and {Lane}, S. and {Lang}, D. and {Lange}, J.~U. and {Lasker}, J. and {Le Guillou}, L. and {Leauthaud}, A. and {Le Van Suu}, A. and {Levi}, Michael E. and {Li}, T.~S. and {Magneville}, C. and {Manera}, M. and {Manser}, Christopher J. and {Marshall}, B. and {Martini}, Paul and {McCollam}, W. and {McDonald}, P. and {Meisner}, Aaron M. and {Mena-Fern{\'a}ndez}, J. and {Meneses-Rizo}, J. and {Mezcua}, M. and {Miller}, T. and {Miquel}, R. and {Montero-Camacho}, P. and {Moon}, J. and {Moustakas}, J. and {Mueller}, E. and {Mu{\~n}oz-Guti{\'e}rrez}, Andrea and {Myers}, Adam D. and {Nadathur}, S. and {Najita}, J. and {Napolitano}, L. and {Neilsen}, E. and {Newman}, Jeffrey A. and {Nie}, J.~D. and {Ning}, Y. and {Niz}, G. and {Norberg}, P. and {Noriega}, Hern{\'a}n E. and {O'Brien}, T. and {Obuljen}, A. and {Palanque-Delabrouille}, N. and {Palmese}, A. and {Zhiwei}, P. and {Pappalardo}, D. and {PENG}, X. and {Percival}, W.~J. and {Perruchot}, S. and {Pogge}, R. and {Poppett}, C. and {Porredon}, A. and {Prada}, F. and {Prochaska}, J. and {Pucha}, R. and {P{\'e}rez-Fern{\'a}ndez}, A. and {P{\'e}rez-R{\`a}fols}, I. and {Rabinowitz}, D. and {Raichoor}, A.},
        title = "{Overview of the Instrumentation for the Dark Energy Spectroscopic Instrument}",
      journal = {\aj},
         year = 2022,
        month = nov,
       volume = {164},
       number = {5},
          eid = {207},
        pages = {207},
          doi = {10.3847/1538-3881/ac882b},
archivePrefix = {arXiv},
       eprint = {2205.10939},
 primaryClass = {astro-ph.IM},
       adsurl = {https://ui.adsabs.harvard.edu/abs/2022AJ....164..207D}
}

@ARTICLE{DESILS,
       author = {{Dey}, Arjun and {Schlegel}, David J. and {Lang}, Dustin and {Blum}, Robert and {Burleigh}, Kaylan and {Fan}, Xiaohui and {Findlay}, Joseph R. and {Finkbeiner}, Doug and {Herrera}, David and {Juneau}, St{\'e}phanie and {Landriau}, Martin and {Levi}, Michael and {McGreer}, Ian and {Meisner}, Aaron and {Myers}, Adam D. and {Moustakas}, John and {Nugent}, Peter and {Patej}, Anna and {Schlafly}, Edward F. and {Walker}, Alistair R. and {Valdes}, Francisco and {Weaver}, Benjamin A. and {Y{\`e}che}, Christophe and {Zou}, Hu and {Zhou}, Xu and {Abareshi}, Behzad and {Abbott}, T.~M.~C. and {Abolfathi}, Bela and {Aguilera}, C. and {Alam}, Shadab and {Allen}, Lori and {Alvarez}, A. and {Annis}, James and {Ansarinejad}, Behzad and {Aubert}, Marie and {Beechert}, Jacqueline and {Bell}, Eric F. and {BenZvi}, Segev Y. and {Beutler}, Florian and {Bielby}, Richard M. and {Bolton}, Adam S. and {Brice{\~n}o}, C{\'e}sar and {Buckley-Geer}, Elizabeth J. and {Butler}, Karen and {Calamida}, Annalisa and {Carlberg}, Raymond G. and {Carter}, Paul and {Casas}, Ricard and {Castander}, Francisco J. and {Choi}, Yumi and {Comparat}, Johan and {Cukanovaite}, Elena and {Delubac}, Timoth{\'e}e and {DeVries}, Kaitlin and {Dey}, Sharmila and {Dhungana}, Govinda and {Dickinson}, Mark and {Ding}, Zhejie and {Donaldson}, John B. and {Duan}, Yutong and {Duckworth}, Christopher J. and {Eftekharzadeh}, Sarah and {Eisenstein}, Daniel J. and {Etourneau}, Thomas and {Fagrelius}, Parker A. and {Farihi}, Jay and {Fitzpatrick}, Mike and {Font-Ribera}, Andreu and {Fulmer}, Leah and {G{\"a}nsicke}, Boris T. and {Gaztanaga}, Enrique and {George}, Koshy and {Gerdes}, David W. and {Gontcho}, Satya Gontcho A. and {Gorgoni}, Claudio and {Green}, Gregory and {Guy}, Julien and {Harmer}, Diane and {Hernandez}, M. and {Honscheid}, Klaus and {Huang}, Lijuan Wendy and {James}, David J. and {Jannuzi}, Buell T. and {Jiang}, Linhua and {Joyce}, Richard and {Karcher}, Armin and {Karkar}, Sonia and {Kehoe}, Robert and {Kneib}, Jean-Paul and {Kueter-Young}, Andrea and {Lan}, Ting-Wen and {Lauer}, Tod R. and {Le Guillou}, Laurent and {Le Van Suu}, Auguste and {Lee}, Jae Hyeon and {Lesser}, Michael and {Perreault Levasseur}, Laurence and {Li}, Ting S. and {Mann}, Justin L. and {Marshall}, Robert and {Mart{\'\i}nez-V{\'a}zquez}, C.~E. and {Martini}, Paul and {du Mas des Bourboux}, H{\'e}lion and {McManus}, Sean and {Meier}, Tobias Gabriel and {M{\'e}nard}, Brice and {Metcalfe}, Nigel and {Mu{\~n}oz-Guti{\'e}rrez}, Andrea and {Najita}, Joan and {Napier}, Kevin and {Narayan}, Gautham and {Newman}, Jeffrey A. and {Nie}, Jundan and {Nord}, Brian and {Norman}, Dara J. and {Olsen}, Knut A.~G. and {Paat}, Anthony and {Palanque-Delabrouille}, Nathalie and {Peng}, Xiyan and {Poppett}, Claire L. and {Poremba}, Megan R. and {Prakash}, Abhishek and {Rabinowitz}, David and {Raichoor}, Anand and {Rezaie}, Mehdi and {Robertson}, A.~N. and {Roe}, Natalie A. and {Ross}, Ashley J. and {Ross}, Nicholas P. and {Rudnick}, Gregory and {Safonova}, Sasha and {Saha}, Abhijit and {S{\'a}nchez}, F. Javier and {Savary}, Elodie and {Schweiker}, Heidi and {Scott}, Adam and {Seo}, Hee-Jong and {Shan}, Huanyuan and {Silva}, David R. and {Slepian}, Zachary and {Soto}, Christian and {Sprayberry}, David and {Staten}, Ryan and {Stillman}, Coley M. and {Stupak}, Robert J. and {Summers}, David L. and {Sien Tie}, Suk and {Tirado}, H. and {Vargas-Maga{\~n}a}, Mariana and {Vivas}, A. Katherina and {Wechsler}, Risa H. and {Williams}, Doug and {Yang}, Jinyi and {Yang}, Qian and {Yapici}, Tolga and {Zaritsky}, Dennis and {Zenteno}, A. and {Zhang}, Kai and {Zhang}, Tianmeng and {Zhou}, Rongpu and {Zhou}, Zhimin},
        title = "{Overview of the DESI Legacy Imaging Surveys}",
      journal = {\aj},
         year = 2019,
        month = may,
       volume = {157},
       number = {5},
          eid = {168},
        pages = {168},
          doi = {10.3847/1538-3881/ab089d},
archivePrefix = {arXiv},
       eprint = {1804.08657},
 primaryClass = {astro-ph.IM},
       adsurl = {https://ui.adsabs.harvard.edu/abs/2019AJ....157..168D}
}

@ARTICLE{Adame2025,
       author = {{Adame}, A.~G. and {Aguilar}, J. and {Ahlen}, S. and {Alam}, S. and {Alexander}, D.~M. and {Alvarez}, M. and {Alves}, O. and {Anand}, A. and {Andrade}, U. and {Armengaud}, E. and {Avila}, S. and {Aviles}, A. and {Awan}, H. and {Bailey}, S. and {Baltay}, C. and {Bault}, A. and {Behera}, J. and {BenZvi}, S. and {Beutler}, F. and {Bianchi}, D. and {Blake}, C. and {Blum}, R. and {Brieden}, S. and {Brodzeller}, A. and {Brooks}, D. and {Brown}, Z. and {Buckley-Geer}, E. and {Burtin}, E. and {Calderon}, R. and {Canning}, R. and {Carnero Rosell}, A. and {Cereskaite}, R. and {Cervantes-Cota}, J.~L. and {Chabanier}, S. and {Chaussidon}, E. and {Chaves-Montero}, J. and {Chen}, S. and {Chen}, X. and {Claybaugh}, T. and {Cole}, S. and {Cuceu}, A. and {Davis}, T.~M. and {Dawson}, K. and {de la Macorra}, A. and {de Mattia}, A. and {Deiosso}, N. and {Demina}, R. and {Dey}, A. and {Dey}, B. and {Ding}, Z. and {Doel}, P. and {Edelstein}, J. and {Eftekharzadeh}, S. and {Eisenstein}, D.~J. and {Elliott}, A. and {Fagrelius}, P. and {Fanning}, K. and {Ferraro}, S. and {Ereza}, J. and {Findlay}, N. and {Flaugher}, B. and {Font-Ribera}, A. and {Forero-S{\'a}nchez}, D. and {Forero-Romero}, J.~E. and {Frenk}, C.~S. and {Garcia-Quintero}, C. and {Gazta{\~n}aga}, E. and {Gil-Mar{\'\i}n}, H. and {Gontcho}, S. Gontcho A. and {Gonzalez-Morales}, A.~X. and {Gonzalez-Perez}, V. and {Gordon}, C. and {Green}, D. and {Gruen}, D. and {Gsponer}, R. and {Gutierrez}, G. and {Guy}, J. and {Hadzhiyska}, B. and {Hahn}, C. and {Hanif}, M.~M.~S. and {Herrera-Alcantar}, H.~K. and {Honscheid}, K. and {Hou}, J. and {Howlett}, C. and {Huterer}, D. and {Ir{\v{s}}i{\v{c}}}, V. and {Ishak}, M. and {Juneau}, S. and {Kara{\c{c}}ayl{\i}}, N.~G. and {Kehoe}, R. and {Kent}, S. and {Kirkby}, D. and {Kitaura}, F.-S. and {Kong}, H. and {Kremin}, A. and {Krolewski}, A. and {Lai}, Y. and {Lan}, T.-W. and {Landriau}, M. and {Lang}, D. and {Lasker}, J. and {Le Goff}, J.~M. and {Le Guillou}, L. and {Leauthaud}, A. and {Levi}, M.~E. and {Li}, T.~S. and {Lodha}, K. and {Magneville}, C. and {Manera}, M. and {Margala}, D. and {Martini}, P. and {Maus}, M. and {McDonald}, P. and {Medina-Varela}, L. and {Meisner}, A. and {Mena-Fern{\'a}ndez}, J. and {Miquel}, R. and {Moon}, J. and {Moore}, S. and {Moustakas}, J. and {Mudur}, N. and {Mueller}, E. and {Mu{\~n}oz-Guti{\'e}rrez}, A. and {Myers}, A.~D. and {Nadathur}, S. and {Napolitano}, L. and {Neveux}, R. and {Newman}, J.~A. and {Nguyen}, N.~M. and {Nie}, J. and {Niz}, G. and {Noriega}, H.~E. and {Padmanabhan}, N. and {Paillas}, E. and {Palanque-Delabrouille}, N. and {Pan}, J. and {Penmetsa}, S. and {Percival}, W.~J. and {Pieri}, M.~M. and {Pinon}, M. and {Poppett}, C. and {Porredon}, A. and {Prada}, F. and {P{\'e}rez-Fern{\'a}ndez}, A. and {P{\'e}rez-R{\`a}fols}, I. and {Rabinowitz}, D. and {Raichoor}, A. and {Ram{\'\i}rez-P{\'e}rez}, C. and {Ramirez-Solano}, S. and {Rashkovetskyi}, M. and {Ravoux}, C. and {Rezaie}, M. and {Rich}, J. and {Rocher}, A. and {Rockosi}, C. and {Roe}, N.~A. and {Rosado-Marin}, A. and {Ross}, A.~J. and {Rossi}, G. and {Ruggeri}, R. and {Ruhlmann-Kleider}, V. and {Samushia}, L. and {Sanchez}, E. and {Saulder}, C. and {Schlafly}, E.~F. and {Schlegel}, D. and {Scholte}, D. and {Schubnell}, M. and {Seo}, H. and {Sharples}, R. and {Silber}, J. and {Slosar}, A. and {Smith}, A. and {Sprayberry}, D. and {Tan}, T. and {Tarl{\'e}}, G. and {Trusov}, S. and {Vaisakh}, R. and {Valcin}, D. and {Valdes}, F. and {Vargas-Maga{\~n}a}, M. and {Verde}, L. and {Walther}, M. and {Wang}, B. and {Wang}, M.~S. and {Weaver}, B.~A. and {Weaverdyck}, N. and {Wechsler}, R.~H. and {Weinberg}, D.~H. and {White}, M. and {Wilson}, M.~J. and {Yu}, J. and {Yu}, Y. and {Yuan}, S. and {Y{\`e}che}, C. and {Zaborowski}, E.~A. and {Zarrouk}, P. and {Zhang}, H. and {Zhao}, C. and {Zhao}, R.},
        title = "{DESI 2024 II: sample definitions, characteristics, and two-point clustering statistics}",
      journal = {\jcap},
         year = 2025,
        month = jul,
       volume = {2025},
       number = {7},
          eid = {017},
        pages = {017},
          doi = {10.1088/1475-7516/2025/07/017},
archivePrefix = {arXiv},
       eprint = {2411.12020},
 primaryClass = {astro-ph.CO},
       adsurl = {https://ui.adsabs.harvard.edu/abs/2025JCAP...07..017A}
}

@ARTICLE{BGS,
       author = {{Hahn}, ChangHoon and {Wilson}, Michael J. and {Ruiz-Macias}, Omar and {Cole}, Shaun and {Weinberg}, David H. and {Moustakas}, John and {Kremin}, Anthony and {Tinker}, Jeremy L. and {Smith}, Alex and {Wechsler}, Risa H. and {Ahlen}, Steven and {Alam}, Shadab and {Bailey}, Stephen and {Brooks}, David and {Cooper}, Andrew P. and {Davis}, Tamara M. and {Dawson}, Kyle and {Dey}, Arjun and {Dey}, Biprateep and {Eftekharzadeh}, Sarah and {Eisenstein}, Daniel J. and {Fanning}, Kevin and {Forero-Romero}, Jaime E. and {Frenk}, Carlos S. and {Gazta{\~n}aga}, Enrique and {A Gontcho}, Satya Gontcho and {Guy}, Julien and {Honscheid}, Klaus and {Ishak}, Mustapha and {Juneau}, St{\'e}phanie and {Kehoe}, Robert and {Kisner}, Theodore and {Lan}, Ting-Wen and {Landriau}, Martin and {Le Guillou}, Laurent and {Levi}, Michael E. and {Magneville}, Christophe and {Martini}, Paul and {Meisner}, Aaron and {Myers}, Adam D. and {Nie}, Jundan and {Norberg}, Peder and {Palanque-Delabrouille}, Nathalie and {Percival}, Will J. and {Poppett}, Claire and {Prada}, Francisco and {Raichoor}, Anand and {Ross}, Ashley J. and {Gaines}, Sasha and {Saulder}, Christoph and {Schlafly}, Eddie and {Schlegel}, David and {Sierra-Porta}, David and {Tarle}, Gregory and {Weaver}, Benjamin A. and {Y{\`e}che}, Christophe and {Zarrouk}, Pauline and {Zhou}, Rongpu and {Zhou}, Zhimin and {Zou}, Hu},
        title = "{The DESI Bright Galaxy Survey: Final Target Selection, Design, and Validation}",
      journal = {\aj},
         year = 2023,
        month = jun,
       volume = {165},
       number = {6},
          eid = {253},
        pages = {253},
          doi = {10.3847/1538-3881/accff8},
archivePrefix = {arXiv},
       eprint = {2208.08512},
 primaryClass = {astro-ph.CO},
       adsurl = {https://ui.adsabs.harvard.edu/abs/2023AJ....165..253H}
}

@article{Zhou_2023pz,
   title={DESI luminous red galaxy samples for cross-correlations},
   volume={2023},
   ISSN={1475-7516},
   url={http://dx.doi.org/10.1088/1475-7516/2023/11/097},
   DOI={10.1088/1475-7516/2023/11/097},
   number={11},
   journal={Journal of Cosmology and Astroparticle Physics},
   publisher={IOP Publishing},
   author={Zhou, Rongpu and Ferraro, Simone and White, Martin and DeRose, Joseph and Sailer, Noah and Aguilar, Jessica and Ahlen, Steven and Bailey, Stephen and Brooks, David and Claybaugh, Todd and Dawson, Kyle and de la Macorra, Axel and Dey, Biprateep and Doel, Peter and Font-Ribera, Andreu and Forero-Romero, Jaime E. and Gontcho A Gontcho, Satya and Guy, Julien and Kremin, Anthony and Lambert, Andrew and Le Guillou, Laurent and Levi, Michael and Magneville, Christophe and Manera, Marc and Meisner, Aaron and Miquel, Ramon and Moustakas, John and Myers, Adam D. and Newman, Jeffrey A. and Nie, Jundan and Percival, Will and Rezaie, Mehdi and Rossi, Graziano and Sanchez, Eusebio and Schlegel, David and Schubnell, Michael and Seo, Hee-Jong and Tarlé, Gregory and Zhou, Zhimin},
   year={2023},
   month=Nov, pages={097} }

@article{Zhou_2023sm,
   title={Target Selection and Validation of DESI Luminous Red Galaxies},
   volume={165},
   ISSN={1538-3881},
   url={http://dx.doi.org/10.3847/1538-3881/aca5fb},
   DOI={10.3847/1538-3881/aca5fb},
   number={2},
   journal={The Astronomical Journal},
   publisher={American Astronomical Society},
   author={Zhou, Rongpu and Dey, Biprateep and Newman, Jeffrey A. and Eisenstein, Daniel J. and Dawson, K. and Bailey, S. and Berti, A. and Guy, J. and Lan, Ting-Wen and Zou, H. and Aguilar, J. and Ahlen, S. and Alam, Shadab and Brooks, D. and de la Macorra, A. and Dey, A. and Dhungana, G. and Fanning, K. and Font-Ribera, A. and Gontcho, S. Gontcho A. and Honscheid, K. and Ishak, Mustapha and Kisner, T. and Kovács, A. and Kremin, A. and Landriau, M. and Levi, Michael E. and Magneville, C. and Manera, Marc and Martini, P. and Meisner, Aaron M. and Miquel, R. and Moustakas, J. and Myers, Adam D. and Nie, Jundan and Palanque-Delabrouille, N. and Percival, W. J. and Poppett, C. and Prada, F. and Raichoor, A. and Ross, A. J. and Schlafly, E. and Schlegel, D. and Schubnell, M. and Tarlé, Gregory and Weaver, B. A. and Wechsler, R. H. and Yéche, Christophe and Zhou, Zhimin},
   year={2023},
   month=Jan, pages={58} }

@misc{wang2025measurementdispersionunicodex2013galaxycrosspowerspectrum,
      title={Measurement of the DispersionxGalaxy Cross-Power Spectrum with the Second CHIME/FRB Catalog}, 
      author={Haochen Wang and Kiyoshi Masui and Shion Andrew and Emmanuel Fonseca and B. M. Gaensler and R. C. Joseph and Victoria M. Kaspi and Bikash Kharel and Adam E. Lanman and Calvin Leung and Lluis Mas-Ribas and Juan Mena-Parra and Kenzie Nimmo and Aaron B. Pearlman and Ue-Li Pen and J. Xavier Prochaska and Ryan Raikman and Kaitlyn Shin and Seth R. Siegel and Kendrick M. Smith and Ingrid H. Stairs},
      year={2025},
      eprint={2506.08932},
      archivePrefix={arXiv},
      primaryClass={astro-ph.CO},
      url={https://arxiv.org/abs/2506.08932}, 
}

@ARTICLE{treecorr,
       author = {{Jarvis}, M. and {Bernstein}, G. and {Jain}, B.},
        title = "{The skewness of the aperture mass statistic}",
      journal = {\mnras},
         year = 2004,
        month = jul,
       volume = {352},
       number = {1},
        pages = {338-352},
          doi = {10.1111/j.1365-2966.2004.07926.x},
archivePrefix = {arXiv},
       eprint = {astro-ph/0307393},
 primaryClass = {astro-ph},
       adsurl = {https://ui.adsabs.harvard.edu/abs/2004MNRAS.352..338J}
}

@ARTICLE{Hartlap2007,
       author = {{Hartlap}, J. and {Simon}, P. and {Schneider}, P.},
        title = "{Why your model parameter confidences might be too optimistic. Unbiased estimation of the inverse covariance matrix}",
      journal = {\aap},
         year = 2007,
        month = mar,
       volume = {464},
       number = {1},
        pages = {399-404},
          doi = {10.1051/0004-6361:20066170},
archivePrefix = {arXiv},
       eprint = {astro-ph/0608064},
 primaryClass = {astro-ph},
       adsurl = {https://ui.adsabs.harvard.edu/abs/2007A&A...464..399H}
}

@article{Foreman_Mackey_2013,
   title={<tt>emcee</tt>: The MCMC Hammer},
   volume={125},
   ISSN={1538-3873},
   url={http://dx.doi.org/10.1086/670067},
   DOI={10.1086/670067},
   number={925},
   journal={Publications of the Astronomical Society of the Pacific},
   publisher={IOP Publishing},
   author={Foreman-Mackey, Daniel and Hogg, David W. and Lang, Dustin and Goodman, Jonathan},
   year={2013},
   month=Mar, pages={306–312} }

@misc{sharma2026signaturessuppressedmatterclustering,
      title={Signatures of Suppressed Matter Clustering revealed by Fast Radio Bursts}, 
      author={Kritti Sharma and Elisabeth Krause and Vikram Ravi and Liam Connor and Dhayaa Anbajagane and Pranjal R. S},
      year={2026},
      eprint={2604.17162},
      archivePrefix={arXiv},
      primaryClass={astro-ph.CO},
      url={https://arxiv.org/abs/2604.17162}, 
}

@article{Popesso_2026,
   title={The hot gas mass fraction in halos: From Milky Way-like groups to massive clusters},
   volume={707},
   ISSN={1432-0746},
   url={http://dx.doi.org/10.1051/0004-6361/202453256},
   DOI={10.1051/0004-6361/202453256},
   journal={Astronomy \& Astrophysics},
   publisher={EDP Sciences},
   author={Popesso, P. and Biviano, A. and Marini, I. and Dolag, K. and Vladutescu-Zopp, S. and Csizi, B. and Biffi, V. and Lamer, G. and Robothan, A. and Bravo, M. and Lovisari, L. and Ettori, S. and Angelinelli, M. and Driver, S. and Toptun, V. and Dev, A. and Mazengo, D. and Merloni, A. and Comparat, J. and Ponti, G. and Mroczkowski, T. and Bulbul, E. and Grandis, S. and Bahar, E.},
   year={2026},
   month=Mar, pages={A362} }

@misc{bigwood2024weaklensingcombinedkinetic,
      title={Weak lensing combined with the kinetic Sunyaev Zel'dovich effect: A study of baryonic feedback}, 
      author={L. Bigwood and A. Amon and A. Schneider and J. Salcido and I. G. McCarthy and C. Preston and D. Sanchez and D. Sijacki and E. Schaan and S. Ferraro and N. Battaglia and A. Chen and S. Dodelson and A. Roodman and A. Pieres and A. Ferte and A. Alarcon and A. Drlica-Wagner and A. Choi and A. Navarro-Alsina and A. Campos and A. J. Ross and A. Carnero Rosell and B. Yin and B. Yanny and C. Sanchez and C. Chang and C. Davis and C. Doux and D. Gruen and E. S. Rykoff and E. M. Huff and E. Sheldon and F. Tarsitano and F. Andrade-Oliveira and G. M. Bernstein and G. Giannini and H. T. Diehl and H. Huang and I. Harrison and I. Sevilla-Noarbe and I. Tutusaus and J. Elvin-Poole and J. McCullough and J. Zuntz and J. Blazek and J. DeRose and J. Cordero and J. Prat and J. Myles and K. Eckert and K. Bechtol and K. Herner and L. F. Secco and M. Gatti and M. Raveri and M. Carrasco Kind and M. R. Becker and M. A. Troxel and M. Jarvis and N. MacCrann and O. Friedrich and O. Alves and P. -F. Leget and R. Chen and R. P. Rollins and R. H. Wechsler and R. A. Gruendl and R. Cawthon and S. Allam and S. L. Bridle and S. Pandey and S. Everett and T. Shin and W. G. Hartley and X. Fang and Y. Zhang and M. Aguena and J. Annis and D. Bacon and E. Bertin and S. Bocquet and D. Brooks and J. Carretero and F. J. Castander and L. N. da Costa and M. E. S. Pereira and J. De Vicente and S. Desai and P. Doel and I. Ferrero and B. Flaugher and J. Frieman and J. Garcia-Bellido and E. Gaztanaga and G. Gutierrez and S. R. Hinton and D. L. Hollowood and K. Honscheid and D. Huterer and D. J. James and K. Kuehn and O. Lahav and S. Lee and J. L. Marshall and J. Mena-Fernandez and R. Miquel and J. Muir and M. Paterno and A. A. Plazas Malagon and A. Porredon and A. K. Romer and S. Samuroff and E. Sanchez and D. Sanchez Cid and M. Smith and M. Soares-Santos and E. Suchyta and M. E. C. Swanson and G. Tarle and C. To and N. Weaverdyck and J. Weller and P. Wiseman and M. Yamamoto},
      year={2024},
      eprint={2404.06098},
      archivePrefix={arXiv},
      primaryClass={astro-ph.CO},
      url={https://arxiv.org/abs/2404.06098}, 
}

@misc{law2024deepsynopticarrayscience,
      title={Deep Synoptic Array Science: First FRB and Host Galaxy Catalog}, 
      author={C. J. Law and K. Sharma and V. Ravi and G. Chen and M. Catha and L. Connor and J. T. Faber and G. Hallinan and C. Harnach and G. Hellbourg and R. Hobbs and D. Hodge and M. Hodges and J. W. Lamb and P. Rasmussen and M. B. Sherman and J. Shi and D. Simard and R. Squillace and S. Weinreb and D. P. Woody and N. Yadlapalli},
      year={2024},
      eprint={2307.03344},
      archivePrefix={arXiv},
      primaryClass={astro-ph.HE},
      url={https://arxiv.org/abs/2307.03344}, 
}

@misc{schneider2025baryonificationalternativehydrodynamicalsimulations,
      title={Baryonification: An alternative to hydrodynamical simulations for cosmological studies}, 
      author={Aurel Schneider and Michael Kovač and Jozef Bucko and Andrina Nicola and Robert Reischke and Sambit K. Giri and Romain Teyssier and Tilman Tröster and Alexandre Refregier and Matthieu Schaller and Joop Schaye},
      year={2025},
      eprint={2507.07892},
      archivePrefix={arXiv},
      primaryClass={astro-ph.CO},
      url={https://arxiv.org/abs/2507.07892}, 
}

@misc{DESIDR1,
      title={Data Release 1 of the Dark Energy Spectroscopic Instrument}, 
      author={DESI Collaboration and M. Abdul Karim and A. G. Adame and D. Aguado and J. Aguilar and S. Ahlen and S. Alam and G. Aldering and D. M. Alexander and R. Alfarsy and L. Allen and C. Allende Prieto and O. Alves and A. Anand and U. Andrade and E. Armengaud and S. Avila and A. Aviles and H. Awan and S. Bailey and A. Baleato Lizancos and O. Ballester and A. Bault and J. Bautista and R. Bean and J. Behera and S. BenZvi and L. Beraldo e Silva and J. R. Bermejo-Climent and F. Beutler and D. Bianchi and C. Blake and R. Blum and A. S. Bolton and M. Bonici and S. Brieden and A. Brodzeller and D. Brooks and E. Buckley-Geer and E. Burtin and A. Byström and R. Canning and A. Carnero Rosell and A. Carr and P. Carrilho and L. Casas and F. J. Castander and R. Cereskaite and J. L. Cervantes-Cota and E. Chaussidon and J. Chaves-Montero and S. Chen and X. Chen and C. Circosta and T. Claybaugh and S. Cole and A. P. Cooper and M. -C. Cousinou and A. Cuceu and T. M. Davis and K. S. Dawson and R. de Belsunce and R. de la Cruz and A. de la Macorra and A. de Mattia and N. Deiosso and J. Della Costa and R. Demina and U. Demirbozan and J. DeRose and A. Dey and B. Dey and J. Ding and Z. Ding and P. Doel and K. Douglass and M. Dowicz and H. Ebina and J. Edelstein and D. J. Eisenstein and W. Elbers and N. Emas and S. Escoffier and P. Fagrelius and X. Fan and K. Fanning and G. Favole and V. A. Fawcett and E. Fernández-García and S. Ferraro and N. Findlay and A. Font-Ribera and J. E. Forero-Romero and D. Forero-Sánchez and C. S. Frenk and B. T. Gänsicke and L. Galbany and J. García-Bellido and C. Garcia-Quintero and L. H. Garrison and E. Gaztañaga and H. Gil-Marín and A. Gloudemans and O. Y. Gnedin and S. Gontcho A Gontcho and D. Gonzalez and A. X. Gonzalez-Morales and V. Gonzalez-Perez and C. Gordon and O. Graur and D. Green and D. Gruen and R. Gsponer and C. Guandalin and G. Gutierrez and J. Guy and C. Hahn and J. J. Han and J. Han and S. He and H. K. Herrera-Alcantar and S. Heydenreich and K. Honscheid and J. Hou and C. Howlett and D. Huterer and V. Iršič and M. Ishak and A. Jacques and L. Jiang and J. Jimenez and Y. P. Jing and B. Joachimi and S. Joudaki and R. Joyce and E. Jullo and S. Juneau and N. G. Karaçaylı and T. Karim and R. Kehoe and S. Kent and A. Khederlarian and D. Kirkby and T. Kisner and F. -S. Kitaura and N. Kizhuprakkat and H. Kong and S. E. Koposov and A. Kremin and A. Krolewski and O. Lahav and Y. Lai and C. Lamman and T. -W. Lan and M. Landriau and D. Lang and J. U. Lange and J. Lasker and J. M. Le Goff and L. Le Guillou and A. Leauthaud and M. E. Levi and S. Li and T. S. Li and W. Liu and K. Lodha and M. Lokken and Y. Luo and Y. Luo and C. Magneville and M. Manera and C. J. Manser and D. Margala and P. Martini and M. Maus and J. McCullough and P. McDonald and G. E. Medina and L. Medina-Varela and A. Meisner and J. Mena-Fernández and A. Menegas and J. Meneses-Rizo and M. Mezcua and R. Miquel and P. Montero-Camacho and J. Moon and J. Moustakas and A. Muñoz-Gutiérrez and D. Muñoz-Santos and A. D. Myers and J. Myles and S. Nadathur and J. Najita and L. Napolitano and J. A. Newman and F. Nikakhtar and R. Nikutta and G. Niz and H. E. Noriega and P. Nugent and N. Padmanabhan and E. Paillas and N. Palanque-Delabrouille and A. Palmese and J. Pan and Z. Pan and D. Parkinson and J. A. Peacock and M. P. Ibanez and W. J. Percival and A. Pérez-Fernández and I. Pérez-Ràfols and P. Peterson and J. Piat and M. M. Pieri and M. Pinon and C. Poppett and A. Porredon and F. Prada and R. Pucha and F. Qin and D. Rabinowitz and A. Raichoor and C. Ramírez-Pérez and S. Ramirez-Solano and M. Rashkovetskyi and C. Ravoux and B. Ried Guachalla and A. H. Riley and A. Rocher and C. Rockosi and J. Rohlf and A. J. Rosado-Marín and A. J. Ross and C. Ross and G. Rossi and R. Ruggeri and V. Ruhlmann-Kleider and C. G. Sabiu and K. Said and N. Sailer and A. Saintonge and Y. Salcedo Hernandez and L. Samushia and E. Sanchez and N. Sanders and N. Sandford and S. Satyavolu and C. Saulder and A. K. Saydjari and E. F. Schlafly and D. Schlegel and D. Scholte and M. Schubnell and A. Semenaite and H. Seo and A. Shafieloo and R. Sharples and J. Silber and F. Sinigaglia and M. Siudek and Z. Slepian and A. Smith and M. Soumagnac and D. Sprayberry and J. Suárez-Pérez and J. Swanson and T. Tan and G. Tarlé and P. Taylor and G. Thomas and R. Tojeiro and R. J. Turner and W. Turner and L. A. Ureña-López and R. Vaisakh and M. Valluri and G. Valogiannis and M. Vargas-Magaña and L. Verde and P. Vielzeuf and M. Walther and B. Wang and M. S. Wang and W. Wang and B. A. Weaver and N. Weaverdyck and R. H. Wechsler and D. H. Weinberg and M. White and A. Whitford and M. Wolfson and J. Yang and C. Yèche and S. Youles and J. Yu and S. Yuan and E. A. Zaborowski and P. Zarrouk and H. Zhang and C. Zhao and R. Zhao and Z. Zheng and C. Zhou and R. Zhou and Y. Zhou and H. Zou and S. Zou and Y. Zu},
      year={2026},
      eprint={2503.14745},
      archivePrefix={arXiv},
      primaryClass={astro-ph.CO},
      url={https://arxiv.org/abs/2503.14745}, 
}

@misc{shirasaki2026crosscorrelatinggalaxiescosmicdispersion,
      title={Cross-correlating galaxies and cosmic dispersion measures: Constraints on the gas-to-halo mass relation from 2MASS galaxies and 133 localized fast radio bursts}, 
      author={Masato Shirasaki and Ryuichi Takahashi and Ken Osato and Kunihito Ioka},
      year={2026},
      eprint={2601.21336},
      archivePrefix={arXiv},
      primaryClass={astro-ph.CO},
      url={https://arxiv.org/abs/2601.21336}, 
}

@article{Schaan_2021,
   title={Atacama Cosmology Telescope: Combined kinematic and thermal Sunyaev-Zel’dovich measurements from BOSS CMASS and LOWZ halos},
   volume={103},
   ISSN={2470-0029},
   url={http://dx.doi.org/10.1103/PhysRevD.103.063513},
   DOI={10.1103/physrevd.103.063513},
   number={6},
   journal={Physical Review D},
   publisher={American Physical Society (APS)},
   author={Schaan, Emmanuel and Ferraro, Simone and Amodeo, Stefania and Battaglia, Nicholas and Aiola, Simone and Austermann, Jason E. and Beall, James A. and Bean, Rachel and Becker, Daniel T. and Bond, Richard J. and Calabrese, Erminia and Calafut, Victoria and Choi, Steve K. and Denison, Edward V. and Devlin, Mark J. and Duff, Shannon M. and Duivenvoorden, Adriaan J. and Dunkley, Jo and Dünner, Rolando and Gallardo, Patricio A. and Guan, Yilun and Han, Dongwon and Hill, J. Colin and Hilton, Gene C. and Hilton, Matt and Hložek, Renée and Hubmayr, Johannes and Huffenberger, Kevin M. and Hughes, John P. and Koopman, Brian J. and MacInnis, Amanda and McMahon, Jeff and Madhavacheril, Mathew S. and Moodley, Kavilan and Mroczkowski, Tony and Naess, Sigurd and Nati, Federico and Newburgh, Laura B. and Niemack, Michael D. and Page, Lyman A. and Partridge, Bruce and Salatino, Maria and Sehgal, Neelima and Schillaci, Alessandro and Sifón, Cristóbal and Smith, Kendrick M. and Spergel, David N. and Staggs, Suzanne and Storer, Emilie R. and Trac, Hy and Ullom, Joel N. and Van Lanen, Jeff and Vale, Leila R. and van Engelen, Alexander and Magaña, Mariana Vargas and Vavagiakis, Eve M. and Wollack, Edward J. and Xu, Zhilei and },
   year={2021},
   month=Mar }

@article{Moster_2012,
   title={Galactic star formation and accretion histories from matching galaxies to dark matter haloes},
   volume={428},
   ISSN={1365-2966},
   url={http://dx.doi.org/10.1093/mnras/sts261},
   DOI={10.1093/mnras/sts261},
   number={4},
   journal={Monthly Notices of the Royal Astronomical Society},
   publisher={Oxford University Press (OUP)},
   author={Moster, Benjamin P. and Naab, Thorsten and White, Simon D. M.},
   year={2012},
   month=Dec, pages={3121–3138} }

@article{Schneider_2015,
doi = {10.1088/1475-7516/2015/12/049},
url = {https://doi.org/10.1088/1475-7516/2015/12/049},
year = {2015},
month = {dec},
publisher = {},
volume = {2015},
number = {12},
pages = {049},
author = {Schneider, Aurel and Teyssier, Romain},
title = {A new method to quantify the effects of baryons on the matter power spectrum},
journal = {Journal of Cosmology and Astroparticle Physics}
}

@article{Schneider_2019,
   title={Quantifying baryon effects on the matter power spectrum and the weak lensing shear correlation},
   volume={2019},
   ISSN={1475-7516},
   url={http://dx.doi.org/10.1088/1475-7516/2019/03/020},
   DOI={10.1088/1475-7516/2019/03/020},
   number={03},
   journal={Journal of Cosmology and Astroparticle Physics},
   publisher={IOP Publishing},
   author={Schneider, Aurel and Teyssier, Romain and Stadel, Joachim and Chisari, Nora Elisa and Brun, Amandine M.C. Le and Amara, Adam and Refregier, Alexandre},
   year={2019},
   month=Mar, pages={020–020} }

@article{Amodeo_2021,
   title={Atacama Cosmology Telescope: Modeling the gas thermodynamics in BOSS CMASS galaxies from kinematic and thermal Sunyaev-Zel’dovich measurements},
   volume={103},
   ISSN={2470-0029},
   url={http://dx.doi.org/10.1103/PhysRevD.103.063514},
   DOI={10.1103/physrevd.103.063514},
   number={6},
   journal={Physical Review D},
   publisher={American Physical Society (APS)},
   author={Amodeo, Stefania and Battaglia, Nicholas and Schaan, Emmanuel and Ferraro, Simone and Moser, Emily and Aiola, Simone and Austermann, Jason E. and Beall, James A. and Bean, Rachel and Becker, Daniel T. and Bond, Richard J. and Calabrese, Erminia and Calafut, Victoria and Choi, Steve K. and Denison, Edward V. and Devlin, Mark and Duff, Shannon M. and Duivenvoorden, Adriaan J. and Dunkley, Jo and Dünner, Rolando and Gallardo, Patricio A. and Hall, Kirsten R. and Han, Dongwon and Hill, J. Colin and Hilton, Gene C. and Hilton, Matt and Hložek, Renée and Hubmayr, Johannes and Huffenberger, Kevin M. and Hughes, John P. and Koopman, Brian J. and MacInnis, Amanda and McMahon, Jeff and Madhavacheril, Mathew S. and Moodley, Kavilan and Mroczkowski, Tony and Naess, Sigurd and Nati, Federico and Newburgh, Laura B. and Niemack, Michael D. and Page, Lyman A. and Partridge, Bruce and Schillaci, Alessandro and Sehgal, Neelima and Sifón, Cristóbal and Spergel, David N. and Staggs, Suzanne and Storer, Emilie R. and Ullom, Joel N. and Vale, Leila R. and van Engelen, Alexander and Van Lanen, Jeff and Vavagiakis, Eve M. and Wollack, Edward J. and Xu, Zhilei},
   year={2021},
   month=Mar }

@misc{siegel2025jointxraykineticsunyaevzeldovich,
      title={Joint X-ray, kinetic Sunyaev-Zeldovich, and weak lensing measurements: toward a consensus picture of efficient gas expulsion from groups and clusters}, 
      author={Jared Siegel and Alexandra Amon and Ian G. McCarthy and Leah Bigwood and Masaya Yamamoto and Esra Bulbul and Jenny E. Greene and Jamie McCullough and Matthieu Schaller and Joop Schaye},
      year={2025},
      eprint={2509.10455},
      archivePrefix={arXiv},
      primaryClass={astro-ph.CO},
      url={https://arxiv.org/abs/2509.10455}, 
}

@misc{siegel2025suppressionmatterpowerspectrum,
      title={The suppression of the matter power spectrum: strong feedback from X-ray gas mass fractions, kSZ effect profiles, and galaxy-galaxy lensing}, 
      author={Jared Siegel and Leah Bigwood and Alexandra Amon and Jamie McCullough and Masaya Yamamoto and Ian G. McCarthy and Matthieu Schaller and Aurel Schneider and Joop Schaye},
      year={2025},
      eprint={2512.02954},
      archivePrefix={arXiv},
      primaryClass={astro-ph.CO},
      url={https://arxiv.org/abs/2512.02954}, 
}

@article{Guachalla_2025,
   title={Backlighting extended gas halos around luminous red galaxies: Kinematic Sunyaev-Zel’dovich effect from DESI Y1 and ACT data},
   volume={112},
   ISSN={2470-0029},
   url={http://dx.doi.org/10.1103/lqbj-wcqj},
   DOI={10.1103/lqbj-wcqj},
   number={10},
   journal={Physical Review D},
   publisher={American Physical Society (APS)},
   author={Guachalla, Bernardita Ried and Schaan, Emmanuel and Hadzhiyska, Boryana and Ferraro, Simone and Aguilar, Jessica N. and Ahlen, Steven and Battaglia, Nicholas and Bianchi, Davide and Bond, Richard and Brooks, David and Claybaugh, Todd and Coulton, William R. and de la Macorra, Axel and Devlin, Mark J. and Dey, Arjun and Doel, Peter and Dunkley, Jo and Fanning, Kevin and Forero-Romero, Jaime and Gaztañaga, Enrique and Gontcho, Satya Gontcho A. and Gutierrez, Gaston and Guy, Julien and Hill, J. Colin and Honscheid, Klaus and Juneau, Stephanie and Kisner, Theodore and Kremin, Anthony and Lambert, Andrew and Landriau, Martin and Le Guillou, Laurent and MacCrann, Niall and Manera, Marc and Meisner, Aaron and Miquel, Ramon and Moodley, Kavilan and Moustakas, John and Mroczkowski, Tony and Myers, Adam D. and Niemack, Michael D. and Niz, Gustavo and Palanque-Delabrouille, Nathalie and Percival, Will and Pérez-Ràfols, Ignasi and Poppett, Claire and Prada, Francisco and Qu, Frank J. and Rossi, Graziano and Sanchez, Eusebio and Schlegel, David and Schubnell, Michael and Seo, Hee-Jong and Sifón, Cristóbal and Spergel, David N. and Sprayberry, David and Tarlé, Gregory and Vargas-Magaña, Mariana and Vavagiakis, Eve M. and Weaver, Benjamin A. and Wollack, Edward J. and Zarrouk, Pauline},
   year={2025},
   month=Nov }

@article{Zhang_2024,
   title={The hot circumgalactic medium in the eROSITA All-Sky Survey: I. X-ray surface brightness profiles},
   volume={690},
   ISSN={1432-0746},
   url={http://dx.doi.org/10.1051/0004-6361/202449412},
   DOI={10.1051/0004-6361/202449412},
   journal={Astronomy \& Astrophysics},
   publisher={EDP Sciences},
   author={Zhang, Yi and Comparat, Johan and Ponti, Gabriele and Merloni, Andrea and Nandra, Kirpal and Haberl, Frank and Locatelli, Nicola and Zhang, Xiaoyuan and Sanders, Jeremy and Zheng, Xueying and Liu, Ang and Popesso, Paola and Liu, Teng and Truong, Nhut and Pillepich, Annalisa and Predehl, Peter and Salvato, Mara and Shreeram, Soumya and Yeung, Michael C. H. and Ni, Qingling},
   year={2024},
   month=Oct, pages={A267} }

@article{Amon_2022,
   title={Dark Energy Survey Year 3 results: Cosmology from cosmic shear and robustness to data calibration},
   volume={105},
   ISSN={2470-0029},
   url={http://dx.doi.org/10.1103/PhysRevD.105.023514},
   DOI={10.1103/physrevd.105.023514},
   number={2},
   journal={Physical Review D},
   publisher={American Physical Society (APS)},
   author={Amon, A. and Gruen, D. and Troxel, M. A. and MacCrann, N. and Dodelson, S. and Choi, A. and Doux, C. and Secco, L. F. and Samuroff, S. and Krause, E. and Cordero, J. and Myles, J. and DeRose, J. and Wechsler, R. H. and Gatti, M. and Navarro-Alsina, A. and Bernstein, G. M. and Jain, B. and Blazek, J. and Alarcon, A. and Ferté, A. and Lemos, P. and Raveri, M. and Campos, A. and Prat, J. and Sánchez, C. and Jarvis, M. and Alves, O. and Andrade-Oliveira, F. and Baxter, E. and Bechtol, K. and Becker, M. R. and Bridle, S. L. and Camacho, H. and Carnero Rosell, A. and Carrasco Kind, M. and Cawthon, R. and Chang, C. and Chen, R. and Chintalapati, P. and Crocce, M. and Davis, C. and Diehl, H. T. and Drlica-Wagner, A. and Eckert, K. and Eifler, T. F. and Elvin-Poole, J. and Everett, S. and Fang, X. and Fosalba, P. and Friedrich, O. and Gaztanaga, E. and Giannini, G. and Gruendl, R. A. and Harrison, I. and Hartley, W. G. and Herner, K. and Huang, H. and Huff, E. M. and Huterer, D. and Kuropatkin, N. and Leget, P. and Liddle, A. R. and McCullough, J. and Muir, J. and Pandey, S. and Park, Y. and Porredon, A. and Refregier, A. and Rollins, R. P. and Roodman, A. and Rosenfeld, R. and Ross, A. J. and Rykoff, E. S. and Sanchez, J. and Sevilla-Noarbe, I. and Sheldon, E. and Shin, T. and Troja, A. and Tutusaus, I. and Tutusaus, I. and Varga, T. N. and Weaverdyck, N. and Yanny, B. and Yin, B. and Zhang, Y. and Zuntz, J. and Aguena, M. and Allam, S. and Annis, J. and Bacon, D. and Bertin, E. and Bhargava, S. and Brooks, D. and Buckley-Geer, E. and Burke, D. L. and Carretero, J. and Costanzi, M. and da Costa, L. N. and Pereira, M. E. S. and De Vicente, J. and Desai, S. and Dietrich, J. P. and Doel, P. and Ferrero, I. and Flaugher, B. and Frieman, J. and García-Bellido, J. and Gaztanaga, E. and Gerdes, D. W. and Giannantonio, T. and Gschwend, J. and Gutierrez, G. and Hinton, S. R. and Hollowood, D. L. and Honscheid, K. and Hoyle, B. and James, D. J. and Kron, R. and Kuehn, K. and Lahav, O. and Lima, M. and Lin, H. and Maia, M. A. G. and Marshall, J. L. and Martini, P. and Melchior, P. and Menanteau, F. and Miquel, R. and Mohr, J. J. and Morgan, R. and Ogando, R. L. C. and Palmese, A. and Paz-Chinchón, F. and Petravick, D. and Pieres, A. and Romer, A. K. and Sanchez, E. and Scarpine, V. and Schubnell, M. and Serrano, S. and Smith, M. and Soares-Santos, M. and Tarle, G. and Thomas, D. and To, C. and Weller, J. and },
   year={2022},
   month=Jan }

@misc{hadzhiyska2025evidencelargebaryonicfeedback,
      title={Evidence for large baryonic feedback at low and intermediate redshifts from kinematic Sunyaev-Zel'dovich observations with ACT and DESI photometric galaxies}, 
      author={B. Hadzhiyska and S. Ferraro and B. Ried Guachalla and E. Schaan and J. Aguilar and N. Battaglia and J. R. Bond and D. Brooks and E. Calabrese and S. K. Choi and T. Claybaugh and W. R. Coulton and K. Dawson and M. Devlin and B. Dey and P. Doel and A. J. Duivenvoorden and J. Dunkley and G. S. Farren and A. Font-Ribera and J. E. Forero-Romero and P. A. Gallardo and E. Gaztañaga and S. Gontcho Gontcho and M. Gralla and L. Le Guillou and G. Gutierrez and J. Guy and J. C. Hill and R. Hložek and K. Honscheid and S. Juneau and T. Kisner and A. Kremin and M. Landriau and R. H. Liu and T. Louis and N. MacCrann and A. de Macorra and M. Madhavacheril and M. Manera and A. Meisner and R. Miquel and K. Moodley and J. Moustakas and T. Mroczkowski and S. Naess and J. Newman and M. D. Niemack and G. Niz and L. Page and N. Palanque-Delabrouille and B. Partridge and W. J. Percival and F. Prada and F. J. Qu and G. Rossi and E. Sanchez and D. Schlegel and M. Schubnell and N. Sehgal and H. Seo and C. Sifón and D. Spergel and D. Sprayberry and S. Staggs and G. Tarlé and C. Vargas and E. M. Vavagiakis and B. A. Weaver and E. J. Wollack and R. Zhou and H. Zou},
      year={2025},
      eprint={2407.07152},
      archivePrefix={arXiv},
      primaryClass={astro-ph.CO},
      url={https://arxiv.org/abs/2407.07152}, 
}

@misc{peeples2019understandingcircumgalacticmediumcritical,
      title={Understanding the circumgalactic medium is critical for understanding galaxy evolution}, 
      author={Molly S. Peeples and Peter Behroozi and Rongmon Bordoloi and Alyson Brooks and James S. Bullock and Joseph N. Burchett and Hsiao-Wen Chen and John Chisholm and Charlotte Christensen and Alison Coil and Lauren Corlies and Aleksandar Diamond-Stanic and Megan Donahue and Claude-André Faucher-Giguère and Henry Ferguson and Drummond Fielding and Andrew J. Fox and David M. French and Steven R. Furlanetto and Mario Gennaro and Karoline M. Gilbert and Erika Hamden and Nimish Hathi and Matthew Hayes and Alaina Henry and J. Christopher Howk and Cameron Hummels and Dušan Kereš and Evan Kirby and Anton M. Koekemoer and Ting-Wen Lan and Lauranne Lanz and David R. Law and Nicolas Lehner and Jennifer M. Lotz and Crystal L. Martin and Kristen McQuinn and Matthew McQuinn and Ferah Munshi and S. Peng Oh and John M. O'Meara and Brian W. O'Shea and Camilla Pacifici and J. E. G. Peek and Marc Postman and Moire Prescott and Mary Putman and Eliot Quataert and Marc Rafelski and Joseph Ribaudo and Kate Rowlands and Kate Rubin and Brett Salmon and Claudia Scarlata and Alice E. Shapley and Raymond Simons and Gregory F. Snyder and Jonathan Stern and Allison L. Strom and Erik Tollerud and Paul Torrey and Grant Tremblay and Todd M. Tripp and Jason Tumlinson and Sarah Tuttle and Frank C. van den Bosch and G. Mark Voit and Q. Daniel Wang and Jessica K. Werk and Benjamin F. Williams and Dennis Zaritsky and Yong Zheng},
      year={2019},
      eprint={1903.05644},
      archivePrefix={arXiv},
      primaryClass={astro-ph.GA},
      url={https://arxiv.org/abs/1903.05644}, 
}

@article{Rafiei_Ravandi_2021,
   title={CHIME/FRB Catalog 1 Results: Statistical Cross-correlations with Large-scale Structure},
   volume={922},
   ISSN={1538-4357},
   url={http://dx.doi.org/10.3847/1538-4357/ac1dab},
   DOI={10.3847/1538-4357/ac1dab},
   number={1},
   journal={The Astrophysical Journal},
   publisher={American Astronomical Society},
   author={Rafiei-Ravandi, Masoud and Smith, Kendrick M. and Li, Dongzi and Masui, Kiyoshi W. and Josephy, Alexander and Dobbs, Matt and Lang, Dustin and Bhardwaj, Mohit and Patel, Chitrang and Bandura, Kevin and Berger, Sabrina and Boyle, P. J. and Brar, Charanjot and Breitman, Daniela and Cassanelli, Tomas and Chawla, Pragya and Adam Dong, Fengqiu and Fonseca, Emmanuel and Gaensler, B. M. and Giri, Utkarsh and Good, Deborah C. and Halpern, Mark and Kaczmarek, Jane and Kaspi, Victoria M. and Leung, Calvin and Lin, Hsiu-Hsien and Mena-Parra, Juan and Meyers, B. W. and Michilli, D. and Münchmeyer, Moritz and Ng, Cherry and Petroff, Emily and Pleunis, Ziggy and Rahman, Mubdi and Sanghavi, Pranav and Scholz, Paul and Shin, Kaitlyn and Stairs, Ingrid H. and Tendulkar, Shriharsh P. and Vanderlinde, Keith and Zwaniga, Andrew},
   year={2021},
   month=Nov, pages={42} }

@misc{takahashi2025measurementangularcrosscorrelationcosmological,
      title={Measurement of angular cross-correlation between the cosmological dispersion measure and the thermal Sunyaev--Zeldovich effect}, 
      author={Ryuichi Takahashi and Kunihito Ioka and Masato Shirasaki and Ken Osato},
      year={2025},
      eprint={2511.02155},
      archivePrefix={arXiv},
      primaryClass={astro-ph.CO},
      url={https://arxiv.org/abs/2511.02155}, 
}

@article{van_Daalen_2011,
   title={The effects of galaxy formation on the matter power spectrum: a challenge for precision cosmology: Galaxy formation and the matter power spectrum},
   volume={415},
   ISSN={0035-8711},
   url={http://dx.doi.org/10.1111/j.1365-2966.2011.18981.x},
   DOI={10.1111/j.1365-2966.2011.18981.x},
   number={4},
   journal={Monthly Notices of the Royal Astronomical Society},
   publisher={Oxford University Press (OUP)},
   author={van Daalen, Marcel P. and Schaye, Joop and Booth, C. M. and Dalla Vecchia, Claudio},
   year={2011},
   month={July}, pages={3649–3665} }

@article{Wayland_2026,
   title={Probing baryonic feedback with fast radio bursts: joint analyses with cosmic shear and galaxy clustering},
   volume={547},
   ISSN={1365-2966},
   url={http://dx.doi.org/10.1093/mnras/stag557},
   DOI={10.1093/mnras/stag557},
   number={4},
   journal={Monthly Notices of the Royal Astronomical Society},
   publisher={Oxford University Press (OUP)},
   author={Wayland, Amy and Alonso, David and Reischke, Robert},
   year={2026},
   month=Mar }

@article{Connor_2023,
   title={Deep Synoptic Array Science: Two Fast Radio Burst Sources in Massive Galaxy Clusters},
   volume={949},
   ISSN={2041-8213},
   url={http://dx.doi.org/10.3847/2041-8213/acd3ea},
   DOI={10.3847/2041-8213/acd3ea},
   number={2},
   journal={The Astrophysical Journal Letters},
   publisher={American Astronomical Society},
   author={Connor, Liam and Ravi, Vikram and Catha, Morgan and Chen, Ge and Faber, Jakob T. and Lamb, James W. and Hallinan, Gregg and Harnach, Charlie and Hellbourg, Greg and Hobbs, Rick and Hodge, David and Hodges, Mark and Law, Casey and Rasmussen, Paul and Sayers, Jack and Sharma, Kritti and Sherman, Myles B. and Shi, Jun and Simard, Dana and Somalwar, Jean and Squillace, Reynier and Weinreb, Sander and Woody, David P. and Yadlapalli, Nitika and },
   year={2023},
   month=May, pages={L26} }

@article{Anna_Thomas_2025,
   title={Evidence for a Hot Galactic Halo around the Andromeda Galaxy Using Fast Radio Bursts along Two Sightlines},
   volume={993},
   ISSN={1538-4357},
   url={http://dx.doi.org/10.3847/1538-4357/ae1014},
   DOI={10.3847/1538-4357/ae1014},
   number={2},
   journal={The Astrophysical Journal},
   publisher={American Astronomical Society},
   author={Anna-Thomas, Reshma and Law, Casey J. and Koch, Eric W. and Gordon, Alexa C. and Sharma, Kritti and Williams, Benjamin F. and Pingel, Nickolas M. and Burke-Spolaor, Sarah and Chen, Zhuo and Stanley, Jordan and Dear, Calvin and Verdi, Frank and Prochaska, J. Xavier and Bower, Geoffrey C. and Chomiuk, Laura and Connor, Liam and Demorest, Paul B. and Fong, Wen-Fai and Nugent, Anya and Walter, Fabian},
   year={2025},
   month=Nov, pages={221} }

@misc{hussaini2025correlationfrbdispersionmeasure,
      title={A Correlation Between FRB Dispersion Measure and Foreground Large-Scale Structure}, 
      author={Maryam Hussaini and Liam Connor and Ralf M. Konietzka and Vikram Ravi and Jakob Faber and Kritti Sharma and Myles Sherman},
      year={2025},
      eprint={2506.04186},
      archivePrefix={arXiv},
      primaryClass={astro-ph.CO},
      url={https://arxiv.org/abs/2506.04186}, 
}

@misc{lanman2026constraininggasmassfractions,
      title={Constraining Gas Mass Fractions in Galaxy Groups and Clusters with the First CHIME/FRB Outrigger}, 
      author={Adam E. Lanman and Sunil Simha and Kiyoshi W. Masui and J. Xavier Prochaska and Rachel Darlinger and Fengqiu Adam Dong and B. M. Gaensler and Ronniy C. Joseph and Jane Kaczmarek and Lordrick Kahinga and Afrokk Khan and Calvin Leung and Lluis Mas-Ribas and Swarali Shivraj Patil and Aaron B. Pearlman and Mawson Sammons and Kaitlyn Shin and Kendrick Smith and Haochen Wang},
      year={2026},
      eprint={2509.07097},
      archivePrefix={arXiv},
      primaryClass={astro-ph.GA},
      url={https://arxiv.org/abs/2509.07097}, 
}

@misc{leung2025nullingbaryonicfeedbackweak,
      title={Nulling baryonic feedback in weak lensing surveys using cross-correlations with fast radio bursts}, 
      author={Calvin Leung and Josh Borrow and Kiyoshi W. Masui and Shion Andrew and Kai-Feng Chen and Joop Schaye and Matthieu Schaller},
      year={2025},
      eprint={2509.19514},
      archivePrefix={arXiv},
      primaryClass={astro-ph.CO},
      url={https://arxiv.org/abs/2509.19514}, 
}

@article{McCarthy_2010,
   title={The case for AGN feedback in galaxy groups: AGN feedback in galaxy groups},
   ISSN={1365-2966},
   url={http://dx.doi.org/10.1111/j.1365-2966.2010.16750.x},
   DOI={10.1111/j.1365-2966.2010.16750.x},
   journal={Monthly Notices of the Royal Astronomical Society},
   publisher={Oxford University Press (OUP)},
   author={McCarthy, I. G. and Schaye, J. and Ponman, T. J. and Bower, R. G. and Booth, C. M. and Vecchia, C. Dalla and Crain, R. A. and Springel, V. and Theuns, T. and Wiersma, R. P. C.},
   year={2010},
   month=May, pages={no–no} }

@misc{reischke2025measurementbaryonicfeedbackfast,
      title={A first measurement of baryonic feedback with Fast Radio Bursts}, 
      author={Robert Reischke and Steffen Hagstotz},
      year={2025},
      eprint={2507.17742},
      archivePrefix={arXiv},
      primaryClass={astro-ph.CO},
      url={https://arxiv.org/abs/2507.17742}, 
}

@misc{sharma2025probingbaryonicfeedbackcosmology,
      title={Probing baryonic feedback and cosmology with 3$\times$2-point statistic of FRBs and galaxies}, 
      author={Kritti Sharma and Elisabeth Krause and Vikram Ravi and Robert Reischke and Liam Connor and Pranjal R. S. and Dhayaa Anbajagane},
      year={2025},
      eprint={2509.05866},
      archivePrefix={arXiv},
      primaryClass={astro-ph.CO},
      url={https://arxiv.org/abs/2509.05866}, 
}

@misc{sharma2025hydrodynamicalsimulationsbasedmodelconnects,
      title={A hydrodynamical simulations-based model that connects the FRB DM--redshift relation to suppression of the matter power spectrum via feedback}, 
      author={Kritti Sharma and Elisabeth Krause and Vikram Ravi and Robert Reischke and Pranjal R. S. and Liam Connor},
      year={2025},
      eprint={2504.18745},
      archivePrefix={arXiv},
      primaryClass={astro-ph.CO},
      url={https://arxiv.org/abs/2504.18745}, 
}

@article{Wu_2023,
   title={A Measurement of Circumgalactic Gas around Nearby Galaxies Using Fast Radio Bursts},
   volume={945},
   ISSN={1538-4357},
   url={http://dx.doi.org/10.3847/1538-4357/acbc7d},
   DOI={10.3847/1538-4357/acbc7d},
   number={2},
   journal={The Astrophysical Journal},
   publisher={American Astronomical Society},
   author={Wu, Xiaohan and McQuinn, Matthew},
   year={2023},
   month=Mar, pages={87} }

@misc{mccarty2026cgmlocaluniversefrbs,
      title={The CGM with local universe FRBs: evidence of strong AGN feedback in a massive elliptical galaxy}, 
      author={Samuel McCarty and Liam Connor and Ralf M. Konietzka},
      year={2026},
      eprint={2602.16781},
      archivePrefix={arXiv},
      primaryClass={astro-ph.GA},
      url={https://arxiv.org/abs/2602.16781}, 
}

@article{Tumlinson_2017,
   title={The Circumgalactic Medium},
   volume={55},
   ISSN={1545-4282},
   url={http://dx.doi.org/10.1146/annurev-astro-091916-055240},
   DOI={10.1146/annurev-astro-091916-055240},
   number={1},
   journal={Annual Review of Astronomy and Astrophysics},
   publisher={Annual Reviews},
   author={Tumlinson, Jason and Peeples, Molly S. and Werk, Jessica K.},
   year={2017},
   month=Aug, pages={389–432} }

@article{EuclidI,
   title={Euclid: I. Overview of the Euclid mission},
   volume={697},
   ISSN={1432-0746},
   url={http://dx.doi.org/10.1051/0004-6361/202450810},
   DOI={10.1051/0004-6361/202450810},
   journal={Astronomy \&amp; Astrophysics},
   publisher={EDP Sciences},
   author={Mellier, Y. and Abdurro’uf and Acevedo Barroso, J. A. and Achúcarro, A. and Adamek, J. and Adam, R. and Addison, G. E. and Aghanim, N. and Aguena, M. and Ajani, V. and Akrami, Y. and Al-Bahlawan, A. and Alavi, A. and Albuquerque, I. S. and Alestas, G. and Alguero, G. and Allaoui, A. and Allen, S. W. and Allevato, V. and Alonso-Tetilla, A. V. and Altieri, B. and Alvarez-Candal, A. and Alvi, S. and Amara, A. and Amendola, L. and Amiaux, J. and Andika, I. T. and Andreon, S. and Andrews, A. and Angora, G. and Angulo, R. E. and Annibali, F. and Anselmi, A. and Anselmi, S. and Arcari, S. and Archidiacono, M. and Aricò, G. and Arnaud, M. and Arnouts, S. and Asgari, M. and Asorey, J. and Atayde, L. and Atek, H. and Atrio-Barandela, F. and Aubert, M. and Aubourg, E. and Auphan, T. and Auricchio, N. and Aussel, B. and Aussel, H. and Avelino, P. P. and Avgoustidis, A. and Avila, S. and Awan, S. and Azzollini, R. and Baccigalupi, C. and Bachelet, E. and Bacon, D. and Baes, M. and Bagley, M. B. and Bahr-Kalus, B. and Balaguera-Antolinez, A. and Balbinot, E. and Balcells, M. and Baldi, M. and Baldry, I. and Balestra, A. and Ballardini, M. and Ballester, O. and Balogh, M. and Bañados, E. and Barbier, R. and Bardelli, S. and Baron, M. and Barreiro, T. and Barrena, R. and Barriere, J.-C. and Barros, B. J. and Barthelemy, A. and Bartolo, N. and Basset, A. and Battaglia, P. and Battisti, A. J. and Baugh, C. M. and Baumont, L. and Bazzanini, L. and Beaulieu, J.-P. and Beckmann, V. and Belikov, A. N. and Bel, J. and Bellagamba, F. and Bella, M. and Bellini, E. and Benabed, K. and Bender, R. and Benevento, G. and Bennett, C. L. and Benson, K. and Bergamini, P. and Bermejo-Climent, J. R. and Bernardeau, F. and Bertacca, D. and Berthe, M. and Berthier, J. and Bethermin, M. and Beutler, F. and Bevillon, C. and Bhargava, S. and Bhatawdekar, R. and Bianchi, D. and Bisigello, L. and Biviano, A. and Blake, R. P. and Blanchard, A. and Blazek, J. and Blot, L. and Bosco, A. and Bodendorf, C. and Boenke, T. and Böhringer, H. and Boldrini, P. and Bolzonella, M. and Bonchi, A. and Bonici, M. and Bonino, D. and Bonino, L. and Bonvin, C. and Bon, W. and Booth, J. T. and Borgani, S. and Borlaff, A. S. and Borsato, E. and Bosco, A. and Bose, B. and Botticella, M. T. and Boucaud, A. and Bouche, F. and Boucher, J. S. and Boutigny, D. and Bouvard, T. and Bouwens, R. and Bouy, H. and Bowler, R. A. A. and Bozza, V. and Bozzo, E. and Branchini, E. and Brando, G. and Brau-Nogue, S. and Brekke, P. and Bremer, M. N. and Brescia, M. and Breton, M.-A. and Brinchmann, J. and Brinckmann, T. and Brockley-Blatt, C. and Brodwin, M. and Brouard, L. and Brown, M. L. and Bruton, S. and Bucko, J. and Buddelmeijer, H. and Buenadicha, G. and Buitrago, F. and Burger, P. and Burigana, C. and Busillo, V. and Busonero, D. and Cabanac, R. and Cabayol-Garcia, L. and Cagliari, M. S. and Caillat, A. and Caillat, L. and Calabrese, M. and Calabro, A. and Calderone, G. and Calura, F. and Camacho Quevedo, B. and Camera, S. and Campos, L. and Cañas-Herrera, G. and Candini, G. P. and Cantiello, M. and Capobianco, V. and Cappellaro, E. and Cappelluti, N. and Cappi, A. and Caputi, K. I. and Cara, C. and Carbone, C. and Cardone, V. F. and Carella, E. and Carlberg, R. G. and Carle, M. and Carminati, L. and Caro, F. and Carrasco, J. M. and Carretero, J. and Carrilho, P. and Carron Duque, J. and Carry, B. and Carvalho, A. and Carvalho, C. S. and Casas, R. and Casas, S. and Casenove, P. and Casey, C. M. and Cassata, P. and Castander, F. J. and Castelao, D. and Castellano, M. and Castiblanco, L. and Castignani, G. and Castro, T. and Cavet, C. and Cavuoti, S. and Chabaud, P.-Y. and Chambers, K. C. and Charles, Y. and Charlot, S. and Chartab, N. and Chary, R. and Chaumeil, F. and Cho, H. and Chon, G. and Ciancetta, E. and Ciliegi, P. and Cimatti, A. and Cimino, M. and Cioni, M.-R. L. and Claydon, R. and Cleland, C. and Clément, B. and Clements, D. L. and Clerc, N. and Clesse, S. and Codis, S. and Cogato, F. and Colbert, J. and Cole, R. E. and Coles, P. and Collett, T. E. and Collins, R. S. and Colodro-Conde, C. and Colombo, C. and Combes, F. and Conforti, V. and Congedo, G. and Conseil, S. and Conselice, C. J. and Contarini, S. and Contini, T. and Conversi, L. and Cooray, A. R. and Copin, Y. and Corasaniti, P.-S. and Corcho-Caballero, P. and Corcione, L. and Cordes, O. and Corpace, O. and Correnti, M. and Costanzi, M. and Costille, A. and Courbin, F. and Courcoult Mifsud, L. and Courtois, H. M. and Cousinou, M.-C. and Covone, G. and Cowell, T. and Cragg, C. and Cresci, G. and Cristiani, S. and Crocce, M. and Cropper, M. and Crouzet, P. E. and Csizi, B. and Cuby, J.-G. and Cucchetti, E. and Cucciati, O. and Cuillandre, J.-C. and Cunha, P. A. C. and Cuozzo, V. and Daddi, E. and D’Addona, M. and Dafonte, C. and Dagoneau, N. and Dalessandro, E. and Dalton, G. B. and D’Amico, G. and Dannerbauer, H. and Danto, P. and Das, I. and Da Silva, A. and da Silva, R. and d’Assignies Doumerg, W. and Daste, G. and Davies, J. E. and Davini, S. and Dayal, P. and de Boer, T. and Decarli, R. and De Caro, B. and Degaudenzi, H. and Degni, G. and de Jong, J. T. A. and de la Bella, L. F. and de la Torre, S. and Delhaise, F. and Delley, D. and Delucchi, G. and De Lucia, G. and Denniston, J. and De Paolis, F. and De Petris, M. and Derosa, A. and Desai, S. and Desjacques, V. and Despali, G. and Desprez, G. and De Vicente-Albendea, J. and Deville, Y. and Dias, J. D. F. and Díaz-Sánchez, A. and Diaz, J. J. and Di Domizio, S. and Diego, J. M. and Di Ferdinando, D. and Di Giorgio, A. M. and Dimauro, P. and Dinis, J. and Dolag, K. and Dolding, C. and Dole, H. and Domínguez Sánchez, H. and Doré, O. and Dournac, F. and Douspis, M. and Dreihahn, H. and Droge, B. and Dryer, B. and Dubath, F. and Duc, P.-A. and Ducret, F. and Duffy, C. and Dufresne, F. and Duncan, C. A. J. and Dupac, X. and Duret, V. and Durrer, R. and Durret, F. and Dusini, S. and Ealet, A. and Eggemeier, A. and Eisenhardt, P. R. M. and Elbaz, D. and Elkhashab, M. Y. and Ellien, A. and Endicott, J. and Enia, A. and Erben, T. and Escartin Vigo, J. A. and Escoffier, S. and Escudero Sanz, I. and Essert, J. and Ettori, S. and Ezziati, M. and Fabbian, G. and Fabricius, M. and Fang, Y. and Farina, A. and Farina, M. and Farinelli, R. and Farrens, S. and Faustini, F. and Feltre, A. and Ferguson, A. M. N. and Ferrando, P. and Ferrari, A. G. and Ferré-Mateu, A. and Ferreira, P. G. and Ferreras, I. and Ferrero, I. and Ferriol, S. and Ferruit, P. and Filleul, D. and Finelli, F. and Finkelstein, S. L. and Finoguenov, A. and Fiorini, B. and Flentge, F. and Focardi, P. and Fonseca, J. and Fontana, A. and Fontanot, F. and Fornari, F. and Fosalba, P. and Fossati, M. and Fotopoulou, S. and Fouchez, D. and Fourmanoit, N. and Frailis, M. and Fraix-Burnet, D. and Franceschi, E. and Franco, A. and Franzetti, P. and Freihoefer, J. and Frenk, C. S. and Frittoli, G. and Frugier, P.-A. and Frusciante, N. and Fumagalli, A. and Fumagalli, M. and Fumana, M. and Fu, Y. and Gabarra, L. and Galeotta, S. and Galluccio, L. and Ganga, K. and Gao, H. and García-Bellido, J. and Garcia, K. and Gardner, J. P. and Garilli, B. and Gaspar-Venancio, L.-M. and Gasparetto, T. and Gautard, V. and Gavazzi, R. and Gaztanaga, E. and Genolet, L. and Genova Santos, R. and Gentile, F. and George, K. and Gerbino, M. and Ghaffari, Z. and Giacomini, F. and Gianotti, F. and Gibb, G. P. S. and Gillard, W. and Gillis, B. and Ginolfi, M. and Giocoli, C. and Girardi, M. and Giri, S. K. and Goh, L. W. K. and Gómez-Alvarez, P. and Gonzalez-Perez, V. and Gonzalez, A. H. and Gonzalez, E. J. and Gonzalez, J. C. and Gouyou Beauchamps, S. and Gozaliasl, G. and Gracia-Carpio, J. and Grandis, S. and Granett, B. R. and Granvik, M. and Grazian, A. and Gregorio, A. and Grenet, C. and Grillo, C. and Grupp, F. and Gruppioni, C. and Gruppuso, A. and Guerbuez, C. and Guerrini, S. and Guidi, M. and Guillard, P. and Gutierrez, C. M. and Guttridge, P. and Guzzo, L. and Gwyn, S. and Haapala, J. and Haase, J. and Haddow, C. R. and Hailey, M. and Hall, A. and Hall, D. and Hamaus, N. and Haridasu, B. S. and Harnois-Déraps, J. and Harper, C. and Hartley, W. G. and Hasinger, G. and Hassani, F. and Hatch, N. A. and Haugan, S. V. H. and Häußler, B. and Heavens, A. and Heisenberg, L. and Helmi, A. and Helou, G. and Hemmati, S. and Henares, K. and Herent, O. and Hernández-Monteagudo, C. and Heuberger, T. and Hewett, P. C. and Heydenreich, S. and Hildebrandt, H. and Hirschmann, M. and Hjorth, J. and Hoar, J. and Hoekstra, H. and Holland, A. D. and Holliman, M. S. and Holmes, W. and Hook, I. and Horeau, B. and Hormuth, F. and Hornstrup, A. and Hosseini, S. and Hu, D. and Hudelot, P. and Hudson, M. J. and Huertas-Company, M. and Huff, E. M. and Hughes, A. C. N. and Humphrey, A. and Hunt, L. K. and Huynh, D. D. and Ibata, R. and Ichikawa, K. and Iglesias-Groth, S. and Ilbert, O. and Ilić, S. and Ingoglia, L. and Iodice, E. and Israel, H. and Israelsson, U. E. and Izzo, L. and Jablonka, P. and Jackson, N. and Jacobson, J. and Jafariyazani, M. and Jahnke, K. and Jain, B. and Jansen, H. and Jarvis, M. J. and Jasche, J. and Jauzac, M. and Jeffrey, N. and Jhabvala, M. and Jimenez-Teja, Y. and Jimenez Muñoz, A. and Joachimi, B. and Johansson, P. H. and Joudaki, S. and Jullo, E. and Kajava, J. J. E. and Kang, Y. and Kannawadi, A. and Kansal, V. and Karagiannis, D. and Kärcher, M. and Kashlinsky, A. and Kazandjian, M. V. and Keck, F. and Keihänen, E. and Kerins, E. and Kermiche, S. and Khalil, A. and Kiessling, A. and Kiiveri, K. and Kilbinger, M. and Kim, J. and King, R. and Kirkpatrick, C. C. and Kitching, T. and Kluge, M. and Knabenhans, M. and Knapen, J. H. and Knebe, A. and Kneib, J.-P. and Kohley, R. and Koopmans, L. V. E. and Koskinen, H. and Koulouridis, E. and Kou, R. and Kovács, A. and Kovačić, I. and Kowalczyk, A. and Koyama, K. and Kraljic, K. and Krause, O. and Kruk, S. and Kubik, B. and Kuchner, U. and Kuijken, K. and Kümmel, M. and Kunz, M. and Kurki-Suonio, H. and Lacasa, F. and Lacey, C. G. and La Franca, F. and Lagarde, N. and Lahav, O. and Laigle, C. and La Marca, A. and La Marle, O. and Lamine, B. and Lam, M. C. and Lançon, A. and Landt, H. and Langer, M. and Lapi, A. and Larcheveque, C. and Larsen, S. S. and Lattanzi, M. and Laudisio, F. and Laugier, D. and Laureijs, R. and Laurent, V. and Lavaux, G. and Lawrenson, A. and Lazanu, A. and Lazeyras, T. and Le Boulc’h, Q. and Le Brun, A. M. C. and Le Brun, V. and Leclercq, F. and Lee, S. and Le Graet, J. and Legrand, L. and Leirvik, K. N. and Le Jeune, M. and Lembo, M. and Le Mignant, D. and Lepinzan, M. D. and Lepori, F. and Le Reun, A. and Leroy, G. and Lesci, G. F. and Lesgourgues, J. and Leuzzi, L. and Levi, M. E. and Liaudat, T. I. and Libet, G. and Liebing, P. and Ligori, S. and Lilje, P. B. and Lin, C.-C. and Linde, D. and Linder, E. and Lindholm, V. and Linke, L. and Li, S.-S. and Liu, S. J. and Lloro, I. and Lobo, F. S. N. and Lodieu, N. and Lombardi, M. and Lombriser, L. and Lonare, P. and Longo, G. and López-Caniego, M. and Lopez Lopez, X. and Lorenzo Alvarez, J. and Loureiro, A. and Loveday, J. and Lusso, E. and Macias-Perez, J. and Maciaszek, T. and Maggio, G. and Magliocchetti, M. and Magnard, F. and Magnier, E. A. and Magro, A. and Mahler, G. and Mainetti, G. and Maino, D. and Maiorano, E. and Maiorano, E. and Malavasi, N. and Mamon, G. A. and Mancini, C. and Mandelbaum, R. and Manera, M. and Manjón-García, A. and Mannucci, F. and Mansutti, O. and Manteiga Outeiro, M. and Maoli, R. and Maraston, C. and Marcin, S. and Marcos-Arenal, P. and Margalef-Bentabol, B. and Marggraf, O. and Marinucci, D. and Marinucci, M. and Markovic, K. and Marleau, F. R. and Marpaud, J. and Martignac, J. and Martín-Fleitas, J. and Martin-Moruno, P. and Martin, E. L. and Martinelli, M. and Martinet, N. and Martin, H. and Martins, C. J. A. P. and Marulli, F. and Massari, D. and Massey, R. and Masters, D. C. and Matarrese, S. and Matsuoka, Y. and Matthew, S. and Maughan, B. J. and Mauri, N. and Maurin, L. and Maurogordato, S. and McCarthy, K. and McConnachie, A. W. and McCracken, H. J. and McDonald, I. and McEwen, J. D. and McPartland, C. J. R. and Medinaceli, E. and Mehta, V. and Mei, S. and Melchior, M. and Melin, J.-B. and Ménard, B. and Mendes, J. and Mendez-Abreu, J. and Meneghetti, M. and Mercurio, A. and Merlin, E. and Metcalf, R. B. and Meylan, G. and Migliaccio, M. and Mignoli, M. and Miller, L. and Miluzio, M. and Milvang-Jensen, B. and Mimoso, J. P. and Miquel, R. and Miyatake, H. and Mobasher, B. and Mohr, J. J. and Monaco, P. and Monguió, M. and Montoro, A. and Mora, A. and Moradinezhad Dizgah, A. and Moresco, M. and Moretti, C. and Morgante, G. and Morisset, N. and Moriya, T. J. and Morris, P. W. and Mortlock, D. J. and Moscardini, L. and Mota, D. F. and Mottet, S. and Moustakas, L. A. and Moutard, T. and Müller, T. and Munari, E. and Murphree, G. and Murray, C. and Murray, N. and Musi, P. and Nadathur, S. and Nagam, B. C. and Nagao, T. and Naidoo, K. and Nakajima, R. and Nally, C. and Natoli, P. and Navarro-Alsina, A. and Navarro Girones, D. and Neissner, C. and Nersesian, A. and Nesseris, S. and Nguyen-Kim, H. N. and Nicastro, L. and Nichol, R. C. and Nielbock, M. and Niemi, S.-M. and Nieto, S. and Nilsson, K. and Noller, J. and Norberg, P. and Nouri-Zonoz, A. and Ntelis, P. and Nucita, A. A. and Nugent, P. and Nunes, N. J. and Nutma, T. and Ocampo, I. and Odier, J. and Oesch, P. A. and Oguri, M. and Magalhaes Oliveira, D. and Onoue, M. and Oosterbroek, T. and Oppizzi, F. and Ordenovic, C. and Osato, K. and Pacaud, F. and Pace, F. and Padilla, C. and Paech, K. and Pagano, L. and Page, M. J. and Palazzi, E. and Paltani, S. and Pamuk, S. and Pandolfi, S. and Paoletti, D. and Paolillo, M. and Papaderos, P. and Pardede, K. and Parimbelli, G. and Parmar, A. and Partmann, C. and Pasian, F. and Passalacqua, F. and Paterson, K. and Patrizii, L. and Pattison, C. and Paulino-Afonso, A. and Paviot, R. and Peacock, J. A. and Pearce, F. R. and Pedersen, K. and Peel, A. and Peletier, R. F. and Pellejero Ibanez, M. and Pello, R. and Penny, M. T. and Percival, W. J. and Perez-Garrido, A. and Perotto, L. and Pettorino, V. and Pezzotta, A. and Pezzuto, S. and Philippon, A. and Pierre, M. and Piersanti, O. and Pietroni, M. and Piga, L. and Pilo, L. and Pires, S. and Pisani, A. and Pizzella, A. and Pizzuti, L. and Plana, C. and Polenta, G. and Pollack, J. E. and Poncet, M. and Pöntinen, M. and Pool, P. and Popa, L. A. and Popa, V. and Popp, J. and Porciani, C. and Porth, L. and Potter, D. and Poulain, M. and Pourtsidou, A. and Pozzetti, L. and Prandoni, I. and Pratt, G. W. and Prezelus, S. and Prieto, E. and Pugno, A. and Quai, S. and Quilley, L. and Racca, G. D. and Raccanelli, A. and Rácz, G. and Radinović, S. and Radovich, M. and Ragagnin, A. and Ragnit, U. and Raison, F. and Ramos-Chernenko, N. and Ranc, C. and Rasera, Y. and Raylet, N. and Rebolo, R. and Refregier, A. and Reimberg, P. and Reiprich, T. H. and Renk, F. and Renzi, A. and Retre, J. and Revaz, Y. and Reylé, C. and Reynolds, L. and Rhodes, J. and Ricci, F. and Ricci, M. and Riccio, G. and Ricken, S. O. and Rissanen, S. and Risso, I. and Rix, H.-W. and Robin, A. C. and Rocca-Volmerange, B. and Rocci, P.-F. and Rodenhuis, M. and Rodighiero, G. and Rodriguez Monroy, M. and Rollins, R. P. and Romanello, M. and Roman, J. and Romelli, E. and Romero-Gomez, M. and Roncarelli, M. and Rosati, P. and Rosset, C. and Rossetti, E. and Roster, W. and Rottgering, H. J. A. and Rozas-Fernández, A. and Ruane, K. and Rubino-Martin, J. A. and Rudolph, A. and Ruppin, F. and Rusholme, B. and Sacquegna, S. and Sáez-Casares, I. and Saga, S. and Saglia, R. and Sahlén, M. and Saifollahi, T. and Sakr, Z. and Salvalaggio, J. and Salvaterra, R. and Salvati, L. and Salvato, M. and Salvignol, J.-C. and Sánchez, A. G. and Sanchez, E. and Sanders, D. B. and Sapone, D. and Saponara, M. and Sarpa, E. and Sarron, F. and Sartori, S. and Sartoris, B. and Sassolas, B. and Sauniere, L. and Sauvage, M. and Sawicki, M. and Scaramella, R. and Scarlata, C. and Scharré, L. and Schaye, J. and Schewtschenko, J. A. and Schindler, J.-T. and Schinnerer, E. and Schirmer, M. and Schmidt, F. and Schmidt, F. and Schmidt, M. and Schneider, A. and Schneider, M. and Schneider, P. and Schöneberg, N. and Schrabback, T. and Schultheis, M. and Schulz, S. and Schuster, N. and Schwartz, J. and Sciotti, D. and Scodeggio, M. and Scognamiglio, D. and Scott, D. and Scottez, V. and Secroun, A. and Sefusatti, E. and Seidel, G. and Seiffert, M. and Sellentin, E. and Selwood, M. and Semboloni, E. and Sereno, M. and Serjeant, S. and Serrano, S. and Setnikar, G. and Shankar, F. and Sharples, R. M. and Short, A. and Shulevski, A. and Shuntov, M. and Sias, M. and Sikkema, G. and Silvestri, A. and Simon, P. and Sirignano, C. and Sirri, G. and Skottfelt, J. and Slezak, E. and Sluse, D. and Smith, G. P. and Smith, L. C. and Smith, R. E. and Smit, S. J. A. and Soldano, F. and Solheim, B. G. B. and Sorce, J. G. and Sorrenti, F. and Soubrie, E. and Spinoglio, L. and Spurio Mancini, A. and Stadel, J. and Stagnaro, L. and Stanco, L. and Stanford, S. A. and Starck, J.-L. and Stassi, P. and Steinwagner, J. and Stern, D. and Stone, C. and Strada, P. and Strafella, F. and Stramaccioni, D. and Surace, C. and Sureau, F. and Suyu, S. H. and Swindells, I. and Szafraniec, M. and Szapudi, I. and Taamoli, S. and Talia, M. and Tallada-Crespí, P. and Tanidis, K. and Tao, C. and Tarrío, P. and Tavagnacco, D. and Taylor, A. N. and Taylor, J. E. and Taylor, P. L. and Teixeira, E. M. and Tenti, M. and Teodoro Idiago, P. and Teplitz, H. I. and Tereno, I. and Tessore, N. and Testa, V. and Testera, G. and Tewes, M. and Teyssier, R. and Theret, N. and Thizy, C. and Thomas, P. D. and Toba, Y. and Toft, S. and Toledo-Moreo, R. and Tolstoy, E. and Tommasi, E. and Torbaniuk, O. and Torradeflot, F. and Tortora, C. and Tosi, S. and Tosti, S. and Trifoglio, M. and Troja, A. and Trombetti, T. and Tronconi, A. and Tsedrik, M. and Tsyganov, A. and Tucci, M. and Tutusaus, I. and Uhlemann, C. and Ulivi, L. and Urbano, M. and Vacher, L. and Vaillon, L. and Valageas, P. and Valdes, I. and Valentijn, E. A. and Valenziano, L. and Valieri, C. and Valiviita, J. and Van den Broeck, M. and Vassallo, T. and Vavrek, R. and Vega-Ferrero, J. and Venemans, B. and Venhola, A. and Ventura, S. and Verdoes Kleijn, G. and Vergani, D. and Verma, A. and Vernizzi, F. and Veropalumbo, A. and Verza, G. and Vescovi, C. and Vibert, D. and Viel, M. and Vielzeuf, P. and Viglione, C. and Viitanen, A. and Villaescusa-Navarro, F. and Vinciguerra, S. and Visticot, F. and Voggel, K. and von Wietersheim-Kramsta, M. and Vriend, W. J. and Wachter, S. and Walmsley, M. and Walth, G. and Walton, D. M. and Walton, N. A. and Wander, M. and Wang, L. and Wang, Y. and Weaver, J. R. and Weller, J. and Wetzstein, M. and Whalen, D. J. and Whittam, I. H. and Widmer, A. and Wiesmann, M. and Wilde, J. and Williams, O. R. and Winther, H.-A. and Wittje, A. and Wong, J. H. W. and Wright, A. H. and Yankelevich, V. and Yeung, H. W. and Yoon, M. and Youles, S. and Yung, L. Y. A. and Zacchei, A. and Zalesky, L. and Zamorani, G. and Zamorano Vitorelli, A. and Zanoni Marc, M. and Zennaro, M. and Zerbi, F. M. and Zinchenko, I. A. and Zoubian, J. and Zucca, E. and Zumalacarregui, M.},
   year={2025},
   month=apr, pages={A1} }

@article{Ivezi_2019,
   title={LSST: From Science Drivers to Reference Design and Anticipated Data Products},
   volume={873},
   ISSN={1538-4357},
   url={http://dx.doi.org/10.3847/1538-4357/ab042c},
   DOI={10.3847/1538-4357/ab042c},
   number={2},
   journal={The Astrophysical Journal},
   publisher={American Astronomical Society},
   author={Ivezic, Zeljko and Kahn, Steven M. and Tyson, J. Anthony and Abel, Bob and Acosta, Emily and Allsman, Robyn and Alonso, David and AlSayyad, Yusra and Anderson, Scott F. and Andrew, John and P. Angel, James Roger and Angeli, George Z. and Ansari, Reza and Antilogus, Pierre and Araujo, Constanza and Armstrong, Robert and Arndt, Kirk T. and Astier, Pierre and Aubourg, Éric and Auza, Nicole and Axelrod, Tim S. and Bard, Deborah J. and Barr, Jeff D. and Barrau, Aurelian and Bartlett, James G. and Bauer, Amanda E. and Bauman, Brian J. and Baumont, Sylvain and Bechtol, Ellen and Bechtol, Keith and Becker, Andrew C. and Becla, Jacek and Beldica, Cristina and Bellavia, Steve and Bianco, Federica B. and Biswas, Rahul and Blanc, Guillaume and Blazek, Jonathan and Blandford, Roger D. and Bloom, Josh S. and Bogart, Joanne and Bond, Tim W. and Booth, Michael T. and Borgland, Anders W. and Borne, Kirk and Bosch, James F. and Boutigny, Dominique and Brackett, Craig A. and Bradshaw, Andrew and Brandt, William Nielsen and Brown, Michael E. and Bullock, James S. and Burchat, Patricia and Burke, David L. and Cagnoli, Gianpietro and Calabrese, Daniel and Callahan, Shawn and Callen, Alice L. and Carlin, Jeffrey L. and Carlson, Erin L. and Chandrasekharan, Srinivasan and Charles-Emerson, Glenaver and Chesley, Steve and Cheu, Elliott C. and Chiang, Hsin-Fang and Chiang, James and Chirino, Carol and Chow, Derek and Ciardi, David R. and Claver, Charles F. and Cohen-Tanugi, Johann and Cockrum, Joseph J. and Coles, Rebecca and Connolly, Andrew J. and Cook, Kem H. and Cooray, Asantha and Covey, Kevin R. and Cribbs, Chris and Cui, Wei and Cutri, Roc and Daly, Philip N. and Daniel, Scott F. and Daruich, Felipe and Daubard, Guillaume and Daues, Greg and Dawson, William and Delgado, Francisco and Dellapenna, Alfred and Peyster, Robert de and Val-Borro, Miguel de and Digel, Seth W. and Doherty, Peter and Dubois, Richard and Dubois-Felsmann, Gregory P. and Durech, Josef and Economou, Frossie and Eifler, Tim and Eracleous, Michael and Emmons, Benjamin L. and Neto, Angelo Fausti and Ferguson, Henry and Figueroa, Enrique and Fisher-Levine, Merlin and Focke, Warren and Foss, Michael D. and Frank, James and Freemon, Michael D. and Gangler, Emmanuel and Gawiser, Eric and Geary, John C. and Gee, Perry and Geha, Marla and Gessner, Charles J. B. and Gibson, Robert R. and Gilmore, D. Kirk and Glanzman, Thomas and Glick, William and Goldina, Tatiana and Goldstein, Daniel A. and Goodenow, Iain and Graham, Melissa L. and Gressler, William J. and Gris, Philippe and Guy, Leanne P. and Guyonnet, Augustin and Haller, Gunther and Harris, Ron and Hascall, Patrick A. and Haupt, Justine and Hernandez, Fabio and Herrmann, Sven and Hileman, Edward and Hoblitt, Joshua and Hodgson, John A. and Hogan, Craig and Howard, James D. and Huang, Dajun and Huffer, Michael E. and Ingraham, Patrick and Innes, Walter R. and Jacoby, Suzanne H. and Jain, Bhuvnesh and Jammes, Fabrice and Jee, M. James and Jenness, Tim and Jernigan, Garrett and Jevremović, Darko and Johns, Kenneth and Johnson, Anthony S. and Johnson, Margaret W. G. and Jones, R. Lynne and Juramy-Gilles, Claire and Jurić, Mario and Kalirai, Jason S. and Kallivayalil, Nitya J. and Kalmbach, Bryce and Kantor, Jeffrey P. and Karst, Pierre and Kasliwal, Mansi M. and Kelly, Heather and Kessler, Richard and Kinnison, Veronica and Kirkby, David and Knox, Lloyd and Kotov, Ivan V. and Krabbendam, Victor L. and Krughoff, K. Simon and Kubánek, Petr and Kuczewski, John and Kulkarni, Shri and Ku, John and Kurita, Nadine R. and Lage, Craig S. and Lambert, Ron and Lange, Travis and Langton, J. Brian and Guillou, Laurent Le and Levine, Deborah and Liang, Ming and Lim, Kian-Tat and Lintott, Chris J. and Long, Kevin E. and Lopez, Margaux and Lotz, Paul J. and Lupton, Robert H. and Lust, Nate B. and MacArthur, Lauren A. and Mahabal, Ashish and Mandelbaum, Rachel and Markiewicz, Thomas W. and Marsh, Darren S. and Marshall, Philip J. and Marshall, Stuart and May, Morgan and McKercher, Robert and McQueen, Michelle and Meyers, Joshua and Migliore, Myriam and Miller, Michelle and Mills, David J. and Miraval, Connor and Moeyens, Joachim and Moolekamp, Fred E. and Monet, David G. and Moniez, Marc and Monkewitz, Serge and Montgomery, Christopher and Morrison, Christopher B. and Mueller, Fritz and Muller, Gary P. and Arancibia, Freddy Muñoz and Neill, Douglas R. and Newbry, Scott P. and Nief, Jean-Yves and Nomerotski, Andrei and Nordby, Martin and O’Connor, Paul and Oliver, John and Olivier, Scot S. and Olsen, Knut and O’Mullane, William and Ortiz, Sandra and Osier, Shawn and Owen, Russell E. and Pain, Reynald and Palecek, Paul E. and Parejko, John K. and Parsons, James B. and Pease, Nathan M. and Peterson, J. Matt and Peterson, John R. and Petravick, Donald L. and Petrick, M. E. Libby and Petry, Cathy E. and Pierfederici, Francesco and Pietrowicz, Stephen and Pike, Rob and Pinto, Philip A. and Plante, Raymond and Plate, Stephen and Plutchak, Joel P. and Price, Paul A. and Prouza, Michael and Radeka, Veljko and Rajagopal, Jayadev and Rasmussen, Andrew P. and Regnault, Nicolas and Reil, Kevin A. and Reiss, David J. and Reuter, Michael A. and Ridgway, Stephen T. and Riot, Vincent J. and Ritz, Steve and Robinson, Sean and Roby, William and Roodman, Aaron and Rosing, Wayne and Roucelle, Cecille and Rumore, Matthew R. and Russo, Stefano and Saha, Abhijit and Sassolas, Benoit and Schalk, Terry L. and Schellart, Pim and Schindler, Rafe H. and Schmidt, Samuel and Schneider, Donald P. and Schneider, Michael D. and Schoening, William and Schumacher, German and Schwamb, Megan E. and Sebag, Jacques and Selvy, Brian and Sembroski, Glenn H. and Seppala, Lynn G. and Serio, Andrew and Serrano, Eduardo and Shaw, Richard A. and Shipsey, Ian and Sick, Jonathan and Silvestri, Nicole and Slater, Colin T. and Smith, J. Allyn and Smith, R. Chris and Sobhani, Shahram and Soldahl, Christine and Storrie-Lombardi, Lisa and Stover, Edward and Strauss, Michael A. and Street, Rachel A. and Stubbs, Christopher W. and Sullivan, Ian S. and Sweeney, Donald and Swinbank, John D. and Szalay, Alexander and Takacs, Peter and Tether, Stephen A. and Thaler, Jon J. and Thayer, John Gregg and Thomas, Sandrine and Thornton, Adam J. and Thukral, Vaikunth and Tice, Jeffrey and Trilling, David E. and Turri, Max and Berg, Richard Van and Berk, Daniel Vanden and Vetter, Kurt and Virieux, Francoise and Vucina, Tomislav and Wahl, William and Walkowicz, Lucianne and Walsh, Brian and Walter, Christopher W. and Wang, Daniel L. and Wang, Shin-Yawn and Warner, Michael and Wiecha, Oliver and Willman, Beth and Winters, Scott E. and Wittman, David and Wolff, Sidney C. and Wood-Vasey, W. Michael and Wu, Xiuqin and Xin, Bo and Yoachim, Peter and Zhan, Hu},
   year={2019},
   month=mar, pages={111} }

@misc{spergel2013widefieldinfraredsurveytelescopeastrophysics,
      title={Wide-Field InfraRed Survey Telescope-Astrophysics Focused Telescope Assets WFIRST-AFTA Final Report}, 
      author={D. Spergel and N. Gehrels and J. Breckinridge and M. Donahue and A. Dressler and B. S. Gaudi and T. Greene and O. Guyon and C. Hirata and J. Kalirai and N. J. Kasdin and W. Moos and S. Perlmutter and M. Postman and B. Rauscher and J. Rhodes and Y. Wang and D. Weinberg and J. Centrella and W. Traub and C. Baltay and J. Colbert and D. Bennett and A. Kiessling and B. Macintosh and J. Merten and M. Mortonson and M. Penny and E. Rozo and D. Savransky and K. Stapelfeldt and Y. Zu and C. Baker and E. Cheng and D. Content and J. Dooley and M. Foote and R. Goullioud and K. Grady and C. Jackson and J. Kruk and M. Levine and M. Melton and C. Peddie and J. Ruffa and S. Shaklan},
      year={2013},
      eprint={1305.5422},
      archivePrefix={arXiv},
      primaryClass={astro-ph.IM},
      url={https://arxiv.org/abs/1305.5422}, 
}

@ARTICLE{SZoriginal,
       author = {{Sunyaev}, R.~A. and {Zeldovich}, Ya. B.},
        title = "{The Observations of Relic Radiation as a Test of the Nature of X-Ray Radiation from the Clusters of Galaxies}",
      journal = {Comments on Astrophysics and Space Physics},
         year = 1972,
        month = nov,
       volume = {4},
        pages = {173},
       adsurl = {https://ui.adsabs.harvard.edu/abs/1972CoASP...4..173S}
}

@misc{qu2026precisionkinematicsunyaevzeldovichmeasurements,
      title={Precision Kinematic Sunyaev--Zel'dovich Measurements Across Halo Mass and Redshift with DESI DR2 and ACT DR6: Part I. Luminous Red Galaxies}, 
      author={F. J. Qu and B. Ried Guachalla and E. Schaan and B. Hadzhiyska and S. Ferraro and J. Aguilar and S. Ahlen and A. Baleato Lizancos and D. Bianchi and D. Brooks and R. Canning and F. J. Castander and E. Chaussidon and T. Claybaugh and A. Cuceu and A. de la Macorra and B. Dey and P. Doel and A. Font-Ribera and J. E. Forero-Romero and E. Gaztañaga and S. Gontcho A Gontcho and G. Gutierrez and H. K. Herrera-Alcantar and K. Honscheid and C. Howlett and D. Huterer and M. Ishak and R. Kehoe and T. Kisner and A. Kremin and O. Lahav and M. Landriau and L. Le Guillou and M. E. Levi and M. Manera and A. Meisner and R. Miquel and S. Nadathur and J. A. Newman and W. J. Percival and I. P'erez-R`afols and G. Rossi and L. Samushia and E. Sanchez and E. F. Schlafly and D. Schlegel and M. Schubnell and H. Seo and J. Silber and D. Sprayberry and G. Tarl'e and B. A. Weaver and R. Zhou},
      year={2026},
      eprint={2604.19744},
      archivePrefix={arXiv},
      primaryClass={astro-ph.CO},
      url={https://arxiv.org/abs/2604.19744}, 
}

@article{Das_2023,
doi = {10.3847/1538-4357/acd764},
url = {https://doi.org/10.3847/1538-4357/acd764},
year = {2023},
month = {jul},
publisher = {The American Astronomical Society},
volume = {951},
number = {2},
pages = {125},
author = {Das, Sanskriti and Chiang, Yi-Kuan and Mathur, Smita},
title = {Thermal Sunyaev–Zel’dovich Effect in the Circumgalactic Medium. I. Detection, and a Surprising Pattern in Self-similarity and Baryon Sufficiency},
journal = {The Astrophysical Journal}
}

@article{Lorimer_2007,
   title={A Bright Millisecond Radio Burst of Extragalactic Origin},
   volume={318},
   ISSN={1095-9203},
   url={http://dx.doi.org/10.1126/science.1147532},
   DOI={10.1126/science.1147532},
   number={5851},
   journal={Science},
   publisher={American Association for the Advancement of Science (AAAS)},
   author={Lorimer, D. R. and Bailes, M. and McLaughlin, M. A. and Narkevic, D. J. and Crawford, F.},
   year={2007},
   month=nov, pages={777–780} }

@article{Petroff_2019,
   title={Fast radio
                            bursts},
   volume={27},
   ISSN={1432-0754},
   url={http://dx.doi.org/10.1007/s00159-019-0116-6},
   DOI={10.1007/s00159-019-0116-6},
   number={1},
   journal={The Astronomy and Astrophysics Review},
   publisher={Springer Science and Business Media LLC},
   author={Petroff, E. and Hessels, J. W. T. and Lorimer, D. R.},
   year={2019},
   month=may }

@article{Cordes_2019,
   title={Fast Radio Bursts: An Extragalactic Enigma},
   volume={57},
   ISSN={1545-4282},
   url={http://dx.doi.org/10.1146/annurev-astro-091918-104501},
   DOI={10.1146/annurev-astro-091918-104501},
   number={1},
   journal={Annual Review of Astronomy and Astrophysics},
   publisher={Annual Reviews},
   author={Cordes, James M. and Chatterjee, Shami},
   year={2019},
   month=aug, pages={417–465} }

@article{Cen_1999,
doi = {10.1086/306949},
url = {https://doi.org/10.1086/306949},
year = {1999},
month = {mar},
publisher = {},
volume = {514},
number = {1},
pages = {1},
author = {Cen, Renyue and Ostriker, Jeremiah P.},
title = {Where Are the Baryons?},
journal = {The Astrophysical Journal}
}

@article{Chisari_2019,
   title={Modelling baryonic feedback for survey cosmology},
   volume={2},
   url={http://dx.doi.org/10.21105/astro.1905.06082},
   DOI={10.21105/astro.1905.06082},
   number={1},
   journal={The Open Journal of Astrophysics},
   publisher={Maynooth University},
   author={Chisari, Nora Elisa and Mead, Alexander J. and Joudaki, Shahab and Ferreira, Pedro G. and Schneider, Aurel and Mohr, Joseph and Tröster, Tilman and Alonso, David and McCarthy, Ian G. and Martin-Alvarez, Sergio and Devriendt, Julien and Slyz, Adrianne and van Daalen, Marcel P.},
   year={2019},
   month=jun }

@article{vanDaalen2020,
    author = {van Daalen, Marcel P and McCarthy, Ian G and Schaye, Joop},
    title = {Exploring the effects of galaxy formation on matter clustering through a library of simulation power spectra},
    journal = {Monthly Notices of the Royal Astronomical Society},
    volume = {491},
    number = {2},
    pages = {2424-2446},
    year = {2020},
    month = {01},
    issn = {0035-8711},
    doi = {10.1093/mnras/stz3199},
    url = {https://doi.org/10.1093/mnras/stz3199},
    eprint = {https://academic.oup.com/mnras/article-pdf/491/2/2424/33728508/stz3199.pdf},
}

@misc{chaussidon2026measurementgalaxyvelocitypowerspectrum,
      title={Measurement of the galaxy-velocity power spectrum of DESI tracers with the kinematic Sunyaev-Zeldovich effect using DESI DR2 and ACT DR6}, 
      author={Edmond Chaussidon and Selim C. Hotinli and Simone Ferraro and Kendrick Smith and Xinyi Chen and J. Aguilar and S. Ahlen and D. Bianchi and D. Brooks and T. Claybaugh and A. Cuceu and A. de la Macorra and B. Dey and P. Doel and A. Font-Ribera and J. E. Forero-Romero and E. Gaztañaga and S. Gontcho A Gontcho and G. Gutierrez and J. Guy and H. K. Herrera-Alcantar and K. Honscheid and C. Howlett and D. Huterer and M. Ishak and R. Joyce and D. Kirkby and A. Kremin and O. Lahav and M. Landriau and L. Le Guillou and M. Manera and A. Meisner and R. Miquel and S. Nadathur and J. A. Newman and N. Palanque-Delabrouille and W. J. Percival and F. Prada and I. Pérez-Ràfols and G. Rossi and L. Samushia and E. Sanchez and D. Schlegel and M. Schubnell and H. Seo and J. Silber and D. Sprayberry and G. Tarlé and B. A. Weaver and C. Yèche and R. Zhou},
      year={2026},
      eprint={2604.04867},
      archivePrefix={arXiv},
      primaryClass={astro-ph.CO},
      url={https://arxiv.org/abs/2604.04867}, 
}

@article{Alonso_2019,
   title={A unified pseudo-<i>C</i>ℓ framework},
   volume={484},
   ISSN={1365-2966},
   url={http://dx.doi.org/10.1093/mnras/stz093},
   DOI={10.1093/mnras/stz093},
   number={3},
   journal={Monthly Notices of the Royal Astronomical Society},
   publisher={Oxford University Press (OUP)},
   author={Alonso, David and Sanchez, Javier and Slosar, Anže and },
   year={2019},
   month=Jan, pages={4127–4151} }
\bibliographystyle{aasjournalv7}

\end{document}